\documentclass{ws-procs975x65}

\begin{document}

\title{\uppercase{Fundamental physics from black holes, neutron stars and gamma-ray bursts}}

\author{REMO RUFFINI}

\address{
ICRANet, Piazzale della Repubblica 10, I-65122 Pescara, Italy\\[3pt]
Dipartimento di Fisica and ICRA, Universit\`a di Roma ``La Sapienza,'' Piazzale Aldo Moro 5, I-00185 Roma, Italy\\[3pt]
ICRANet, Universit\'e de Nice Sophia Antipolis, Grand Ch\^ateau, BP 2135, 28, avenue de Valrose, 06103 NICE CEDEX 2, France\\[3pt]
E-mail: ruffini@icra.it
}

\begin{abstract}
Gamma-ray bursts (GRBs) and supernovae (SNe) bring new perspectives to the study of neutron stars and white dwarfs, as well as opening new branches of theoretical physics and astrophysics.
\end{abstract}

\bodymatter

\section{Introduction}\label{sec:intro}

I dedicate this talk to John Archibald Wheeler recalling some crucial moments in our collaboration and focusing on two of the most energetic and transient phenomena in the universe: supernovae (SNe) and gamma-ray bursts (GRBs). From the knowledge being gained daily on these astrophysical systems a new physical and astrophysical understanding is emerging. The physics of neutron stars and black holes is rapidly evolving, new more complex astrophysical scenarios previously unimaginable are being developed and new domains of fundamental physics and astrophysics are being opened to scrutiny.

\begin{figure}[t]
\centering
\includegraphics[width=0.70\hsize,clip]{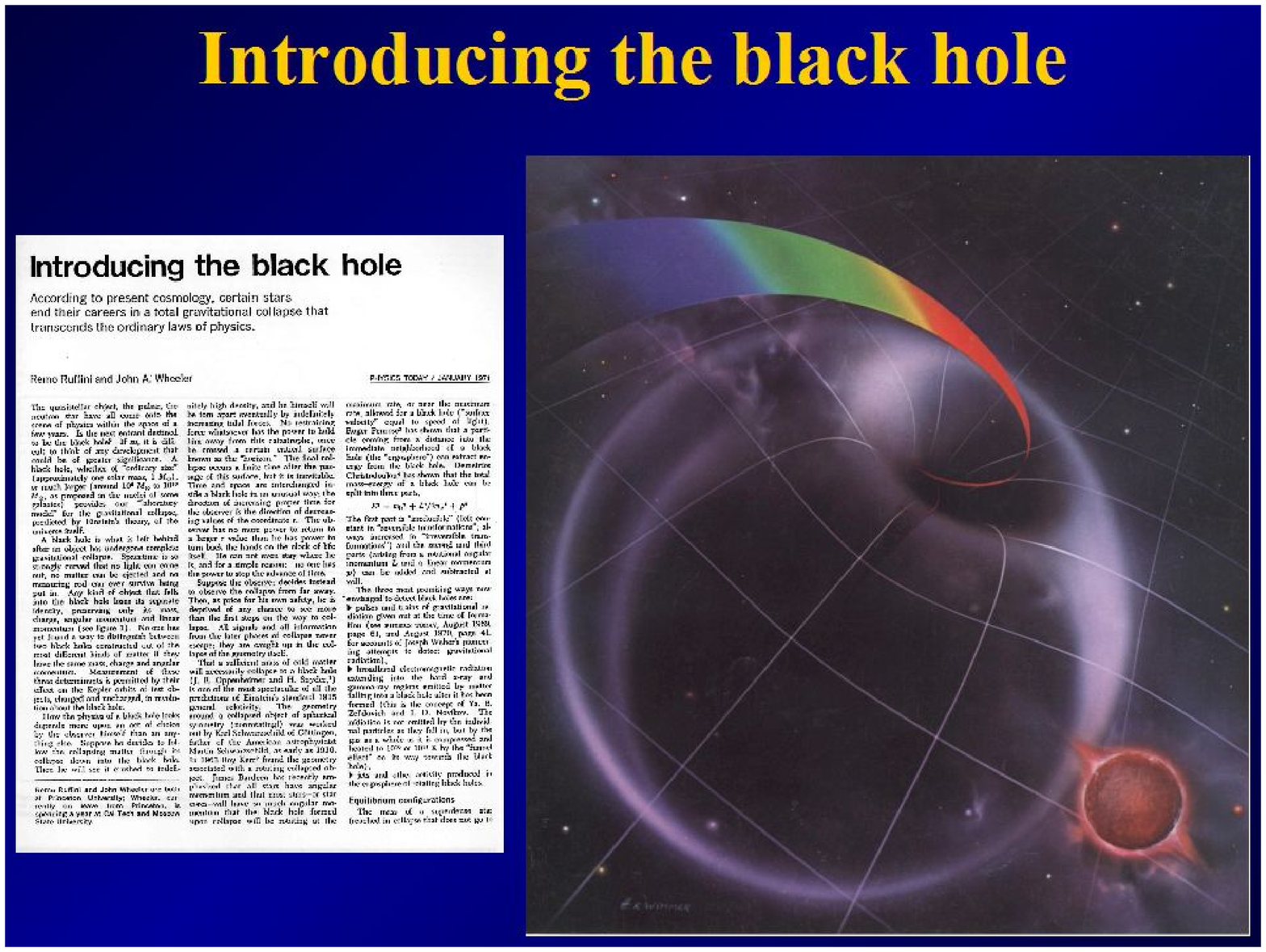}
\includegraphics[width=0.70\hsize,clip]{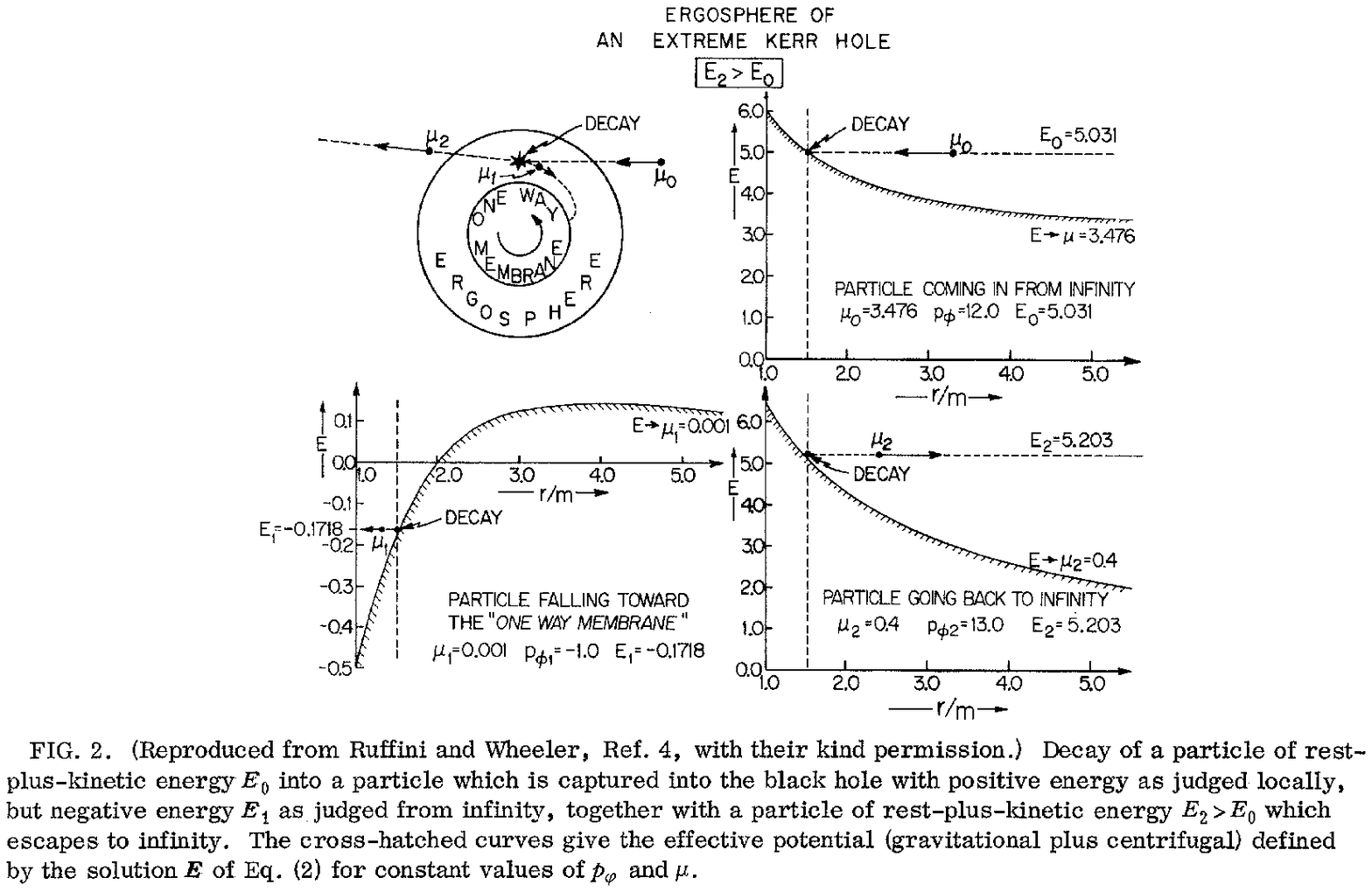}
\caption{\textbf{Above:} It was in the writing of our ESRO report \cite{1971ESRSP..52...45R} with Johnny, which later became the first chapters of our book on ``Black Holes,
Gravitational Waves and Cosmology'' \cite{1974bhgw.book.....R} that some of the crucial new ideas on the last phases of gravitational collapse leading to a black hole were formulated. They were summarized in the celebrated article ``Introducing the 
Black Hole'' \cite{1971PhT....24a..30R}. There, the crucial new concept of a ``black hole'', as a physical and astrophysical system and not just as an analytic solution of the Einstein equations, 
was presented for the first time. \textbf{Below:} The particle decay process in the field of a black hole: 
figure and original caption reproduced from 
Ref.~\refcite{1970PhRvL..25.1596C}.}
\label{fig_ibh}
\end{figure}

\begin{figure}[t]
\centering
\includegraphics[width=0.59\hsize,clip]{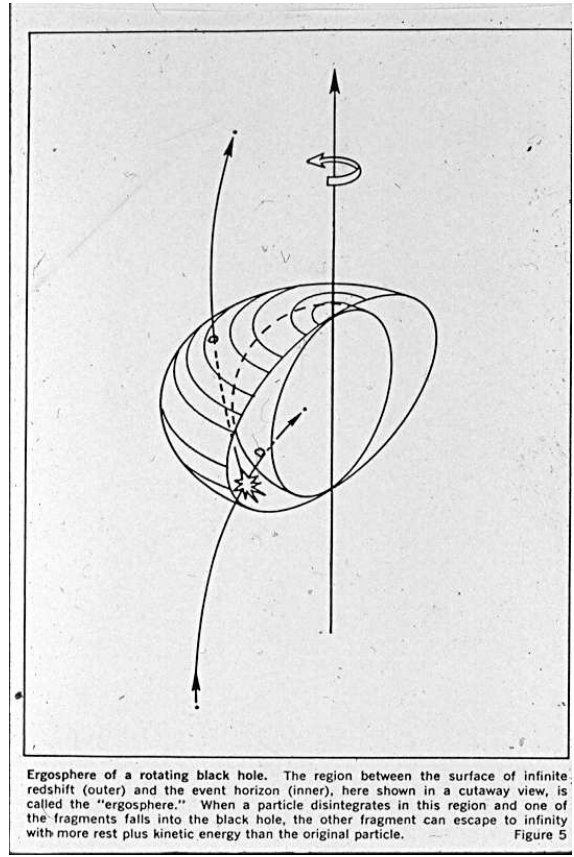}
\caption{The concept of the ``ergosphere'' was introduced \cite{1971PhT....24a..30R} as well as the concept of extraction of rotational and electromagnetic energy from a black hole with their corresponding thermodynamical analogies, further formalized in Refs.~\refcite{1970PhRvL..25.1596C,1971PhRvD...4.3552C}.}
\label{fig_irrmass}
\end{figure}

\begin{figure}[t]
\centering
\includegraphics[width=\hsize,clip]{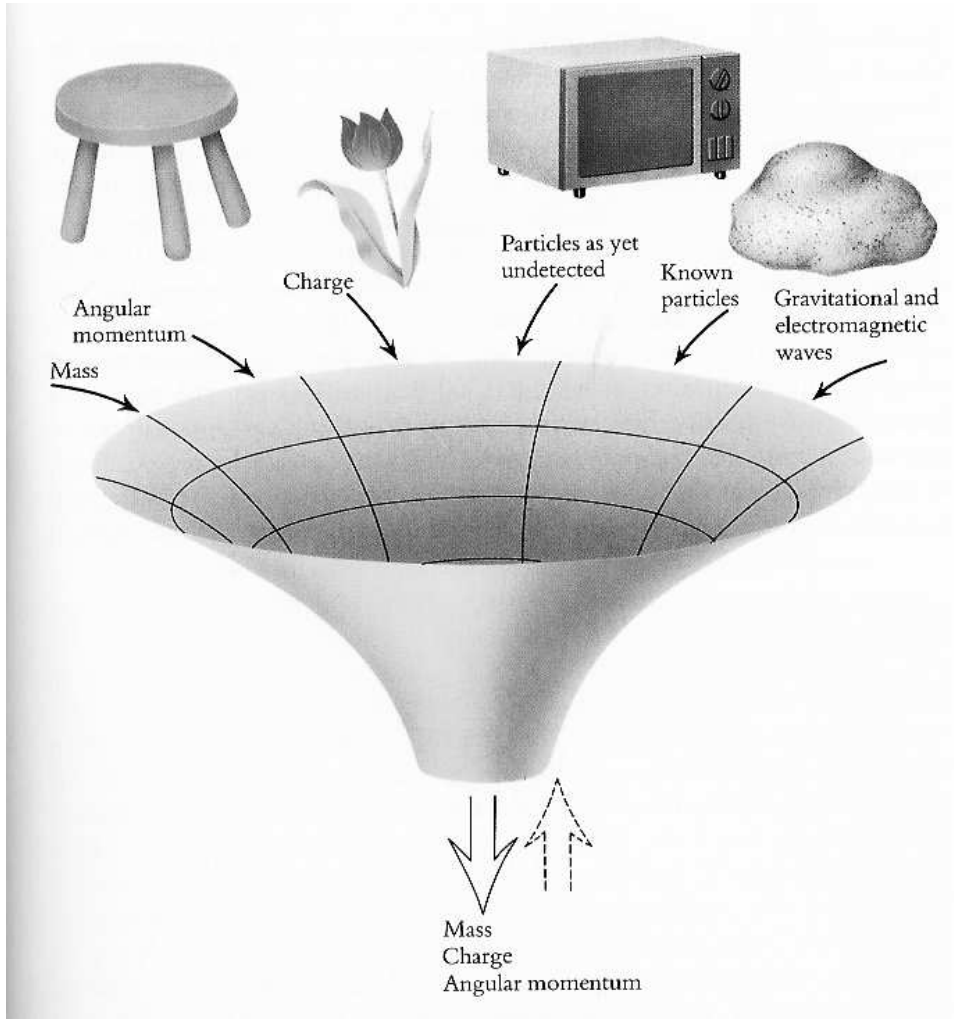}
\caption{Wheeler described another crucial property of black holes in a very provocative pictorial representation illustrating the concept of the black hole uniqueness, namely the unprecedented situation in physics that the final output of the process of gravitational collapse was only dependent on three parameters, mass, charge, and angular momentum, independent of the original details of the precursor astrophysical source. We shall see shortly how this uniqueness concept plays a fundamental role in GRBs.}
\label{fig_uniqueness}
\end{figure}

\begin{figure}[t]
\centering
\includegraphics[width=0.7\hsize,clip]{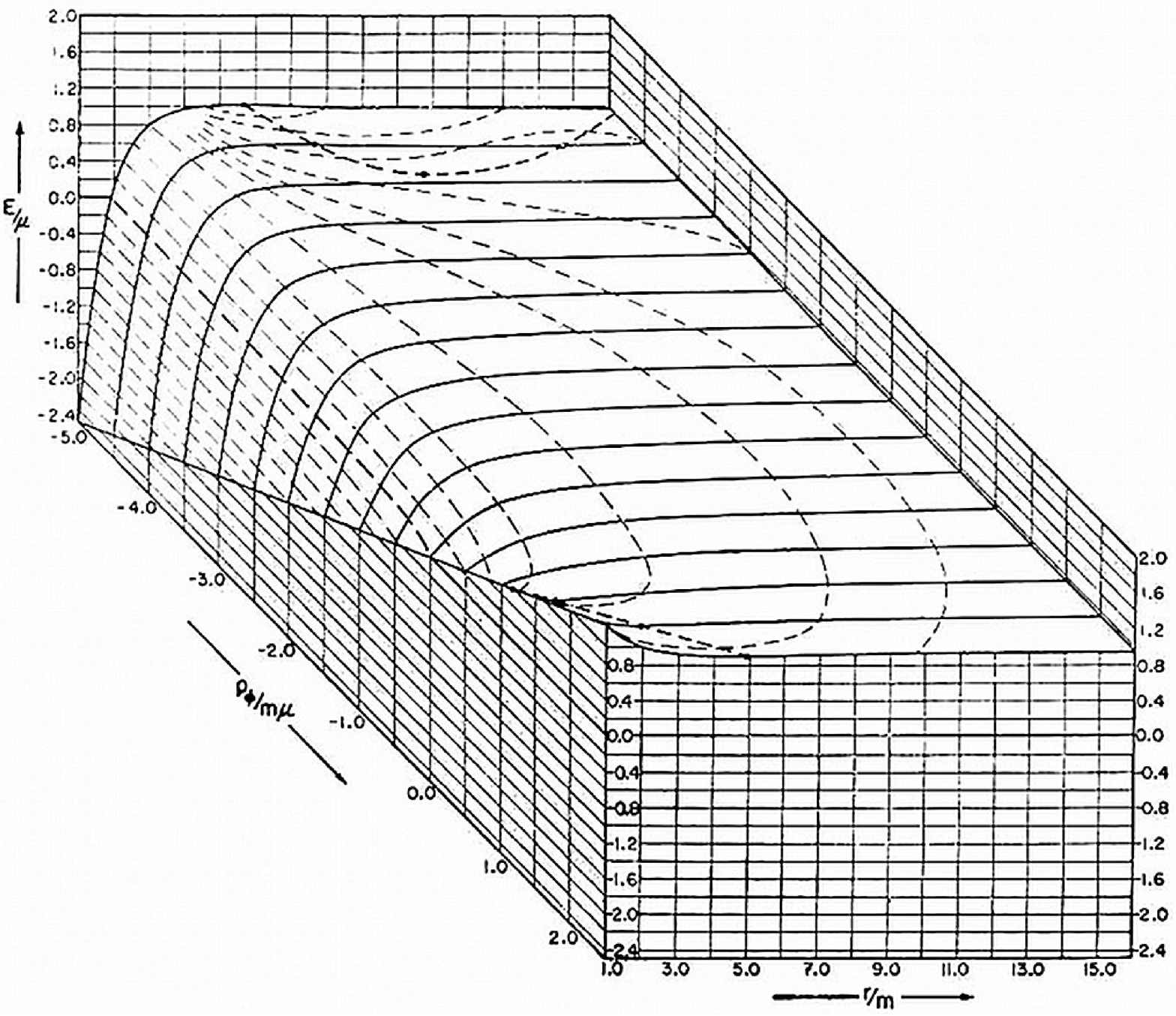}
\includegraphics[width=0.7\hsize,clip]{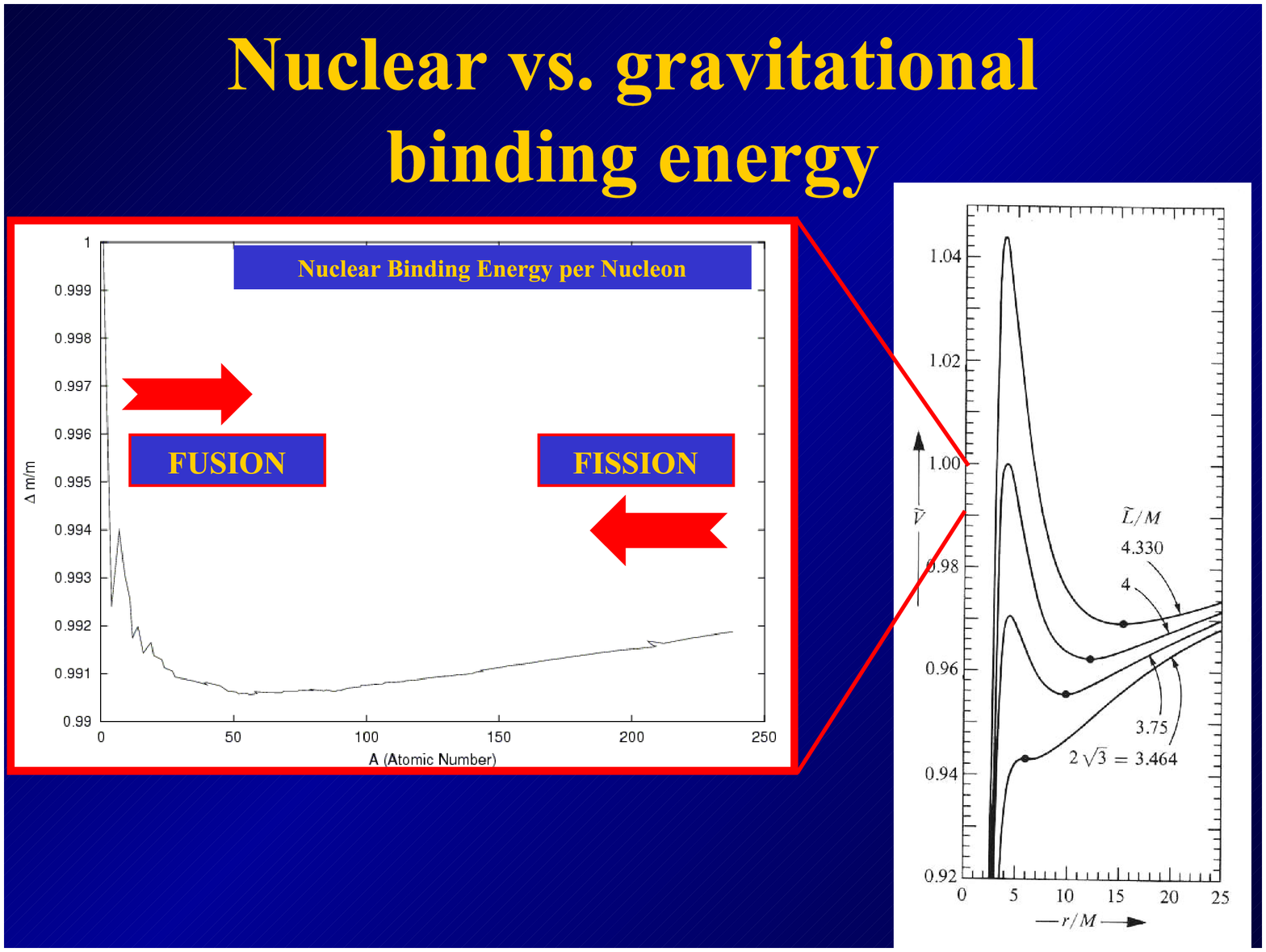}
\caption{\textbf{Above:} The effective potential experienced by a test particle moving in the equatorial plane of an extreme Kerr black hole. For corotating orbits, with positive values of the angular momentum, the maximum binding of $42.35$\% of the rest mass of the test particle is reached at the horizon. Details in Refs.~\refcite{1974bhgw.book.....R,1975ctf..book.....L}. \textbf{Below:} Nuclear vs.\ gravitational binding energy in a Schwarzschild black hole compared and contrasted. The gravitational binding energy in the Kerr case is even bigger. See Ref.~\refcite{1974bhgw.book.....R}, also quoted in Ref.~\refcite{1975ctf..book.....L}.}
\label{effpot}
\end{figure}

I will also outline some of the ideas I advanced with Wheeler in our book ``Black Holes Gravitational Waves and Cosmology'' \cite{1974bhgw.book.....R} and in our article ``Introducing the Black Hole'' \cite{1971PhT....24a..30R} (see Figs.~\ref{fig_ibh}, \ref{fig_irrmass}, \ref{fig_uniqueness}, \ref{effpot}): they are coming to full fruition and new additional topics need urgent attention.

It has been traditionally established that the understanding of an astrophysical phenomenon is only reached when its energetic aspects have been properly identified. This has certainly occurred in three cases of clear success in the 
understanding
of: 
\begin{enumerate}
\item 
the stellar evolution, which was reached after 
thermonuclear reactions were fully tested in 
Earth-bound
experiments and 
their results
applied to astrophysics (see 
Refs.~\refcite{p20,e20,1929ZPhy...52..496G,1929ZPhy...54..656A,vw37,vw38,1939PhRv...55..103B,1957RvMP...29..547B});
\item
pulsars, reached as soon as the rotational energy of neutron stars was recognized as the energy source of these phenomena (see 
Refs.~\refcite{1966ARA&A...4..393W,1969ApJ...155L.107F,1968Natur.217..709H});
\item 
the nature of binary X-ray sources, achieved as soon as the role of the gravitational energy in the fully general relativistic regime was established; the already well known results for a Schwarzschild metric were extended to the Kerr metric (see Fig.~\ref{effpot}, Ref.~\refcite{1974bhgw.book.....R}, also quoted in Ref.~\refcite{1975ctf..book.....L}, and Refs.~\refcite{1978pans.proc.....G,2003IJMPA..18.3127G}); these results were applied to the study of neutron stars and black holes in binary X-ray sources (see Ref.~\refcite{1974asgr.proc..349R}).
\end{enumerate}

If we turn from these very well understood and long duration phenomena to the most energetic and transient ones such as SNe and GRBs, it is well accepted
that gravitational energy in the general relativistic regime
plays the principal role in their explanation, namely the energy 
originating in the process of gravitational collapse either to a neutron star or to a black hole. Such a  release of gravitational energy in 
supernovae is necessary to trigger the release of thermonuclear energy processes as well as the copious emission of neutrinos leading to the formation of a neutron star. Similarly, the more extreme general relativistic conditions created in the collapse to a black hole trigger the vacuum polarization quantum process with a vast production of 
electron-positron pairs 
leading to the creation of a GRB. Nevertheless, some open issues remain in both cases. 

\subsection{Some open issues on SNe}

\begin{figure}[t]
\centering
\includegraphics[width=\hsize,clip]{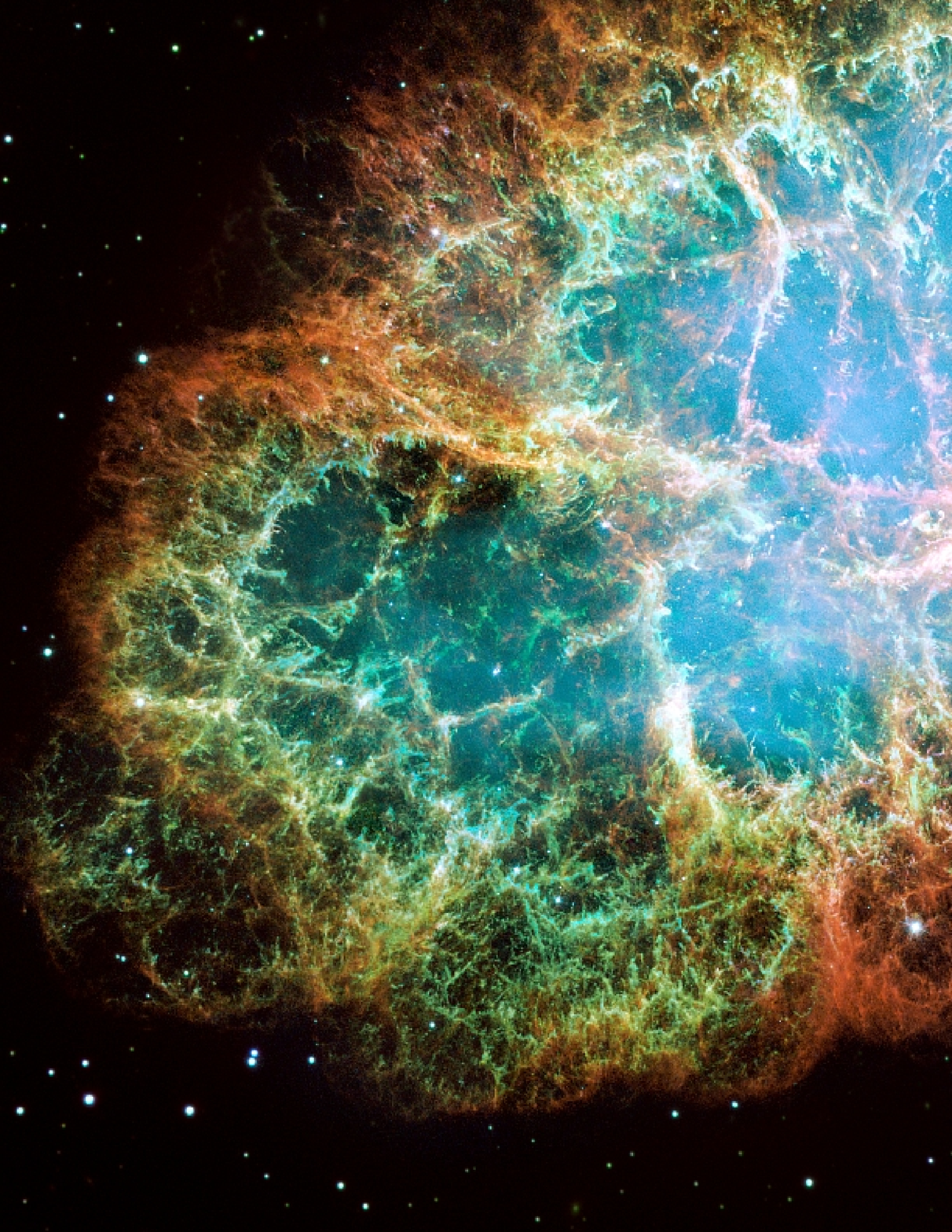}
\caption{The Crab Nebula has certainly been one of the most complex systems ever analyzed by 
the human mind.
Introduced by Messier in his famous catalog of nebulae \cite{messier} as M1, it was discovered by John Bevis and the name 
``Crab Nebula'' was attributed by William Parsons, 3rd Earl of Rosse. The first images of its filamentary structure were obtained by Baade \cite{1942ApJ....96..188B}. It was thanks to the fortunate interaction between Jaan Oort \cite{1942PASP...54...95M} and the sinologist J.J. Duyvendak \cite{1942PASP...54...91D} that it was possible to identify the Crab nebula as the remnant of the SN 
which
exploded in 1054 A.D. This event was recorded in the Chinese chronicles of the Sung dynasty and in the Japanese records Mei-Getsuki. This fascinating history has been reconstructed in the book by I.S. Shklovsky on SNe \cite{1968QB895.S6213....}.}
\label{fig_crab}
\end{figure}

For the case of SNe there is no better prototype then the Crab Nebula see Fig.~\ref{fig_crab}.

The preliminary understanding of the role of a neutron star in the process of gravitational collapse leading to a supernova and the fundamental role of supernovae in the creation of cosmic rays were presented  in a short and epochal paper by Baade and Zwicky \cite{1934PNAS...20..254B}. Following the work of George Gamow \cite{g38}, Oppenheimer and Volkoff \cite{1939PhRv...55..374O} gave the first general relativistic computations for neutron stars and introduced the existence of a  critical mass against gravitational collapse. The continued gravitational collapse, solely described by 
general relativity for a star  with mass larger than the critical mass, was then developed by Oppenheimer and Snyder \cite{1939PhRv...56..455O}. Gamow and Schoenberg \cite{1941PhRv...59..539G} identified the  fundamental role of the neutrino and antineutrino emission in order to dissipate the enormous thermal energy developed in the early phases of gravitational collapse. The neutrino-antineutrino emission, through the URCA process, introduced the essential cooling needed for the formation of a neutron star.

In a pioneering work, Hoyle and Fowler \cite{1960ApJ...132..565H} described the more complex thermonuclear process occurring in the gravitational collapse to a neutron star. They clearly differentiate two different possibilities: the gravitational collapse starting from a white dwarf, with a mass larger than the Stoner-Chandrasekhar critical mass \cite{2010arXiv1012.0154R}, leading to a type I SN; or, alternatively, the gravitational collapse starting from the larger core of a more massive star, leading to a type II SN. The very extensive work on thermonuclear processes preceding and occurring at the onset of gravitational collapse has been exhaustively summarized in the classic book of David Arnett \cite{1996sunu.book.....A}. The observations of the supernova 1987 A together with the associated neutrinos was a particularly important step in the verification of this very complex scenario \cite{1989ARA&A..27..629A}.

There are nevertheless still open questions about the release of the gravitational energy: still unexplained is the process of the expulsion of the SN remnant. 
We still lack an understanding of such a fundamental process. The possibility of explaining such 
an expulsion process by the very large flux of neutrinos emitted during the cooling of the collapsing core has been advocated by Jim Wilson and his collaborators \cite{1985ApJ...295...14B,1985nuas.conf..422W,1986NYASA.470..267W}. But it is a matter of fact that still today the theory is  unable to explain this expulsion process \cite{1996sunu.book.....A}.

My collaborators and I, working in the field of relativistic astrophysics, 
are currently focused on seeing whether the way out of this impasse might be found in some crucial properties of neutron stars which have been neglected so far in a oversimplified approach. This crucial idea may well also have been missed in the extremely complex and time consuming numerical simulations.

There is no doubt that there is plenty of gravitational energy 
available 
in the process of gravitational collapse. The gravitational energy can be 
estimated simply
by (see e.g.\ Ref.~\refcite{1967aits.book.....C} and references therein):
\begin{equation}
E = \frac{3}{5}\frac{GM^2}{R}\,,
\label{eg1}
\end{equation}
where $M$ and $R$ are 
the mass and radius of the neutron star. If one assumes for the final radius of the collapsing star
\begin{equation}
R \sim \alpha \frac{GM}{c^2}\,,
\label{eq_2}
\end{equation}
then the gravitational energy is
\begin{equation}
E \sim \frac{1}{\alpha} Mc^2\,.
\label{eg2}
\end{equation}
For $10 \lesssim \alpha \lesssim 10^2$, typical for a neutron star, one readily obtains gravitational energies on the order of $10^{52}$--$10^{53}$ erg, which can well explain the observed energy of 
SNe in the range $10^{49}$--$10^{51}$ erg. In order to utilize this gravitational energy, either for the expulsion of the remnant or for the observed electromagnetic or particle energy flux emitted during the process of gravitational collapse, it appears essential to \emph{minimize} the kinetic energy of the implosion of the core and transform the gravitational energy into an alternative form of energy with the necessary explosive power. There are two obvious mechanisms for obtaining this minimization process 
for the kinetic energy of the implosion: either the presence of rotation or the presence of new electrodynamical processes. The role of rotational energy is clearly revealed by the rotational energy of the newly formed neutron star, which is then released on time scales of thousands to millions of years in pulsars.

Turning to alternative instantaneous braking mechanisms occurring during the process of gravitational collapse, we are currently exploring the possibility of using electrodynamical processes. In our analysis, even the electrodynamics of the neutron star configurations of equilibrium, especially in the interface of the crust to the core of the neutron star, needs further scrutiny. Until now the condition of local charge neutrality in the neutron star configuration of equilibrium has been adopted mainly for mathematical simplicity (see Sec.~\ref{sec:loc_glob}). The current stringent energy release requirements, both in SN events and in GRBs, as we will see shortly, demand a deeper analysis of the electrodynamical properties during the gravitational collapse process. A good starting point is the study of the equilibrium of a neutron star imposing global but not local charge neutrality. In such cores overcritical electric fields can develop and a large amount of electron-positron plasma can in principle be created during the process of their gravitational collapse, leading to explosive effects. This field of research will be summarized briefly in the conclusions.

\begin{figure}[t]
\centering
\includegraphics[width=0.49\hsize,clip]{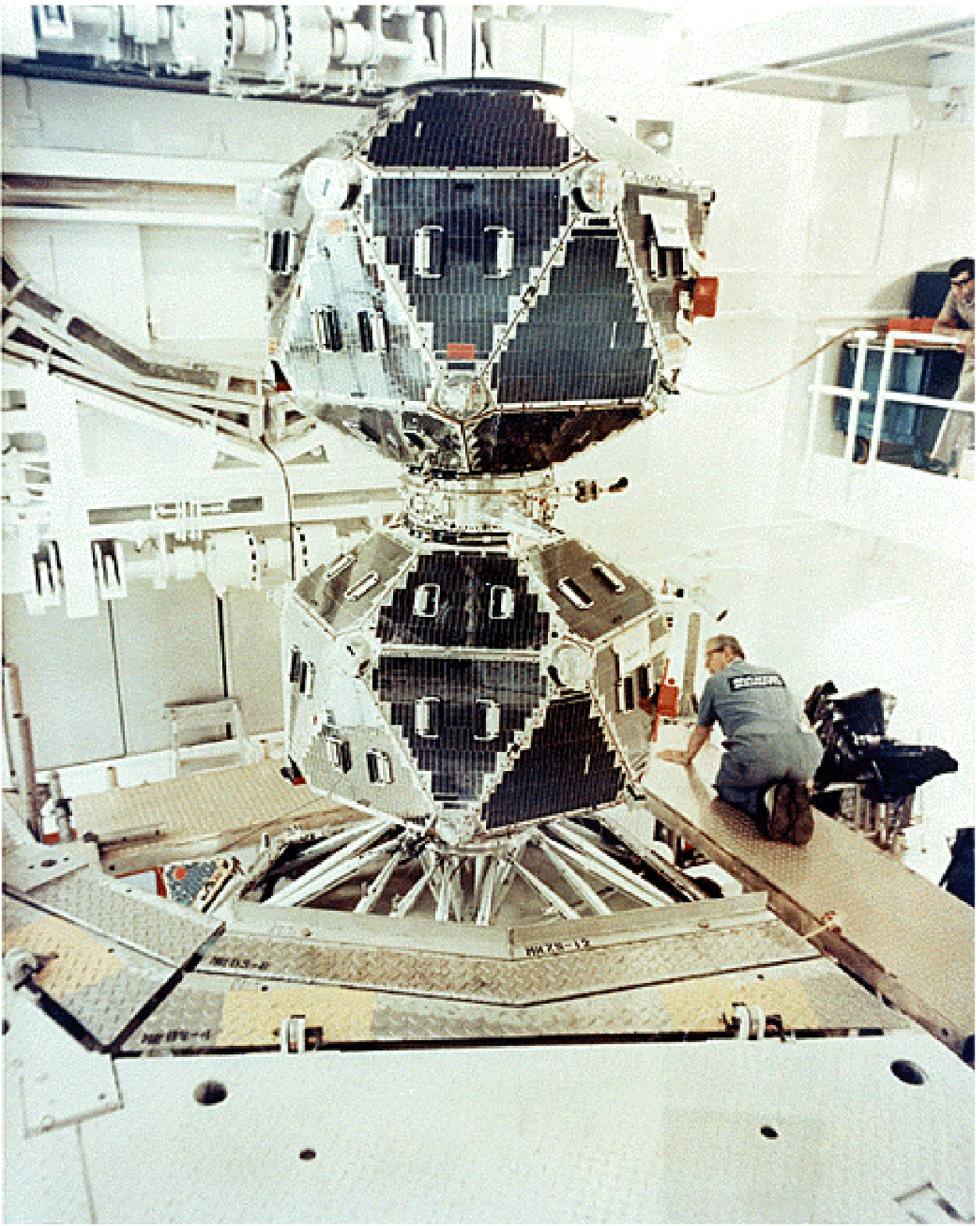}
\includegraphics[width=0.49\hsize,clip]{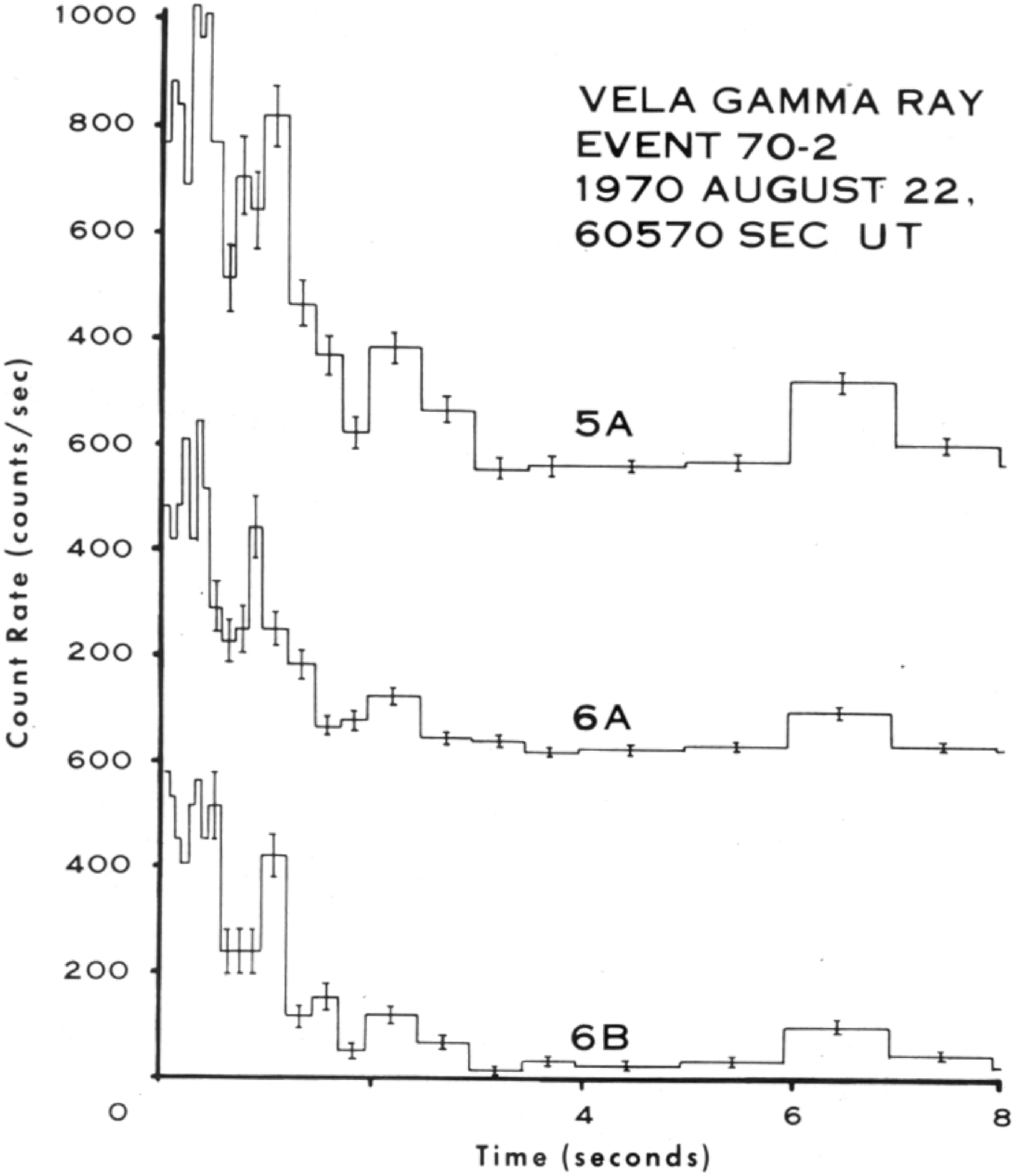}
\caption{\textbf{Left:} the Vela 5A and 5B satellites. \textbf{Right:}
a typical event as recorded by three of the Vela satellites. Details in Ref.~\refcite{1975ASSL...48.....G} by I. Strong.}
\label{fig_vela}
\end{figure}

\subsection{GRBs: The first steps of the fireshell vs.\ fireball scenario and the three paradigms}

For introducing the topic of GRBs there is no better illustration of this momentous research then the images of the original Vela Satellites (see Fig.~\ref{fig_vela}). The discovery of GRBs was a fortunate outcome of one of the most alarming episodes of the cold war between the United States of America and the 
then Soviet Union. The mission was inspired by an unconventional proposal by our colleague Yacob Borisovich Zel'dovich \cite{2010AIPC.1205....1R,ruKerr,2009AIPC.1132..199R}. The first public announcement of the GRB discovery occurred in a session I organized with Herb Gursky at the AAAS meeting in San Francisco \cite{1975ASSL...48...47S}. On the meager observational evidence of the results presented in San Francisco, I worked out a model for GRBs with Thibault Damour, based on the possibility of explaining the origin of their gamma and X-ray fluxes by the process of vacuum polarization ``a la Heisenberg-Euler-Schwinger'' occurring around a black hole endowed with electromagnetic structure \cite{1975PhRvL..35..463D}. In that paper, written a few months after the announcement of the GRB discovery, we clearly pointed out three aspects:
\begin{enumerate}
\item The role of the extractable electromagnetic energy of the black hole, or, more simply, the ``blackholic'' energy \cite{ruKerr}, as the GRB energy source, implicit in the mass-energy formula of the black hole given by Christodoulou and Ruffini \cite{1971PhRvD...4.3552C}:
\begin{equation}
\left\{
\begin{array}{l}
m^2 = \left(m_{ir} + \frac{e^2}{4m_{ir}}\right)^2 + \frac{L^2}{4m_{ir}^2}\, ,\\[6pt]
S = 16 \pi m_{ir}^2\, ,\\[6pt]
\frac{L^2}{4m_{ir}^4} + \frac{e^4}{16m_{ir}^4} \leq 1\, ,\\[6pt]
\delta S = 32 \pi m_{ir} \delta m_{ir} \ge 0\, ,
\end{array}
\right.
\label{irrmass}
\end{equation}
where $m$ is the total mass-energy of the black hole, $m_{ir}$ the irreducible mass, $S$ the surface area and $e$ and $L$ are the black hole charge and angular momentum in geometrical units ($G=c=1$). From this formula it follows that the amount of energy stored in a black hole, and in principle extractable by reversible transformations during the process of gravitational collapse, could be as high as $29\%$ of the rotational energy and $50\%$ of the electromagnetic energy of the black hole. The existence of what became known later as the ``blackholic energy'' extraction process (see Ref.~\refcite{ruKerr}) had its beginnings here. In order to have an instantaneous explosive process, we directed our attention to the electromagnetic energy component in the mass-energy formula. We expected that the extraction of rotational energy, the other form of the blackholic energy, would take place on a much longer time scale, on the order of millions of years, e.g.\ in active galactic nuclei.
\item The creation of an electron-positron pair plasma by the vacuum polarization process in the field of a Kerr-Newman black hole, representing the actual process of energy extraction from the black hole; these processes of $e^+e^-$ pair creation do approach the reversible transformations of a black hole introduced in Ref.~\refcite{1971PhRvD...4.3552C}.
\item The identification of a typical ``energy scale'' 
on the order of $\sim 10^{54} M_{BH}/M_\odot$ erg, where $M_{BH}$ is the black hole mass, as the typical energy of a GRB to be expected in our model.
\end{enumerate}
The theoretical background of this work was well grounded:
\begin{itemize}
\item on the mathematical side the solution of the Einstein-Maxwell equations found by Roy Kerr \cite{1963PhRvL..11..237K}, and by Ted Newman and collaborators \cite{1965JMP.....6..918N}, as well as, on the physical side, the mass-energy formula of the Kerr-Newman black holes (see Eq.~\ref{irrmass});
\item on the quantum physics side, the Heisenberg-Euler \cite{1936ZPhy...98..714H} and Schwinger \cite{1951PhRv...82..664S} formalism for the creation of an $e^+e^-$ pair plasma by the vacuum polarization process; the basic equations in general relativity have been properly identified in Ref.~\refcite{1975PhRvL..35..463D} and fully reviewed in Ref.~\refcite{2010PhR...487....1R};
\item on the observational side, after the discovery by the Vela satellites a large momentum was gained in the international community making GRBs a main object of the observations of a large variety of dedicated missions worldwide. One of the major issues addressed at the time was to know the actual location and distance of these sources in the universe. Just knowing their flux on the Earth, in the absence of such a distance determination, it was not possible to know their energetics. The GRB energies, in fact, could vary over an enormous range of values depending on their location in the solar system, in the galaxy, or in the distant extragalactic universe. Consequently, an enormous number of models existed \cite{ruKl}. This problem was finally overcome by the epochal observations by the BeppoSAX satellite, unequivocally confirming the typical ``energy scale'' I had previously introduced with T. Damour\cite{1997Natur.387..783C}.
\end{itemize}

\begin{figure}[t]
\centering
\includegraphics[width=0.49\hsize,clip]{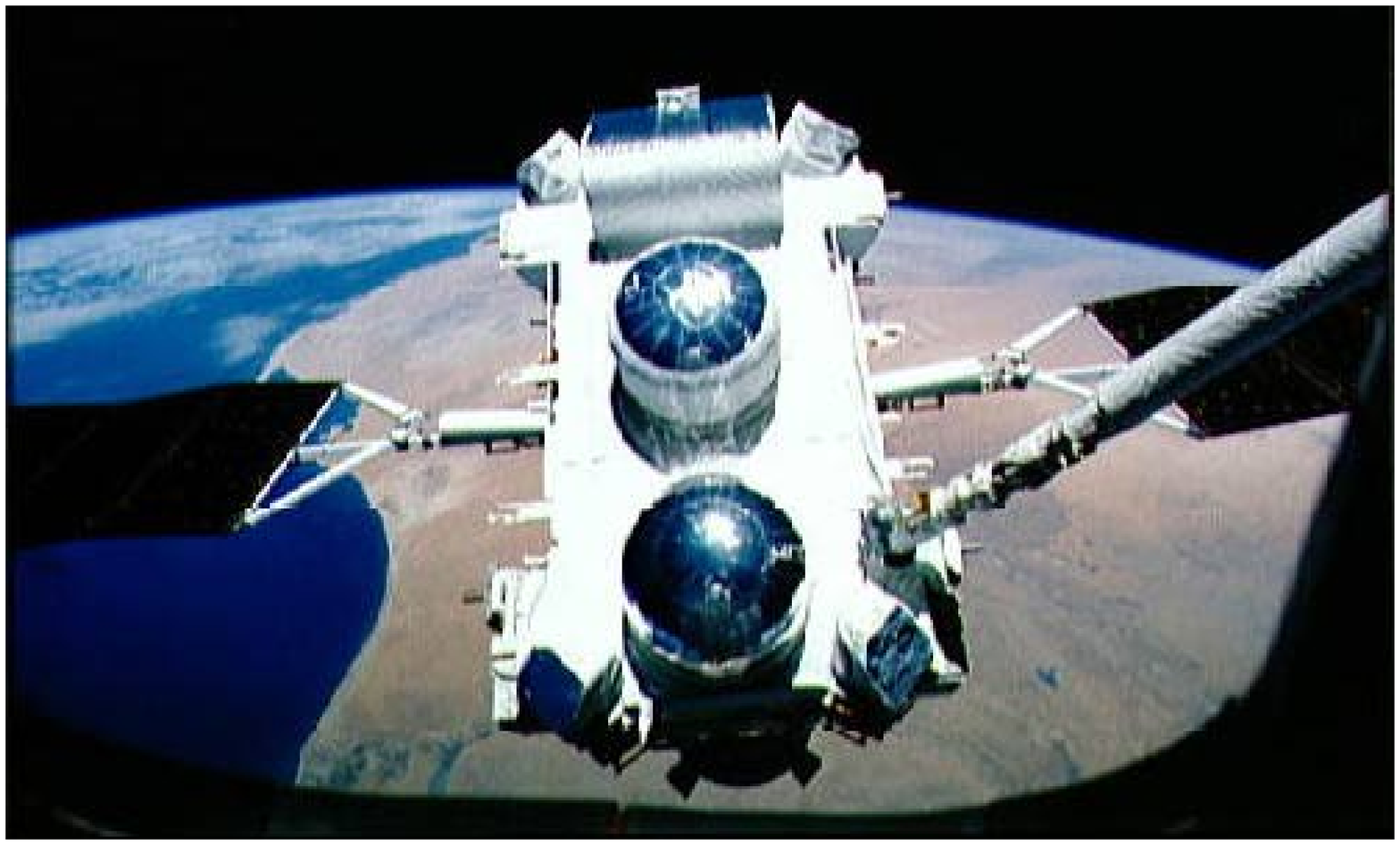}
\includegraphics[height=0.49\hsize,angle=90,clip]{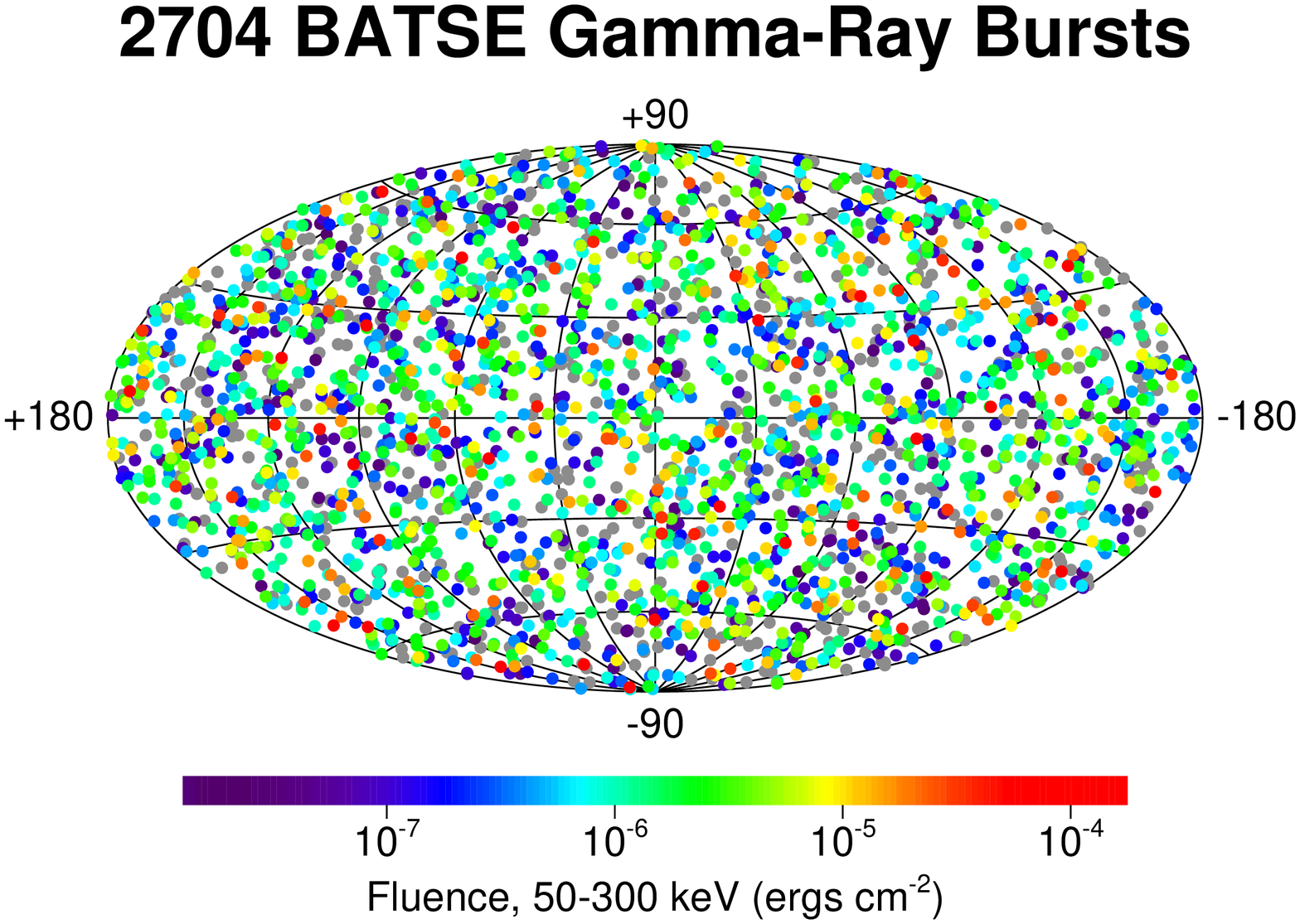}
\caption{The Compton Gamma-Ray Observatory satellite and the position in the sky of the observed GRBs in galactic coordinate. Different colors correspond to different intensities at the detector. There is almost perfect isotropy, both in the spatial and in the energetic distributions.}
\label{fig_isotr}
\end{figure}

\begin{figure}[t]
\centering
\includegraphics[width=\hsize,clip]{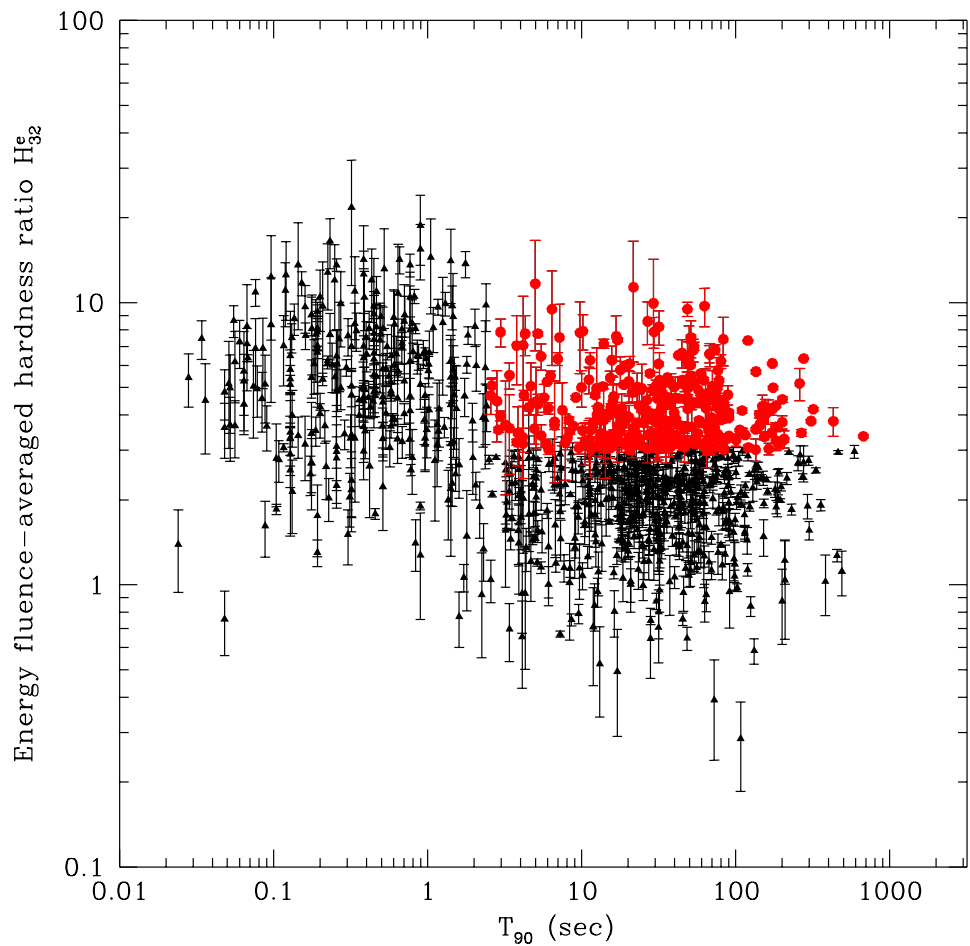}
\caption{The energy fluence-averaged hardness ratio for short ($T < 1$) and long ($T> 1$ s) GRBs are represented. Reproduced with the kind permission of M. Tavani, from Ref.~\refcite{1998ApJ...497L..21T} where the details are given.}
\label{fig_tavani}
\end{figure}

The Compton Gamma-Ray Observatory satellite, with the eight BATSE detectors, was the first of many very successful missions devoted to the study of GRBs. The outcome of these early observations led to two remarkable discoveries: the substantial isotropy of the distribution of the sources in the sky (see Fig.~\ref{fig_isotr}), as well as a clear separation of two families of events: the short GRBs, with observed duration shorter than $\sim 1$ seconds, and the long GRBs, corresponding to the complementary case (see Fig.~\ref{fig_tavani}).

\begin{figure}[t]
\centering
\includegraphics[width=\hsize,clip]{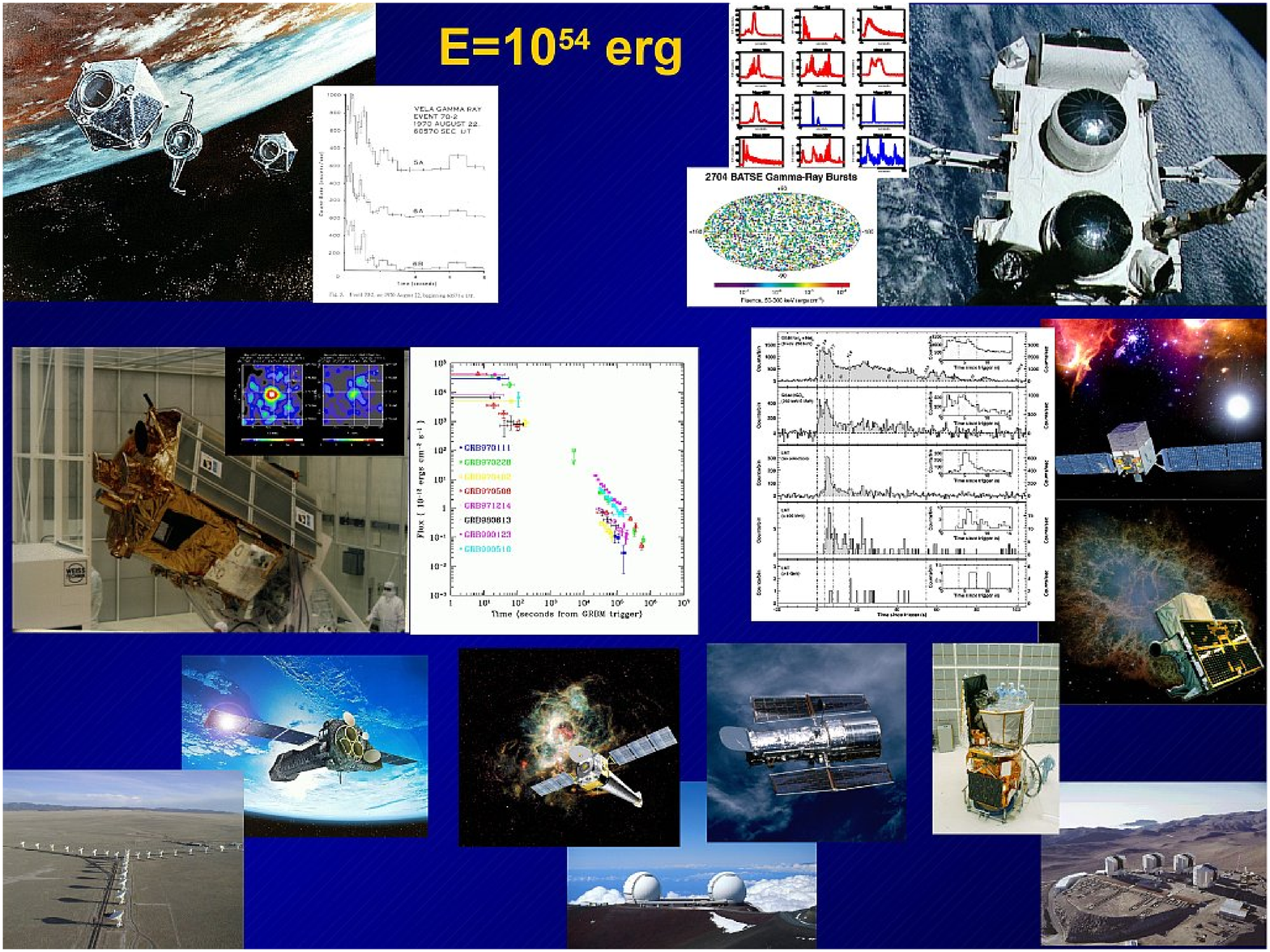}
\caption{The large flotilla of space observatories which followed the Vela satellites is here represented by the sequence of the Compton satellite and the fundamental contribution by the BeppoSAX satellite which, through its discovery of the afterglow, has allowed the determination of a sharper position 
for GRBs and consequently the possibility to identify their optical and radio counterparts with many space and Earth-based telescopes. This has allowed the determination of their cosmological distance and of their isotropic equivalent energy on the order of $10^{54}$ erg. Currently, the satellites XMM, Chandra, Swift, Fermi and Agile are giving new fundamental contributions on the energy, time variability and spectra of GRBs.}
\label{fig_satelliti}
\end{figure}

The number of dedicated missions soon increased (see Fig.~\ref{fig_satelliti}). The crucial discovery came from the Italian-Dutch satellite BeppoSAX \cite{1997Natur.387..783C}, jointly operating, for the first time, in both the X-ray and gamma-ray domain. This allowed the ``prompt radiation'' observed in gamma-rays to be associated with a long-lasting ($10^6$--$10^7$ s) ``afterglow'' observed in X-rays. This extremely important discovery has led to a better understanding of the nature of the GRB sources and offered the essential tool for improving their localization in the sky. It triggered a coordinated and complementary set of observations in the optical and radio energy bands: in turn, the optical identification, made possible by the timely development of the most powerful family of optical telescopes at Mauna Kea and at the VLT in Chile, allowed the determination of the GRB distance and therefore of their energetics (see Fig.~\ref{fig_satelliti}). This led to distances of the sources with $z$ in the range between $z \sim 0.0084$ and $z \sim 8.2$ \cite{2009ARA&A..47..567G,2009Natur.461.1254T,2009Natur.461.1258S}, and corresponding values of the energetics all the way up to the order of $10^{54}$ erg, exactly in the range we had predicted with Thibault Damour 25 years earlier.

We soon returned to our model by introducing three major new conceptual developments:
\begin{enumerate}
\item The identification of the region around the black hole where an overcritical electric field could occur, leading to the creation of $e^+e^-$ pairs out of the vacuum. The $e^+e^-$ were assumed to be in thermal equilibrium; the first identification was performed around a Reissner-Nordstr{\o}m black hole. From the Greek name ``$\delta\acute{\upsilon}\alpha\varsigma$, $\delta\acute{\upsilon}\alpha\delta{o}\varsigma$'' for ``pairs'', the name ``dyadosphere'' was introduced \cite{1998bhhe.conf..167R,1998A&A...338L..87P} for this region, characterized by a total energy $E_{tot}^{e^\pm}$, a mean energy of the pair created on the order of $1$--$4$ MeV and a characteristic radius $10^8$--$10^9$ cm. For details see Ref.~\refcite{1998A&A...338L..87P}.
\item 
The study with Jim Wilson of the dynamics of the $e^+e^-$ pairs as the fundamental acceleration process of GRBs, reaching bulk Lorentz gamma factors $\gamma \sim 10^2$--$10^3$ (the pair-electromagnetic pulse, PEM pulse). This theoretical result was obtained in Ref.~\refcite{1999A&A...350..334R} (for details see Sec.~\ref{sec_canonical}).
\item 
Still with Jim Wilson, the study of the acceleration process of the optically thick $e^+e^-$-baryon plasma (the pair-electromagnetic-baryon pulse, PEMB pulse); this theoretical result was published in Ref.~\refcite{2000A&A...359..855R}. This analysis introduced the new concept of plasma baryon loading defined by $B=M_{B}c^{2}/E_{tot}^{e^\pm}$, where $M_B$ is the total mass of the baryons. To our surprise, we found a maximum value of the baryon loading for the existence of the electron-positron acceleration process $B \leq 10^{-2}$ (for details see Sec.~\ref{sec_canonical}).
\end{enumerate}

Some of these considerations have been consistent with and others in contrast to other articles in the contemporary literature:
\begin{enumerate}
\item 
The role of the $e^+e^-$ plasma, originally presented in our work with Damour \cite{1975PhRvL..35..463D}, was soon generally adopted, with and without references! \cite{1986ApJ...308L..47G,1990ApJ...365L..55S,1993ApJ...405..278M,1993ApJ...415..181M,1993MNRAS.263..861P,1994ApJ...430L..93R};
\item 
There has been a substantial difference between our approach, purporting an initial thermal distribution of the $e^+e^-$ plasma in the dyadosphere, and the work by Cavallo and Rees \cite{1978MNRAS.183..359C}, purporting the total annihilation of the $e^+e^-$ pairs originating in a process of gravitational collapse;
\item 
The model ``a la'' Cavallo-Rees has led to the concept of a ``fireball,'' a very hot cavity originating from the annihilation of the $e^+e^-$ pairs pushing on the surrounding circumstellar matter. In contrast, our model has led to the alternative concept of the ``fireshell'' in which the gradual annihilation of the $e^+e^-$ pairs lead to a self-acceleration of an optically thick shell \cite{1999A&A...350..334R,2000A&A...359..855R}. The cavity inside such a shell is practically at zero temperature, and the $e^+e^-$ pairs gradually annihilate lasting until the fireshell becomes optically thin \cite{2000A&A...359..855R}.
\end{enumerate}

To make this general situation even more exciting, a totally unexpected phenomena has been observed: the discovery of a GRB family that is relatively weak ($10^{49}$--$10^{51}$ erg) and close ($z < 0.17$), coincident in space and time with a SN event (see e.g.\ Ref.~\refcite{Mosca_Orale} and references therein).

The ``majority'' point of view reached consensus on three main points:
\renewcommand{\theenumi}{\Roman{enumi}}
\begin{enumerate}
\item 
That short GRBs originate from binary mergers of white dwarfs, neutron stars and/or black holes in all possible combinations \cite{2006RPPh...69.2259M,2009ARA&A..47..567G}.
\item 
That long GRBs originate from the collapse of massive stars and the observed spikes in their light curves originate in the prolonged activity of an inner engine \cite{1996ApJ...473..998F,1997ApJ...485..270S,1999ApJ...512..683F,1999PhR...314..575P}. In order to make the energy requirement less stringent they postulated the existence of a jet structure \cite{1999ApJ...519L..17S,1999ApJ...526..707P}; in particular, the existence of a standard energetics 
for all GRBs of $\sim 10^{51}$ erg \cite{2001ApJ...562L..55F} was also claimed.
\item
The observations of SNe associated only with long GRBs was considered a clear support for the idea that all long GRBs originate from supernovae in the collapse of very massive stars: the ``collapsar'' model \cite{1993ApJ...405..273W,2006ARA&A..44..507W}.
\end{enumerate}
\renewcommand{\theenumi}{\arabic{enumi}}

Some of aspects of our work have offered a different point of view. We expressed them in three basic paradigms:
\begin{enumerate}
\item The relative space-time transformation (RSTT) paradigm \cite{2001ApJ...555L.107R}: the first paradigm emphasizes the relevance of having the correct space-time parametrization of the source, which necessarily implies the knowledge of the equations of motion of the system and its entire worldline. This procedure was in contrast with the current practice of describing the GRB nature by a piecewise analysis.
\item The interpretation of the burst structure (IBS) paradigm \cite{2001ApJ...555L.113R}: the second paradigm emphasizes the structure of the canonical GRB as composed of a proper GRB (P-GRB) and an extended afterglow, whose characteristics are mainly dominated by the two parameters of the total energy $E_{tot}$ and of the baryon loading $B=M_Bc^2/E_{tot}$ of the $e^+e^-$ plasma (where $M_B$ is the mass of the baryon loading): in the limit of $B \to 0$ the canonical GRB would result in a short GRB and, in the opposite limit of $B \to 10^{-2}$, which is the maximum possible value \cite{2000A&A...359..855R}, it would result in a long GRB. Our model assumes spherical symmetry.
\item The GRB-supernova time sequence (GSTS) paradigm \cite{2001ApJ...555L.117R}: the third paradigm introduces the concept of ``induced gravitational collapse,'' which considered the effect of a GRB triggering the late phase evolution of a highly evolved companion star leading to a supernova (SN).
\end{enumerate}

As we will see, the consistent application of these paradigms has led to conclusions quite different from the ones (I), (II), (III) formulated by the majority, especially with reference to short GRBs and to the collapsar model.

Before introducing the recent developments in our model, I outline some general considerations out of first principles establishing an upper limit on the black hole mass and, consequently, on the energetics of GRBs. I also recall a scenario outlined with John Wheeler.

\section{General considerations on GRB energetics}\label{sec:ener}

\begin{figure}[t]
\centering
\includegraphics[width=\hsize, clip]{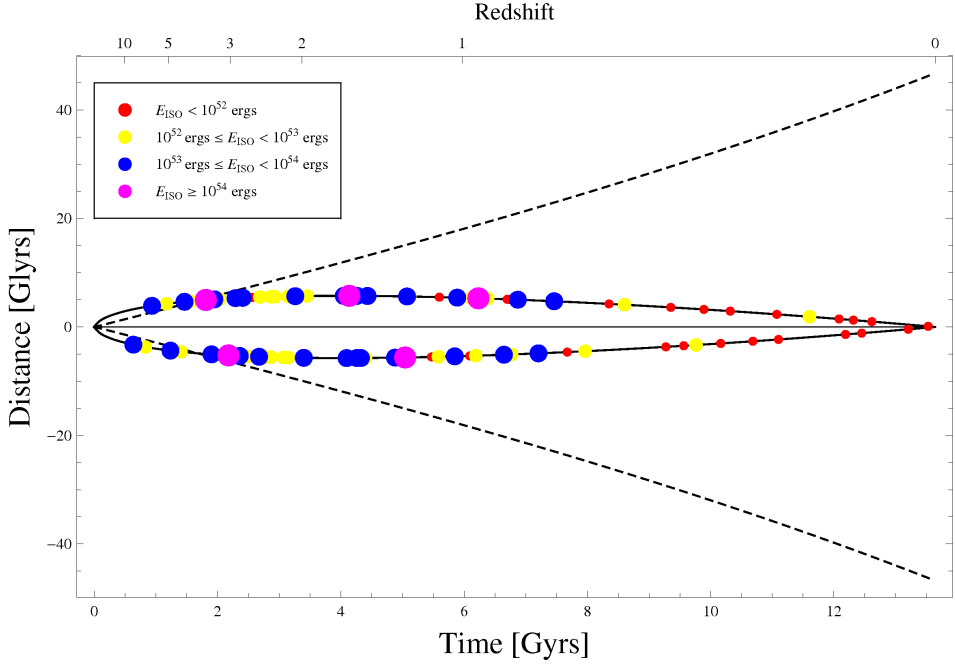}
\caption{A selected number of GRBs with energy in the range $10^{49}$--$10^{55}$ erg are represented on our past light cone as a function of their cosmological redshift and of the time from the big bang. The cosmological horizon is represented by the dashed curve.}
\label{fig_grbs_horizon_new}
\end{figure}

The observations of GRBs, made possible by the large flotilla (see Fig.~\ref{fig_satelliti}) of space observatories in the gamma-ray, X-ray and visible energy ranges as well as by Earth-bound telescopes in the visible and radio energies, has led to the discovery of a GRB almost every day. We exemplify in Fig.~\ref{fig_grbs_horizon_new} some recent representative observations of GRBs, with their energetics 
falling into the range $10^{49}$--$10^{55}$ erg.

\subsection{The upper limit to GRB energetics}

It is appropriate to return to an order of magnitude estimate to express the necessity of having a black hole of less than $\sim 10 M_\odot$ in order to explain the energetics of GRBs. From the expression for the gravitational energy in the process of gravitational collapse, given by Eq.~(\ref{eg1}), and assuming that the dimension of the gravitationally collapsed object is on the order of $R=\alpha GM/c^2$, where $\alpha$ is on the order of $1$ or $2$, typical of a black hole, the gravitational energy available is on the order of $Mc^2/\alpha$. This simple order of magnitude estimate is in agreement with the rigorous energetics defined by the mass-energy formula of the black hole (see Eq.~\ref{irrmass}). If one takes the average density of the collapsing object as:
\begin{equation}
\rho_{cc} = M/[(4/3)\pi \alpha^3 (G/c^2)^3M^3]\, ,
\label{rho_cc}
\end{equation}
and expresses it in terms of the mass $M^\star \equiv M_{pl}^3/m_n^2$, one obtains:
\begin{equation}
\rho_{cc} = m_n/[(4/3)\pi \alpha^3 (M/M^\star)^2 (\hbar^3/c^3m_n^3)]\, .
\label{rho_cc2}
\end{equation}
If one takes the nuclear density
\begin{equation}
\rho_n \equiv A m_n/[(4/3)\pi (\hbar^3/c^3m_\pi^3)]\, ,
\label{rho_n}
\end{equation}
in order to have a vacuum polarization process occur, we must have $\rho_{cc} \geq \rho_n$. This gives an upper limit to a black hole mass of:
\begin{equation}
M_{BH} \leq (m_n/m_\pi)^{3/2} / (M^\star\sqrt{A^2 \alpha^3}) \sim 1/A\alpha^3/2 \sim 9M_{\odot}\, ,
\label{m_bh}
\end{equation}
The collapsing core should have a mass on the order of $10M_{\odot}$. Correspondingly there exists an absolute upper limit to the GRB energetics at $E_{GRB} \lesssim 10^{55}$ erg.

\subsection{The role of the Kerr-Newman solution and open issues in gravitational collapse}

The great advantage of the existence of the Kerr-Newman metric is that we have an exact mathematical solution of the Einstein-Maxwell equations in which we can probe the order of magnitude estimate of the amount of energy released, the spectral distribution of the radiation, and the observed characteristic time of the process at infinity to be expected in a gravitational collapse. It is clear, however, that the use of a Kerr-Newman solution is made just for mathematical and idealized physical convenience. The real GRB astrophysical description will not have an already formed black hole. It will need, in reality, the detailed description of the process of gravitational collapse to a black hole: possibly the most difficult process to be described in physics and astrophysics. For understanding the black holes as energy sources it is essential to identify the transient process leading to building up the electromagnetic structure of the black hole forming process during the gravitational collapse. The build-up process of an overcritical electric field is ended in a sudden discharge by the quantum process leading to the formation of $e^+e^-$ pairs\cite{2009AIPC.1132..199R,2010PhR...487....1R}. When it comes to the detailed description of such a phenomenon, the necessity of broadening our current knowledge to new theoretical fields appears to be necessary. Even the simplest concept of a neutron star has to be generalized by properly accounting for new electrodynamical processes \cite{2008pint.conf..207R} (see Sec.~\ref{sec:loc_glob}). To analyze their dynamical evolution in the process of gravitational collapse is still a formidable theoretical challenge due to the interconnections of general relativity and all the relativistic field theories. The exploration of this much more complex physical and astrophysical reality and the understanding of this ``terra incognita'' is certainly one of the most exciting fields of current research. What we can certainly ascertain since now is that we are obtaining a more unified physical picture extending from the microphysical world of nuclear physics all the way to general relativistic astrophysical systems taking into due account all fundamental interactions. This novel scientific experience will likely influence also a new approach to cosmology as soon as the cosmological density overcomes nuclear densities.

On the other hand the formation of the horizon of the final almost Schwarzschild black hole, occurring after the annihilation of the electromagnetic structure of the Kerr-Newman solution into the $e^+e^-$ plasma creation, is certainly mathematically interesting but physically of smaller relevance since, by its own definition, no blackholic energy is available and it is therefore only indirectly observable by some accretion phenomena.

\begin{figure}[t]
\centering
\includegraphics[width=0.9\hsize,clip]{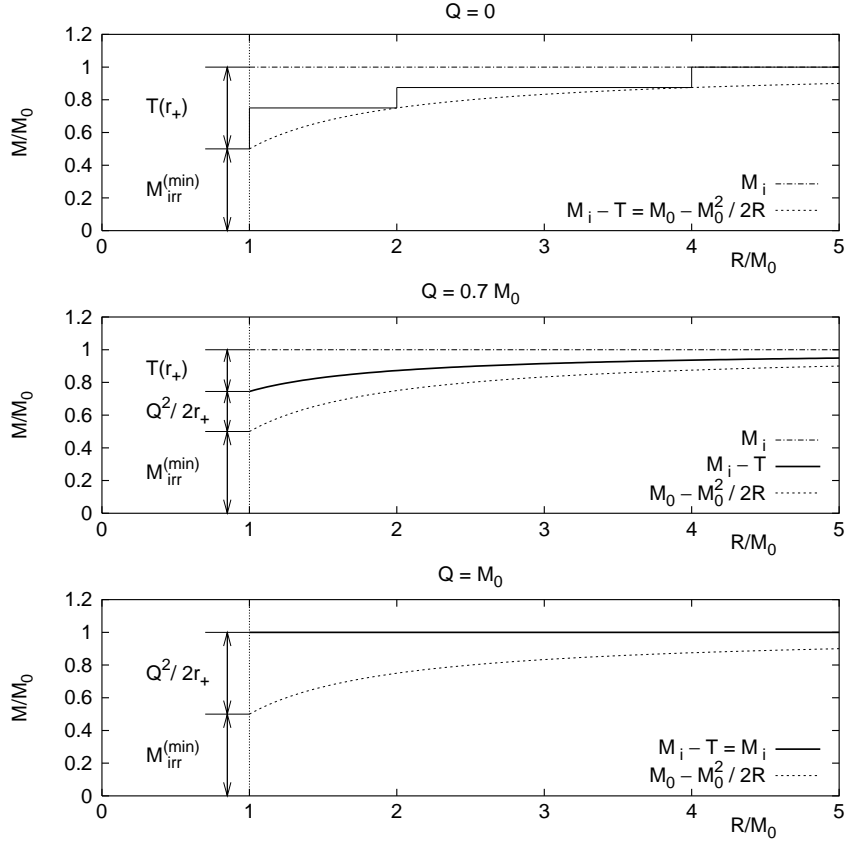}
\caption{Energetics of a shell such that $M_{\mathrm{i}}=M_{0}$, for selected values of the charge. In the first diagram $Q=0$; the dotted-dashed line represents the total energy for a gravitational collapse without any braking process as a function of the radius $R$ of the shell; the solid step function line represents a collapse with suitable braking of the kinetic energy of implosion at selected radii; the dotted line represents the rest mass energy plus the gravitational binding energy. In the second and third diagram $Q/M_{0}=0.7$, $Q/M_{0}=1$ respectively; the dotted-dashed and the dotted lines have the same meaning as above; the solid lines represent the total energy minus the kinetic energy. The region between the solid line and the dotted line corresponds to the stored electromagnetic energy. The region between the dotted-dashed line and the solid line corresponds to the kinetic energy of collapse. In all the cases the sum of the kinetic energy and the electromagnetic energy at the horizon is 50\% of $M_{0}$. Both the electromagnetic and the kinetic energy are in principle extractable. It is most remarkable that the same underlying process occurs in the three cases: the role of the electromagnetic interaction is twofold: a) to reduce the kinetic energy of implosion by the Coulomb repulsion of the shell; b) to store such an energy in the region around the black hole. The stored electromagnetic energy is extractable as shown in Ref.~\refcite{2002PhLB..545..233R} and leads to the pair creation process if the field is overcritical. See also Ref.~\refcite{2005AIPC..782...42R}.}
\label{fig3l2}
\end{figure}

Progress toward the goal of understanding the dynamics of gravitational collapse is being made (see e.g.\ Ref.~\refcite{2010PhR...487....1R}). The basic relevance of the Coulomb structure in reducing the kinetic energy of gravitational collapse and in creating an electrodynamical structure around the black hole has been highlighted in a set of papers using simplified models (see Fig.~\ref{fig3l2} as well as Ref.~\refcite{2009AIPC.1132..199R} and references therein).

The general scenario can be summarized as follows: we start with an initial configuration having global but not local charge neutrality. In the process of gravitational collapse of such
a configuration there is an increase of the electrodynamical structure. The formation of overcritical electric fields leads to the electron-positron plasma creation and the disappearance of all the electromagnetic structure \cite{2010PhR...487....1R}. Finally, a neutral black hole is left over by the process of gravitational collapse. An estimate on the order of magnitude of this process has been simulated by considering the decay of an already formed Kerr-Newman black hole endowed with an overcritical electric field \cite{2010PhR...487....1R}, to an almost Schwarzschild black hole without any electromagnetic structure.

\subsection{The black hole uniqueness theorem in GRBs}

\begin{figure}[t]
\centering
\includegraphics[width=\hsize,clip]{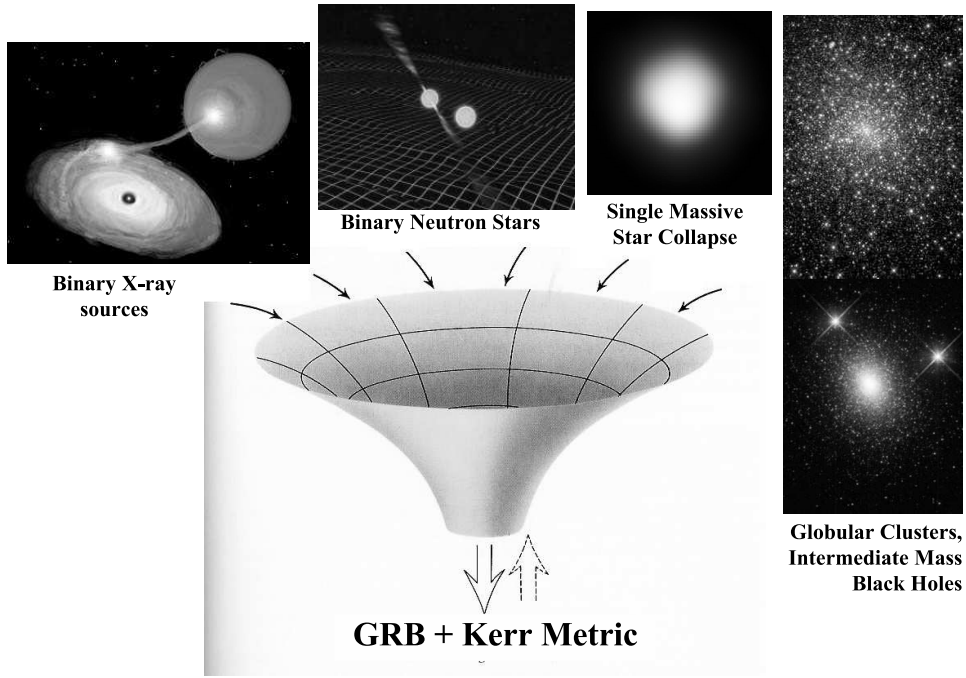}
\caption{The black hole uniqueness theorem represented by John Wheeler in Fig.~\ref{fig_uniqueness} is here applied to the case of GRB emission.}
\label{fig_uniqueness_2}
\end{figure}

In Fig.~\ref{fig_uniqueness} we have recalled the exotic way of illustrating the black hole uniqueness theorem advanced by John Wheeler and expressed in his characteristic pictorial language. Indeed, GRBs offer a most grandiose scenario to check and test the validity of this uniqueness theorem. As we have mentioned above, for reasons of energetics, the GRB energy emission process occurs in the latest phases of gravitational collapse, when the black hole horizon is approached. The GRB progenitors can originate in a vast range of possibilities: from X-ray binaries, binary neutron stars, single star collapse, black holes in the core of globular clusters, or intermediate mass black holes. As recalled above, the process of emission by GRBs occurs in the latest phases of gravitational collapse, when the black hole horizon is approached: there the majority of energy emission occurs, and this emission will be quite independent of the nature of the source. It will be uniquely characterized by the mass, the angular momentum and the electrodynamical structure of the collapsing core. In this sense, we can afford to make a unique model for all GRBs. The observed GRB properties will be only different in the time scale and spectral and energetic properties of the black hole ``dyadotorus'', which will be  defined shortly
(see Fig.~\ref{fig_uniqueness_2}). The effective electrodynamical structure of the collapsing core, idealized in our model as an already formed black hole, will generally depend on a self-consistent treatment of the strong, weak, electromagnetic and gravitational interaction of the collapsing core, and will also depend on the fermion statistical properties (see Sec.~\ref{sec:loc_glob}). The essential information on this collapsing phase will be encoded in the structure, spectrum and time evolution of the P-GRB or of the short GRBs, see Sec.~\ref{sec_canonical}.

\begin{figure}[t]
\centering
\includegraphics[width=\hsize,clip]{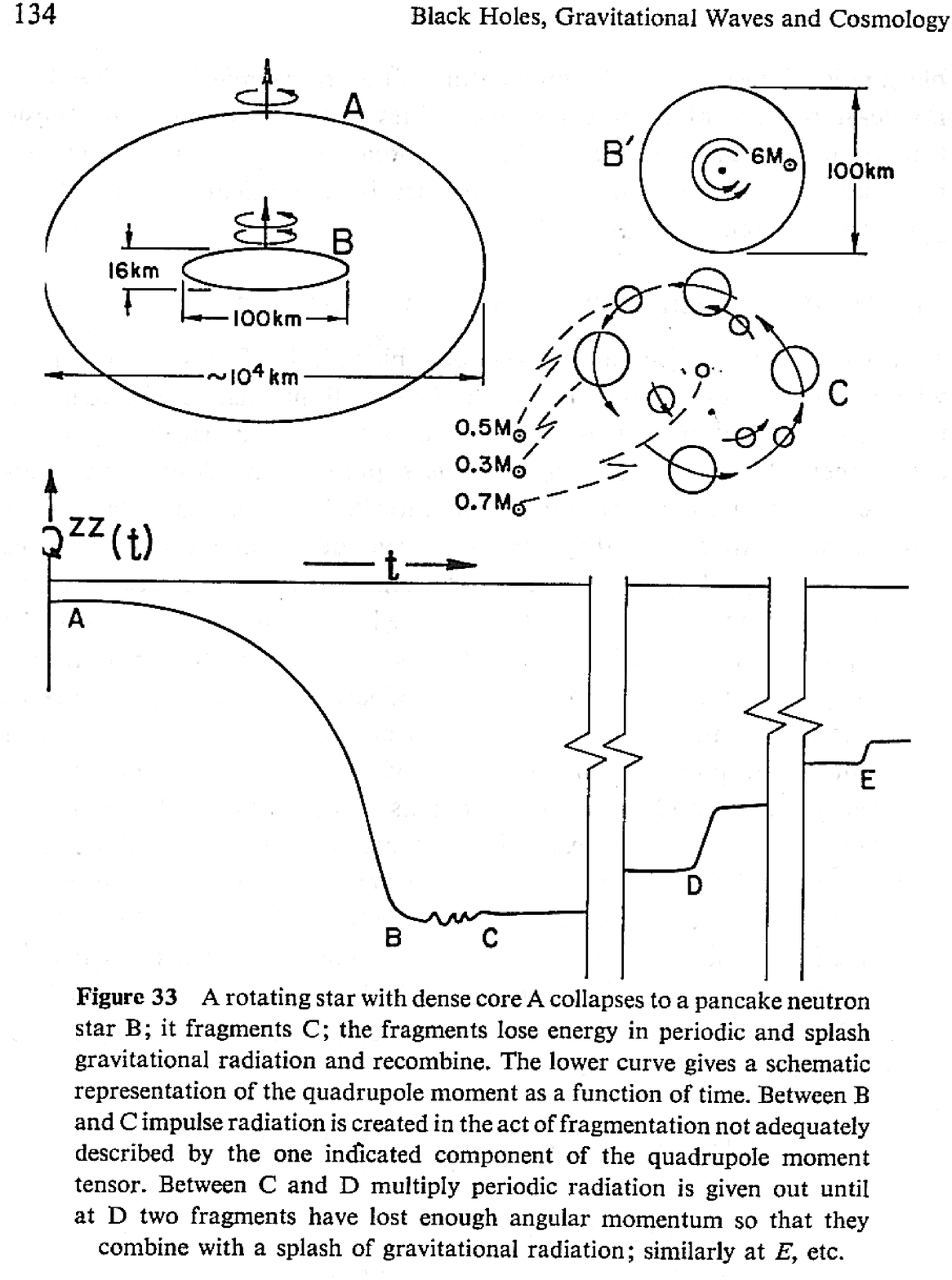}
\caption{Reproduced from Ref.~\refcite{1971ESRSP..52...45R}. See also on the self-gravitating rotating configurations Ref.~\refcite{2002PhRvD..65d4019F} in collaboration with Simonetta Filippi and references therein.}
\label{fig_pag_134}
\end{figure}

I finally recall the pursuit and plunge scenario (see Fig.~\ref{fig_pag_134}), which may become relevant in the study of possible multiple GRB phenomena.

\subsection{Summary of next sections}

I am going to outline in the next sections some of the most significant moments in our research:
\begin{enumerate}
\item the ``canonical'' GRB predicted by the fireshell model;
\item some basic progress in the theoretical understanding of fundamental physical processes motivated by the study of GRBs;
\item a new class of GRBs identified thanks to the application of the above-mentioned three paradigms.
\end{enumerate}

\section{The canonical GRB scenario in the fireshell model}\label{sec_canonical}

\begin{figure}[t]
\centering
\includegraphics[width=\hsize,clip]{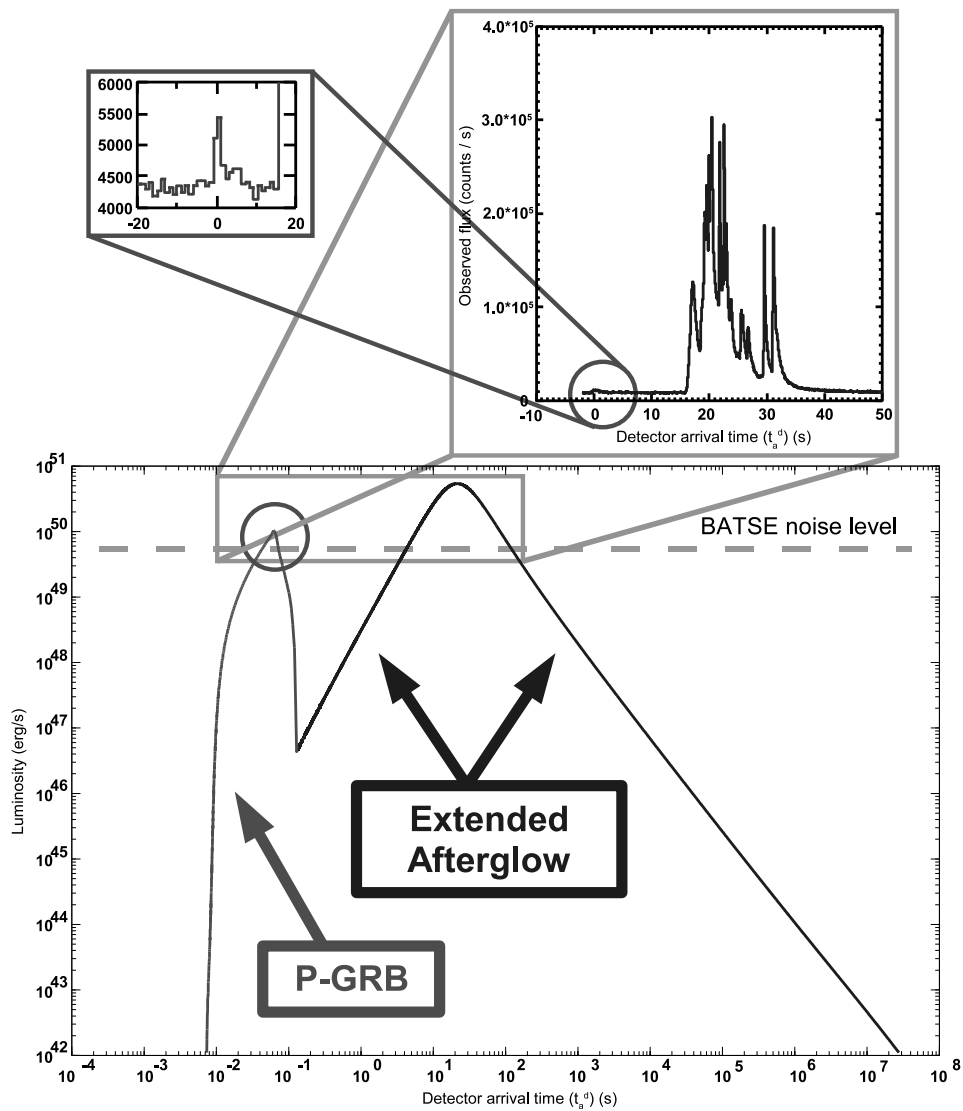}
\caption{The ``canonical GRB'' light curve  computed theoretically for GRB 991216. The prompt emission observed by BATSE is identified with the peak of the extended afterglow, while the small precursor is identified with the P-GRB. For this source we have $E_{e^\pm}^{tot} = 4.83\times 10^{53}$ erg, $B\simeq 3.0\times 10^{-3}$ and $\langle n_{cbm} \rangle \sim 1.0$ particles/cm$^3$. Details in Refs.~\refcite{2001ApJ...555L.113R,2002ApJ...581L..19R,2007AIPC..910...55R}.}
\label{canonical_991216_fig}
\end{figure}

We assume, within the fireshell model, that all GRBs originate from an optically thick $e^+e^-$ plasma with total energy $E_{tot}^{e^\pm}$ in the range $10^{49}$--$10^{54}$ erg, a temperature $T$ in the range $1$--$4$ MeV and typical radii $10^8$--$10^9$ cm \cite{1998A&A...338L..87P}. Such an $e^+e^-$ plasma has been widely adopted in the current literature (see e.g.\ Refs.~\refcite{2005RvMP...76.1143P,2006RPPh...69.2259M} and references therein).

The dynamics of the GRB develops in two very different phases: the first one, optically thick, and the second one, optically thin.

In the initial optically thick phase, after an early expansion, the $e^+e^-$-photon plasma reaches thermal equilibrium with the engulfed baryonic matter $M_B$ described by the dimensionless parameter $B=M_{B}c^{2}/E_{tot}^{e^\pm}$ that must be in the range $B < 10^{-2}$ \cite{1999A&A...350..334R,2000A&A...359..855R}. The fireshell composed of $e^+e^-$-photon-baryon plasma self-accelerates to ultrarelativistic velocities and finally reaches the transparency condition. A flash of radiation is then emitted. This is the P-GRB \cite{2001ApJ...555L.113R}. Different current theoretical treatments of these early expansion phases of GRBs are compared and contrasted in Refs.~\refcite{brvx06} and \refcite{2008AIPC.1065..219R}. The amount of energy radiated in the P-GRB is only a fraction of the initial energy $E_{tot}^{e^\pm}$.

At the decoupling between matter and radiation at the transparency point, the optically thin phase starts, characterized by a baryonic and leptonic matter fireshell endowed with ultrarelativistic Lorentz gamma factors ($200 < \gamma < 2000$). Such an accelerated optically thin fireshell gives rise to a multi-wavelength emission by inelastic collisions with the circumburst medium (CBM). This is the extended afterglow. It has three different regimes: a rising part, a peak and a decaying tail.
We therefore define a ``canonical GRB'' light curve with two sharply different components (see Fig.~\ref{canonical_991216_fig}) \cite{2001ApJ...555L.113R,2007AIPC..910...55R,2007A&A...474L..13B,2008AIPC..966...12B,2008AIPC.1000..305B}: 1) the P-GRB and 2) the extended afterglow. What is usually called ``prompt emission'' in the current literature, in our canonical GRB scenario is therefore composed of the P-GRB together with the rising part and the peak of the extended afterglow. The unjustified mixing of these two components, which originates from different physical processes, leads to difficulties in the current models of GRBs.

\subsection{The optically thick phase}

\begin{figure}[t]
\begin{minipage}{\hsize}
\centering
\includegraphics[width=0.85\hsize,clip]{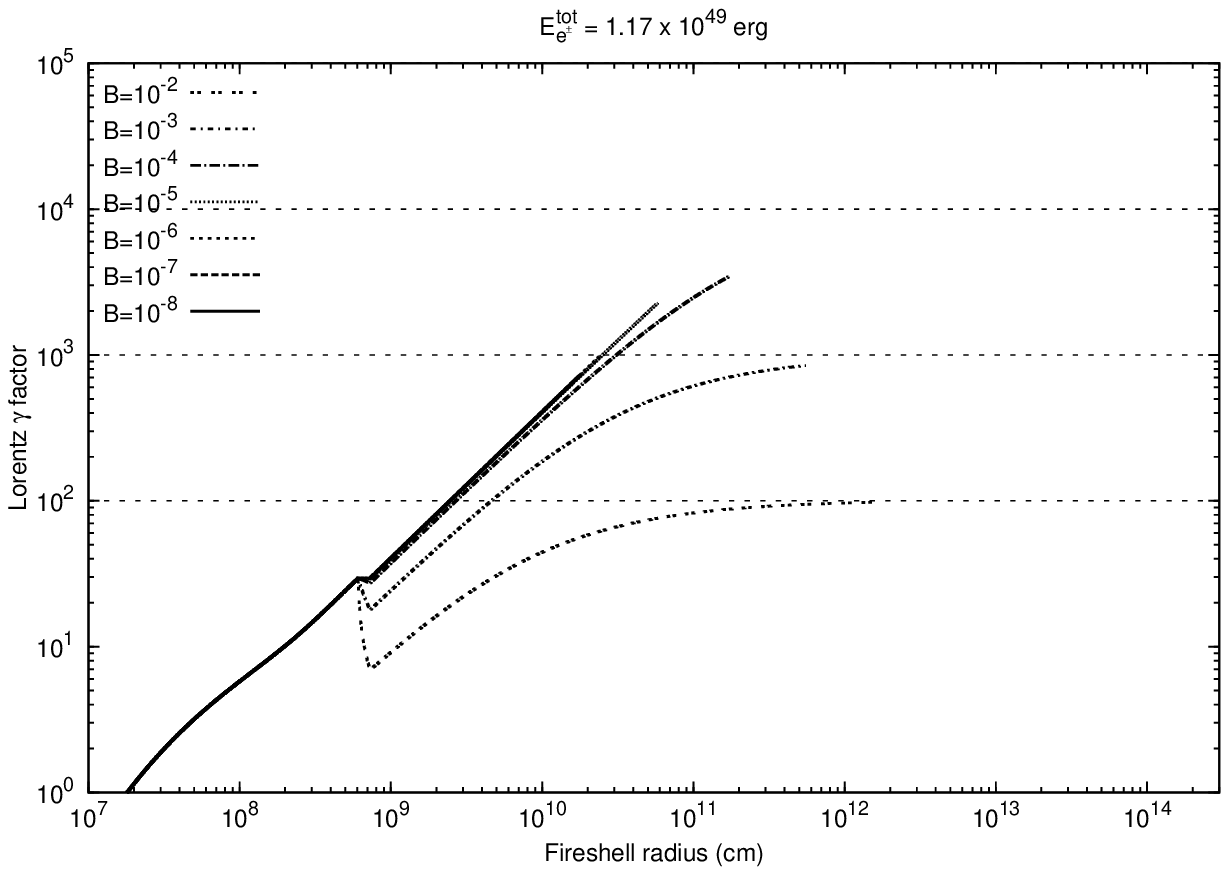}\\
\includegraphics[width=0.85\hsize,clip]{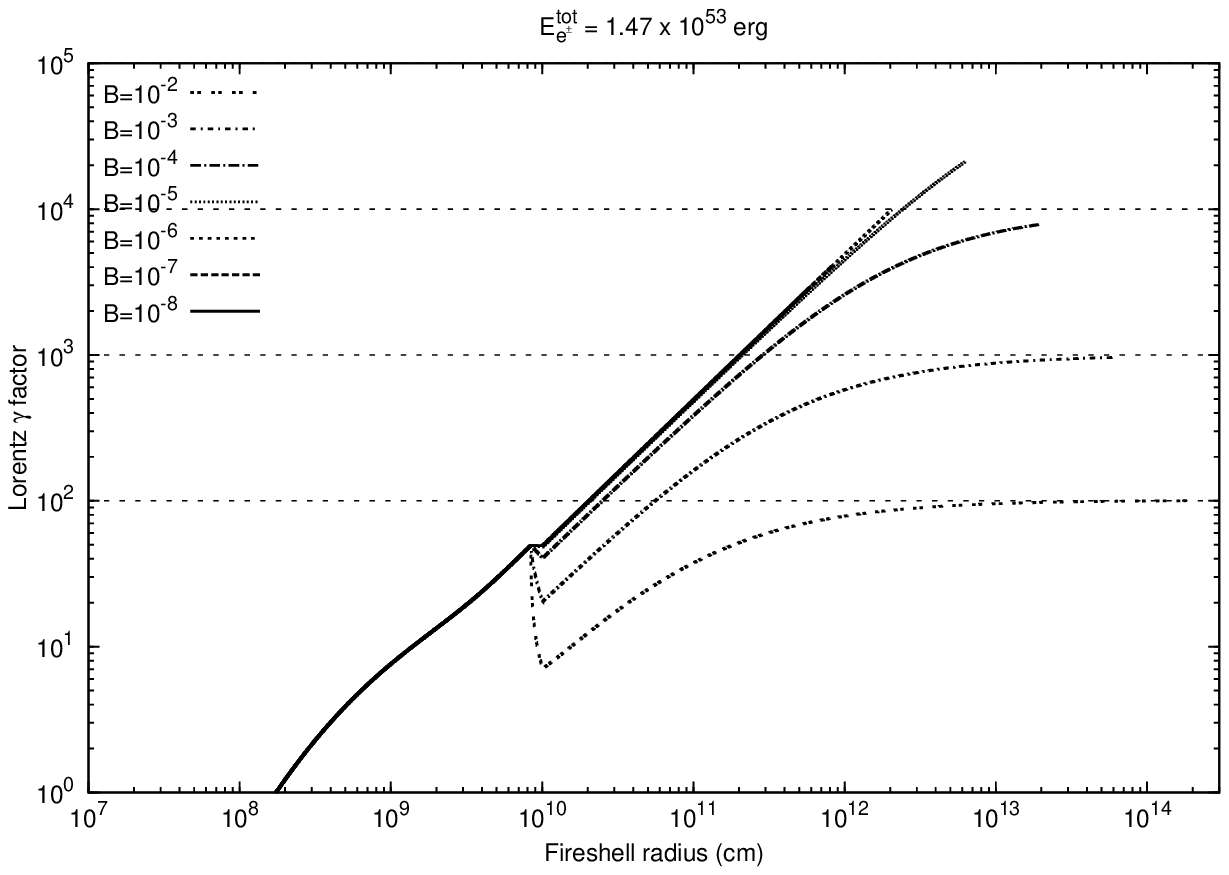}
\end{minipage}
\caption{The Lorentz gamma factor of the expanding fireshell is plotted as a function of its external radius for $7$ different values of the fireshell baryon loading $B$, ranging from $B=10^{-8}$ and $B=10^{-2}$, and two selected limiting values of the total energy $E_{e^\pm}^{tot}$ of the $e^+e^-$ plasma: $E_{e^\pm}^{tot} = 1.17\times 10^{49}$ erg (upper panel) and $E_{e^\pm}^{tot} = 1.47\times 10^{53}$ erg (lower panel). The asymptotic values $\gamma \to 1/B$ are also plotted (dashed horizontal lines). The lines are plotted up to when the fireshell transparency is reached. For details see  Ref.~\refcite{2000A&A...359..855R}.}
\label{MultiGamma}
\end{figure}

In Fig.~\ref{MultiGamma} we recall the evolution of the optically thick fireshell Lorentz gamma factor as a function of the external radius for $7$ different values of the fireshell baryon loading $B$ and two selected limiting values of the total energy $E_{e^\pm}^{tot}$ of the $e^+e^-$ plasma. We can clearly identify three different eras:
\begin{enumerate}
\item \textbf{Era I:} The fireshell is made up only of electrons, positrons and photons in thermodynamic equilibrium (the ``pair-electromagnetic pulse'', or PEM pulse for short). It self-accelerates and begins its expansion into vacuum, because the environment has been cleared by the black hole collapse. The Lorentz gamma factor increases with radius and the dynamics can be described by the energy conservation and the condition of adiabatic expansion \cite{1999A&A...350..334R,brvx06}.
\item \textbf{Era II:} The fireshell impacts with the non-collapsed bayonic remnants and engulfs them. The Lorentz gamma factor drops. The dynamics of this era can be described by imposing energy and momentum conservation during the fully inelastic collision between the fireshell and the baryonic remnant. For the fireshell solution to be still valid, $B \lesssim 10^{-2}$ must hold\cite{2000A&A...359..855R}.\\
\item \textbf{Era III:} The fireshell is now made up of electrons, positrons, baryons and photons in thermodynamic equilibrium (the ``pair-electromagnetic-baryonic pulse'', or PEMB pulse for short). It self-accelerates again and the Lorentz gamma factor increases again with radius up until when the transparency condition is reached, going to an asymptotic value $\gamma_{asym} = 1/B$. If $B \sim 10^{-2}$ the transparency condition is reached when $\gamma \sim \gamma_{asym}$. On the other hand, when $B < 10^{-2}$, the transparency condition is reached much before $\gamma$ reaches its asymptotic value \cite{2000A&A...359..855R,2001ApJ...555L.113R}. In this era the contribution of the rate equation  starts to be crucial in describing the annihilation of the $e^+e^-$ pairs:
\begin{equation}
\frac{\partial}{\partial t} N_{e^\pm} = - N_{e^\pm} \frac{1}{\cal V}\frac{\partial{\cal V}}{\partial t} + \overline{\sigma v} \frac{1}{\gamma^2}\left(N_{e^\pm}^2(T)-N_{e^\pm}^2\right)
\,,
\end{equation}
where $N_{e^\pm}$ is the number of $e^+e^-$ pairs and $N_{e^\pm}(T)$ is the number of $e^+e^-$ pairs at thermal equilibrium at temperature $T$ \cite{2000A&A...359..855R,brvx06}.
\end{enumerate}
In the ``fireball'' model in the current literature the baryons are usually considered to be present in the plasma from the very beginning. In other words, in the fireball dynamics there is only one era corresponding to the Era III above \cite{1993ApJ...415..181M,1990ApJ...365L..55S,1993MNRAS.263..861P,brvx06}. Moreover, the rate equation is usually neglected, and this affects the reaching of the transparency condition. A detailed comparison between the different approaches is given in Ref.~\refcite{brvx06}.

\subsection{The transparency point}

At the transparency point, the value of the $B$ parameter governs the ratio between the energetics of the P-GRB and the kinetic energy of the baryonic and leptonic components giving rise to the extended afterglow. It governs as well the time separation between the peak luminosities of the P-GRB and of the extended afterglow\cite{2001ApJ...555L.113R,2008AIPC.1065..219R}.

\begin{figure}[t]
\centering
\includegraphics[width=0.49\hsize,clip]{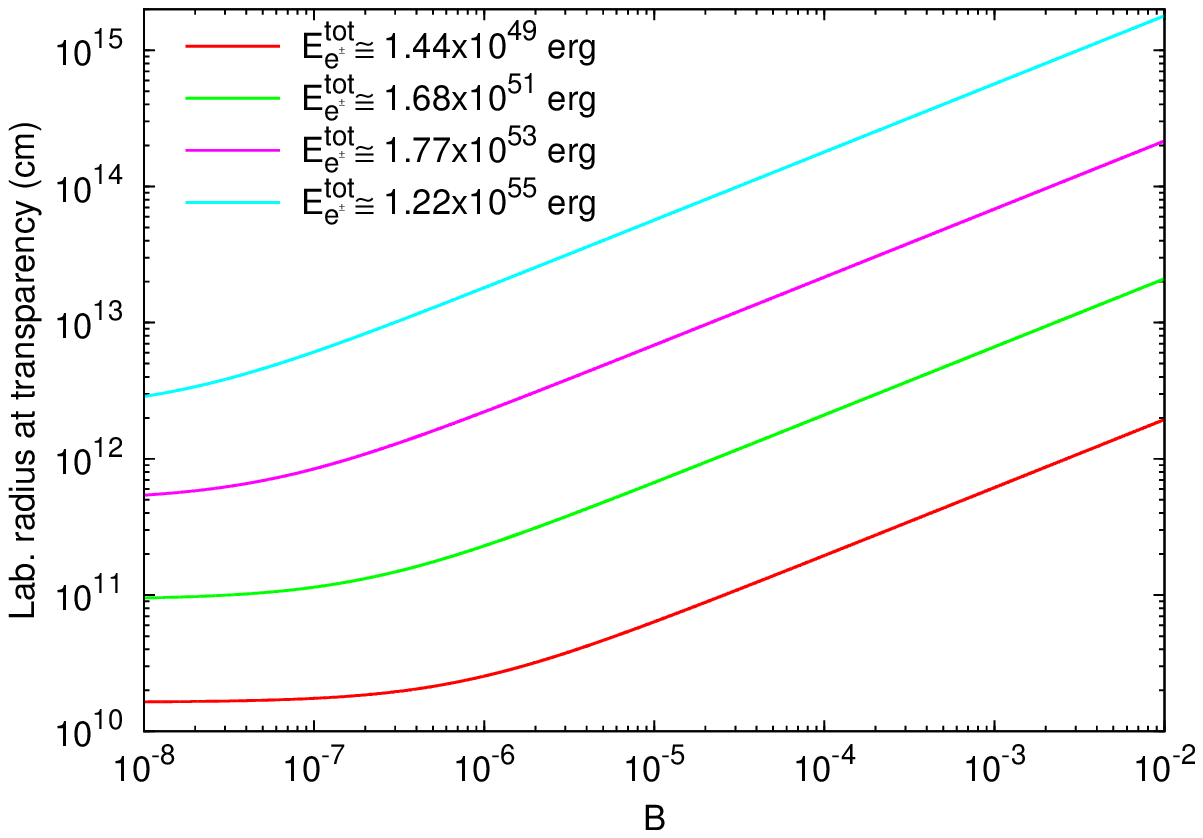}
\includegraphics[width=0.49\hsize,clip]{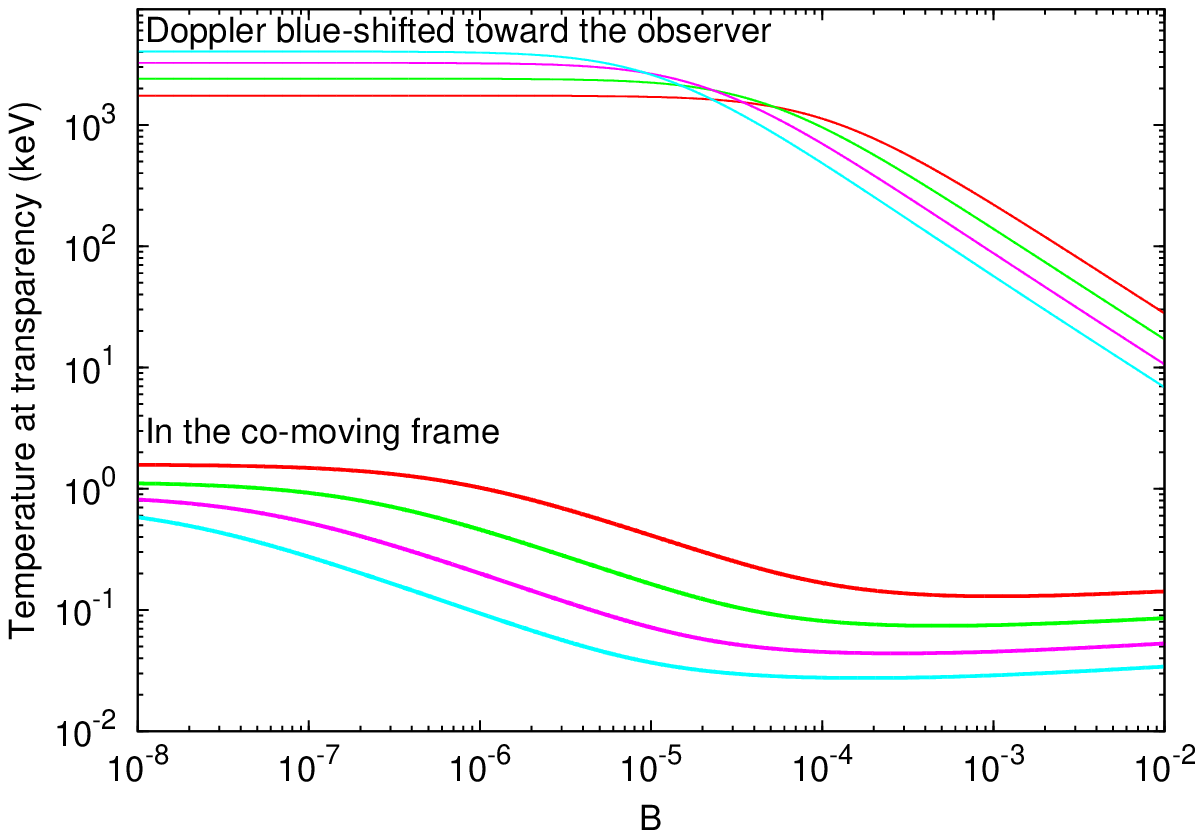}\\
\includegraphics[width=0.49\hsize,clip]{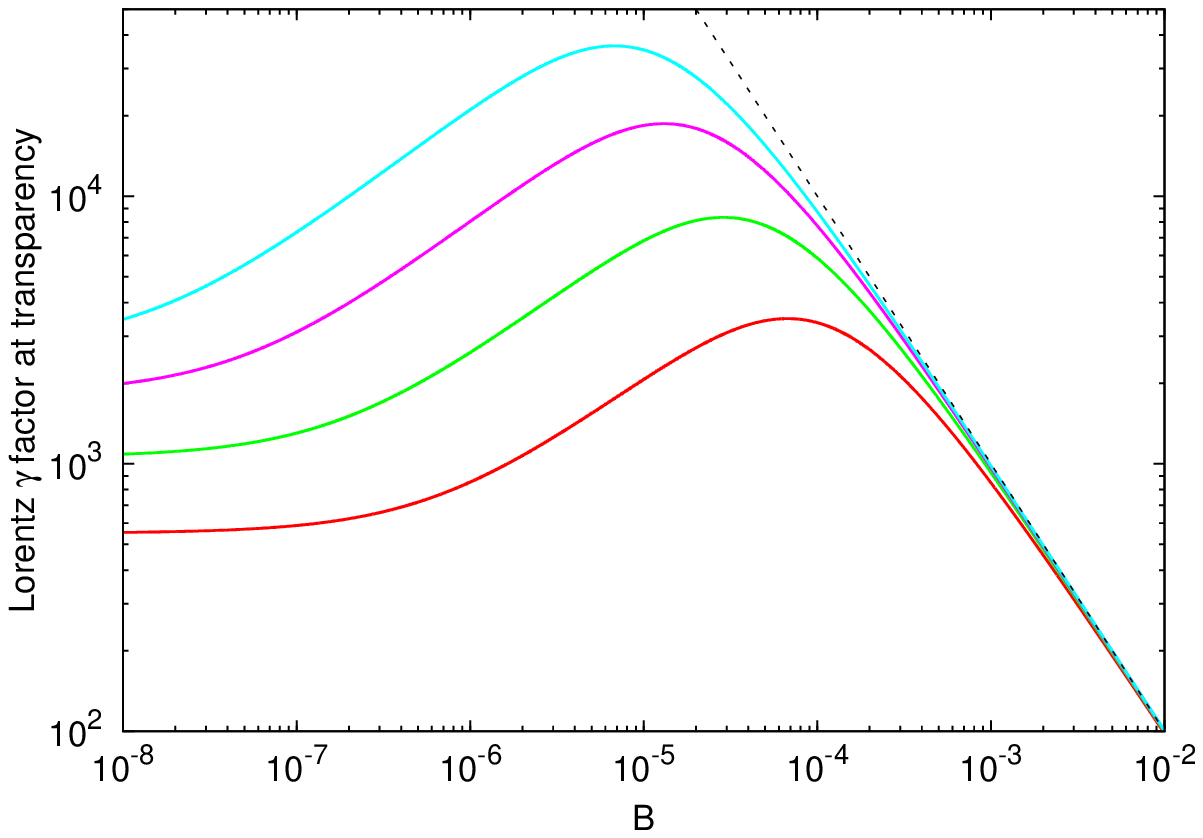}
\includegraphics[width=0.49\hsize,clip]{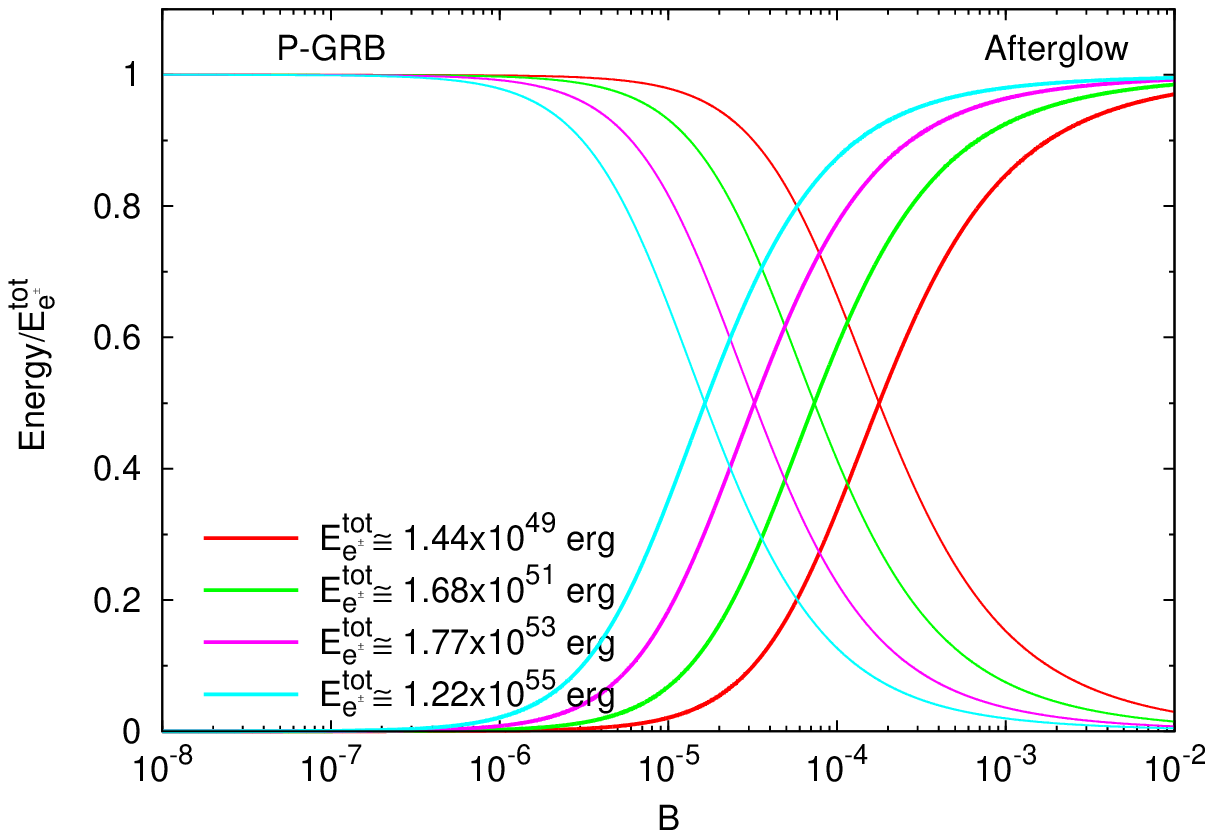}
\caption{At the fireshell transparency point, for $4$ different values of $E^{tot}_{e^\pm}$, we plot as a function of $B$: (upper left) the fireshell radius in the laboratory frame; (upper right) the fireshell temperature in the co-moving frame $T_\circ^{com}$ (thicker lines) and the one Doppler blue-shifted along the line of sight toward the observer in the source cosmological rest frame $T_\circ^{obs}$ (thinner lines); (lower left) the fireshell Lorentz gamma factor $\gamma_\circ$ together with the asymptotic value $\gamma_\circ = 1/B$; (lower right) the energy radiated in the P-GRB (thinner lines, rising when $B$ decreases) and the one converted into baryonic kinetic energy and later emitted in the extended afterglow (thicker lines, rising when $B$ increases), in units of $E_{e^\pm}^{tot}$. For details see Refs.~\refcite{2000A&A...359..855R,2001ApJ...555L.113R}.}
\label{ftemp-fgamma-bcross}
\end{figure}

By solving the rate equation we have evaluated the evolution of the temperature during the fireshell expansion, all the way up to when the transparency condition is reached \cite{1999A&A...350..334R,2000A&A...359..855R}.

In the upper left panel of Fig.~\ref{ftemp-fgamma-bcross} we plot, as a function of $B$, the fireshell radius at the transparency point. The plot is drawn for four different values of $E_{e^\pm}^{tot}$ in the interval $[10^{49}, 10^{55}]$ erg  which encompasses the observed isotropic GRB energies.

In the upper right panel of Fig.~\ref{ftemp-fgamma-bcross} we plot, as a function of $B$, the fireshell temperature $T_\circ$ at the transparency point, i.e.\ the temperature of the P-GRB radiation. The plot is drawn for the same four different values of $E_{e^\pm}^{tot}$ of the upper panel. We plot both the value in the co-moving frame $T_\circ^{com}$ and the value Doppler blue-shifted toward the observed $T_\circ^{obs} = (1+\beta_\circ) \gamma_\circ T_\circ^{com}$, where $\beta_\circ$ is the fireshell speed at the transparency point in units of $c$ \cite{2000A&A...359..855R}.

In the lower left panel of Fig.~\ref{ftemp-fgamma-bcross} we plot, as a function of $B$, the fireshell Lorentz gamma factor at the transparency point $\gamma_\circ$. The plot is drawn for the same four different values of $E_{e^\pm}^{tot}$ as in the upper panel. Also plotted is the asymptotic value $\gamma_\circ = 1/B$, which corresponds to the condition when the entire initial internal energy of the plasma $E_{e^\pm}^{tot}$ has been converted into kinetic energy of the baryons \cite{2000A&A...359..855R}. We see that such an asymptotic value is approached for $B \to 10^{-2}$. We see also that, if $E_{e^\pm}^{tot}$ increases, the maximum values of $\gamma_\circ$ are higher and they are reached for lower values of $B$.

\begin{figure}[t]
\centering
\includegraphics[width=\hsize,clip]{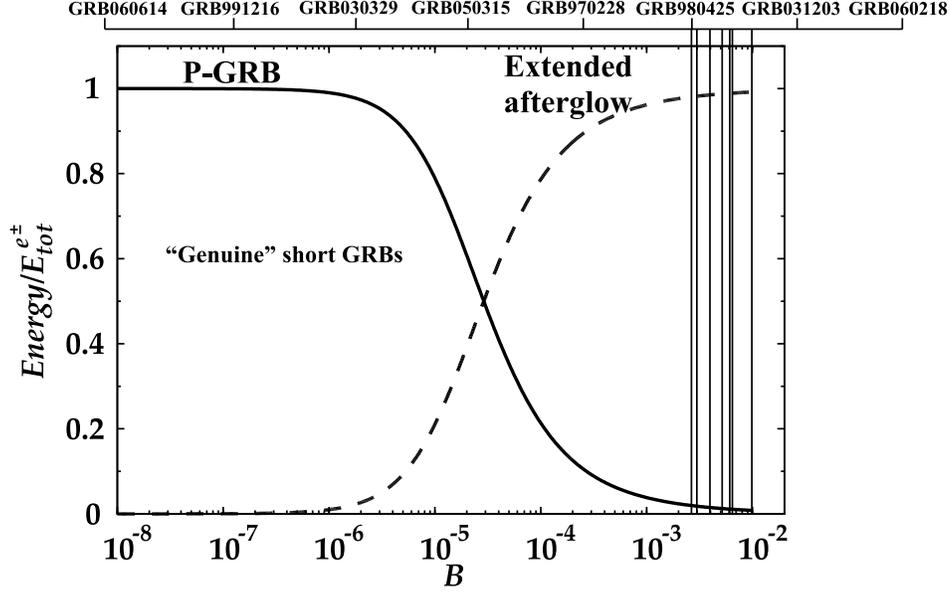}
\caption{Here the energies emitted in the P-GRB (solid line) and in the extended afterglow (dashed line), in units of the total energy of the plasma, are plotted as functions of the $B$ parameter for a typical value of $E^{tot}_{e^\pm} \sim 10^{53}$ erg (see lower panel of Fig.~\ref{ftemp-fgamma-bcross}). When $B \lesssim 10^{-5}$, the P-GRB becomes predominant over the extended afterglow, giving rise to a ``genuine'' short GRB. The figure also shows the values of the $B$ parameter corresponding to some GRBs we analyzed, all belonging to the class of long GRBs.}
\label{f2}
\end{figure}

In the lower right panel of Fig.~\ref{ftemp-fgamma-bcross} we plot, as a function of $B$, the total energy radiated at the transparency point in the P-GRB and that converted into baryonic and leptonic kinetic energy and later emitted in the extended afterglow. The plot is drawn for the same four different values of $E_{e^\pm}^{tot}$ as in the upper panel and middle panels. We see that for $B \lesssim 10^{-5}$ the total energy emitted in the P-GRB is always larger than that emitted in the extended afterglow. The limit $B \rightarrow 0$ gives rise to a ``genuine'' short GRB (see also Fig.~\ref{f2}), namely a GRB whose prompt emission is dominated by the P-GRB and which is followed by a very tiny extended afterglow, if any at all. On the other hand, for $3.0\times 10^{-4} \lesssim B < 10^{-2}$ the total energy emitted in the P-GRB is always smaller than the one emitted in the extended afterglow. If it is not below the instrumental threshold and if $n_{cbm}\sim 1$ particle/cm$^3$, the P-GRB can be observed in this case as a small pulse preceding the main GRB event (which coincides with the peak of the extended afterglow), i.e.\ as a GRB ``precursor'' \cite{2001ApJ...555L.113R,2003AIPC..668...16R,2008AIPC.1000..305B,2008AIPC.1065..219R,2007A&A...474L..13B,2008AIPC.1065..223B}. 

Particularly relevant for the new era of the Agile and Fermi satellites are the GRBs with $B < 10^{-3}$. In this case the P-GRB emission has an observed temperature up to $10^{3}$ keV or higher. This high-energy emission has been unobservable by the Swift satellite.

We must emphasize that all the above estimates have been done for a Reissner-Nordstr\"{o}m black hole endowed with an overcritical electric field. Some differences should exist in the structure of the P-GRB if, instead of a dyadosphere, the collapse gives birth to a dyadotorus. In addition, in the actual process of gravitational collapse the dyadosphere-dyadotorus formation will not occur in an asymptotically flat space. Its boundaries will be characterized by the characteristic processes of gravitational collapse. There is therefore the very exciting possibility that the actual details of the process of gravitational collapse can be inferred in principle from the structure of the P-GRB or of the ``genuine'' short GRBs: from their duration, spectra and time variability.

\subsection{The optically thin phase}

The dynamics of the optically thin fireshell of baryonic matter propagating in the CBM has been obtained from the relativistic conservation laws of energy and momentum (see e.g.\ Ref.~\refcite{2005ApJ...620L..23B}). Such conservation laws are used both in our approach and in others in the current literature (see e.g.\ Refs.~\refcite{1999PhR...314..575P,1999ApJ...512..699C,2005ApJ...620L..23B,2005ApJ...633L..13B,2007AIPC..910...55R}). The main difference is that in the current literature an ultra-relativistic approximation, following the Blandford-McKee \cite{1976PhFl...19.1130B} self-similar solution, is widely adopted, leading to a  simple constant-index power-law relation between the Lorentz gamma factor of the optically thin ``fireshell'' and its radius:
\begin{equation}
\gamma\propto r^{-a}\, ,
\label{gr0}
\end{equation}
with $a=3$ in the fully radiative case and $a=3/2$ in the adiabatic case \cite{1999PhR...314..575P,2005ApJ...633L..13B}. On the contrary, we use the exact solutions of the equations of motion of the fireshell \cite{2004ApJ...605L...1B,2005ApJ...620L..23B,2005ApJ...633L..13B,2006ApJ...644L.105B,2007AIPC..910...55R}. A detailed comparison between the equations used in the two approaches has been presented in Refs.~\refcite{2004ApJ...605L...1B,2005ApJ...620L..23B,2005ApJ...633L..13B,2006ApJ...644L.105B}. In particular, Ref.~\refcite{2005ApJ...633L..13B} shows that the regime represented in Eq.~(\ref{gr0}) is reached only asymptotically when $\gamma_\circ \gg \gamma \gg 1$ in the fully radiative regime and $\gamma_\circ^2 \gg \gamma^2 \gg 1$ in the adiabatic regime, where $\gamma_\circ$ is the initial Lorentz gamma factor of the optically thin fireshell.

\begin{figure}[t]
\centering
\includegraphics[width=0.75\hsize,clip]{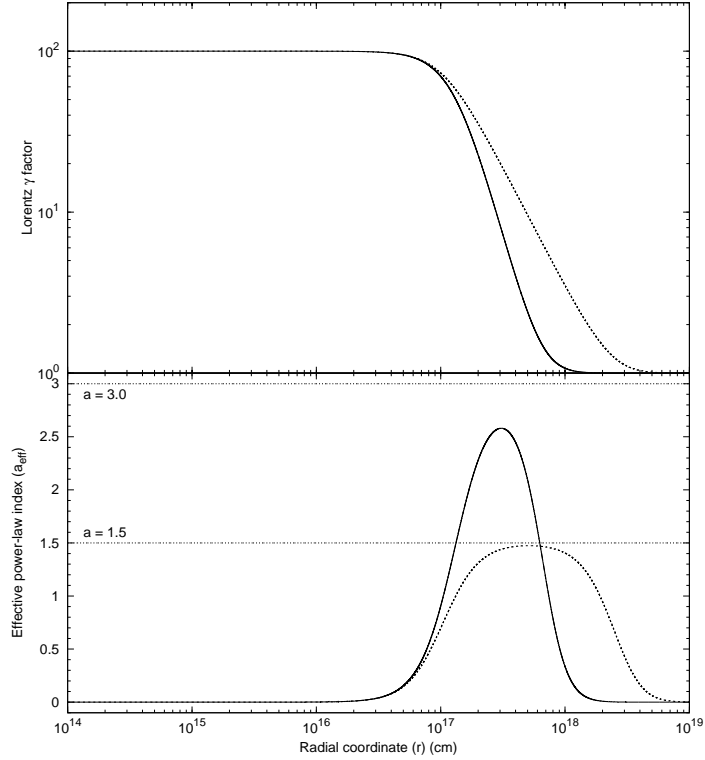}
\caption{In the upper panel, the analytic behavior of the Lorentz gamma factor during the extended afterglow era is plotted versus the radial coordinate of the expanding optically thin fireshell in the fully radiative case (solid line) and in the adiabatic case (dotted line) starting from $\gamma_\circ = 10^2$ and the same initial conditions as the GRB 991216 \cite{2005ApJ...633L..13B}. In the lower panel are plotted the corresponding values of the ``effective'' power-law index $a_{eff}$ (see Eq.~(\ref{eff_a})), which is clearly not constant but highly varying and systematically lower than the constant values $3$ and $3/2$ purported in the current literature (horizontal thin dotted lines).}
\label{gdir_a_comp_rad-ad_mg11}
\end{figure}

In Fig.~\ref{gdir_a_comp_rad-ad_mg11} we show the differences between the two approaches. In the upper panel there are plotted the exact solutions for the fireshell dynamics in the fully radiative and adiabatic cases. In the lower panel we plot the corresponding ``effective'' power-law index $a_{eff}$, defined as the index of the power-law tangent to the exact solution \cite{2005ApJ...633L..13B}:
\begin{equation}
a_{eff} = - \frac{d\ln\gamma}{d\ln r}\, . \label{eff_a}
\end{equation}
Such an ``effective'' power-law index of the exact solution smoothly varies from $0$ to a maximum value which is always smaller than $3$ or $3/2$, in the fully radiative and adiabatic cases respectively, and finally decreases back to $0$ (see Fig.~\ref{gdir_a_comp_rad-ad_mg11}).

\subsection{Extended afterglow luminosity and spectra}

The extended afterglow luminosity in the different energy bands is governed by two quantities associated with the environment. Within the fireshell model, these are the effective CBM density profile, $n_{cbm}$, and the ratio between the effective emitting area $A_{eff}$ and the total area $A_{tot}$ of the expanding baryonic and leptonic shell, ${\cal R}= A_{eff}/A_{tot}$. This last parameter takes into account the CBM filamentary structure \cite{2004IJMPD..13..843R,2005IJMPD..14...97R} and the possible occurrence of fragmentation in the shell \cite{2007A&A...471L..29D}.

Within the ``fireshell'' model, in addition to the determination of the baryon loading, it is therefore possible to infer a detailed description of the CBM, its average density and its porosity and filamentary structure, all the way from the black hole horizon out to a distance $r \lesssim 10^{17}$ cm. Typical dimensions of these clouds of overdense material in the CBM are on the order of $10^{15}$--$10^{16}$ cm and they show a density contrast on the order of $\Delta n_{cbm}/\langle n_{cbm} \rangle \sim 10$. A fascinating possibility for the origin of these clouds has been suggested by David Arnett \cite{LesHouches}, namely that these clouds correspond to matter ejected in the latest phases of the thermonuclear evolution of the progenitor star, leading to the black hole formation originating the GRB. The interaction between the baryons and leptons of the relativistically expanding shell with these CBM clouds corresponds to the spikes in the gamma and X-ray light curve of the prompt emission. This description is missing in the traditional model based on synchrotron emission. In fact, the attempt to use the internal shock model for the prompt emission (see e.g.\ Refs.~\refcite{1994ApJ...430L..93R,2005RvMP...76.1143P,2006RPPh...69.2259M}) only applies to regions where $r > 10^{17}$ cm \cite{2008MNRAS.384...33K}.

In our hypothesis, the emission from the baryonic and leptonic matter shell is spherically symmetric. This allows us to assume, in a first approximation, a modeling of the CBM distribution by thin spherical shells and consequently to consider just its radial dependence \cite{2002ApJ...581L..19R}.

For simplicity and in order to have an estimate of the energetics, the emission process is postulated to be thermal in the co-moving frame of the shell \cite{2004IJMPD..13..843R}. The observed GRB nonthermal spectral shape is due to the convolution of an infinite number of thermal spectra with different temperatures and different Lorentz and Doppler factors. Such a convolution is to be performed over the surfaces of constant arrival time of the photons at the detector (equitemporal surfaces, EQTSs; see e.g.\ Ref.~\refcite{2005ApJ...620L..23B}) encompassing the entire observation time interval\cite{2005ApJ...634L..29B}. We are currently considering additional effects as an alternative to a purely thermal spectrum \cite{2010AIPC.1279..406P,pa10}.

\begin{figure}[t]
\centering
\includegraphics[width=0.49\hsize,clip]{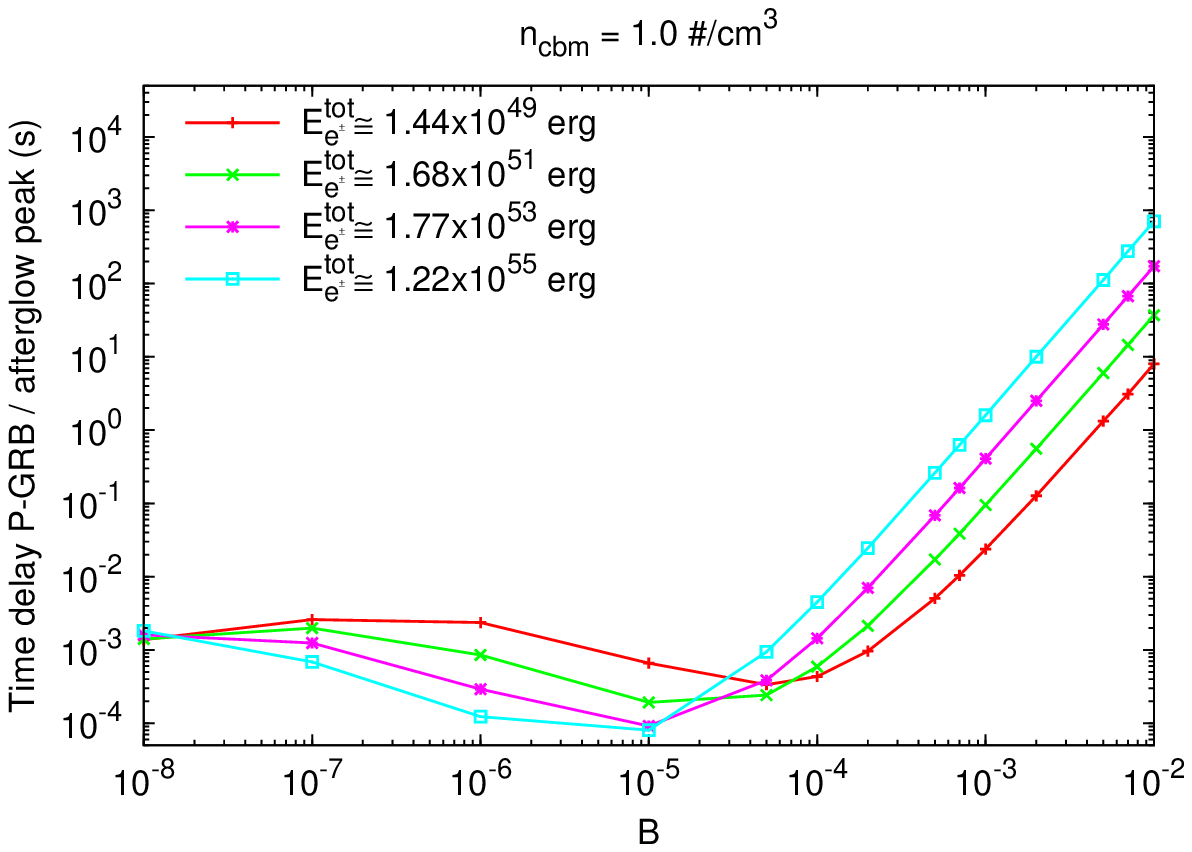}
\includegraphics[width=0.49\hsize,clip]{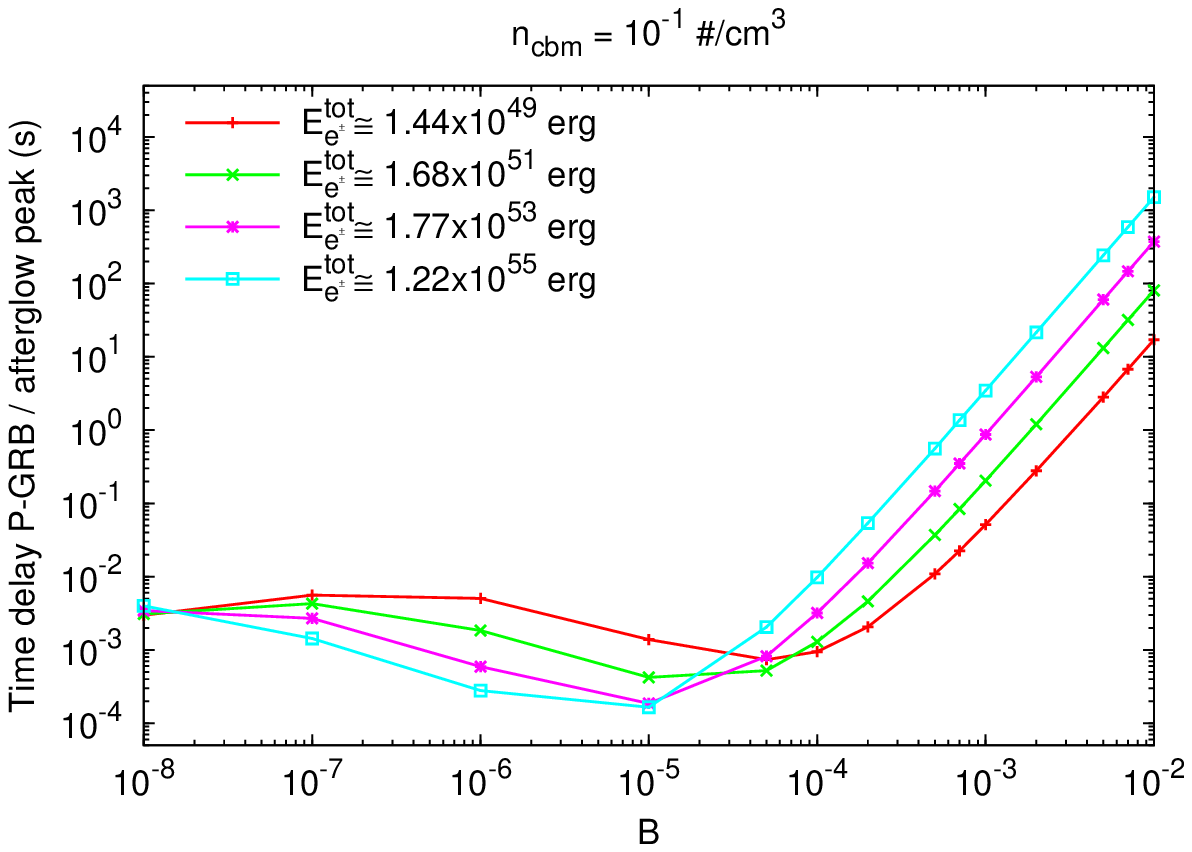}\\
\includegraphics[width=0.49\hsize,clip]{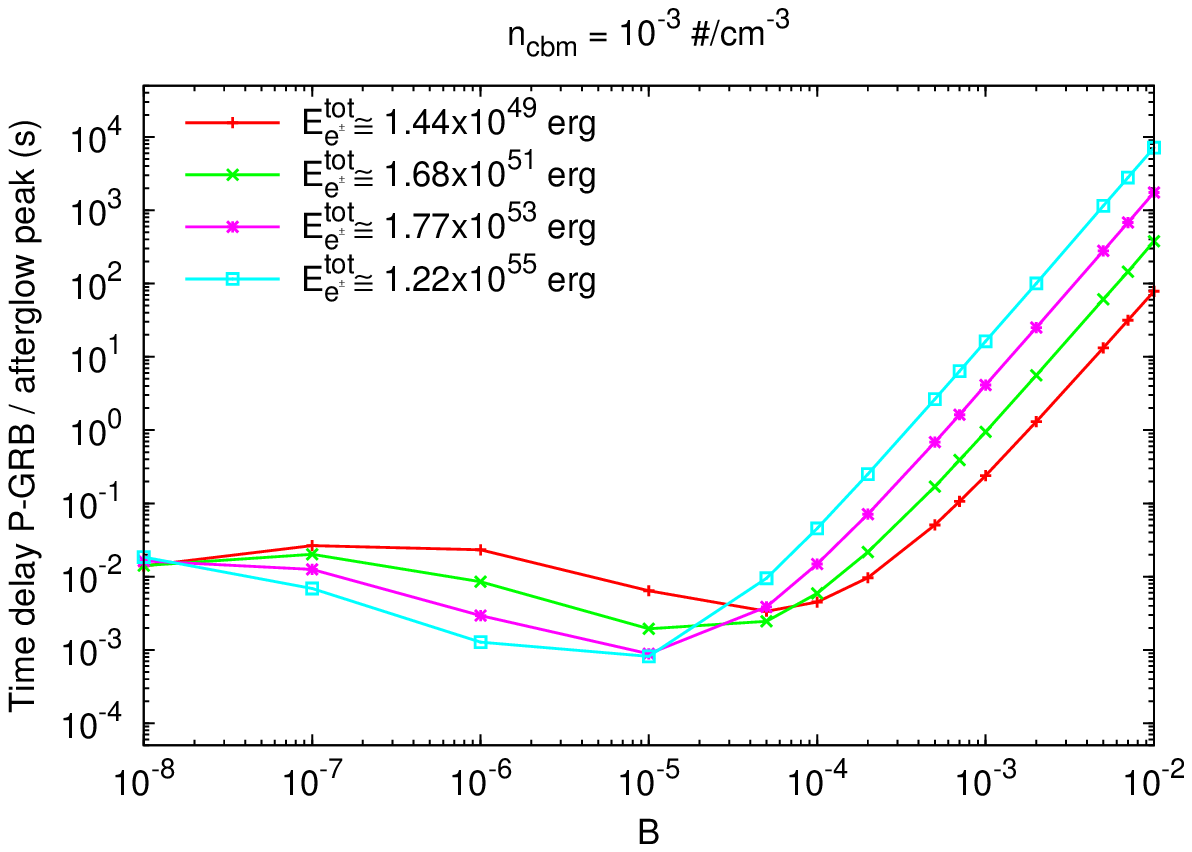}
\includegraphics[width=0.49\hsize,clip]{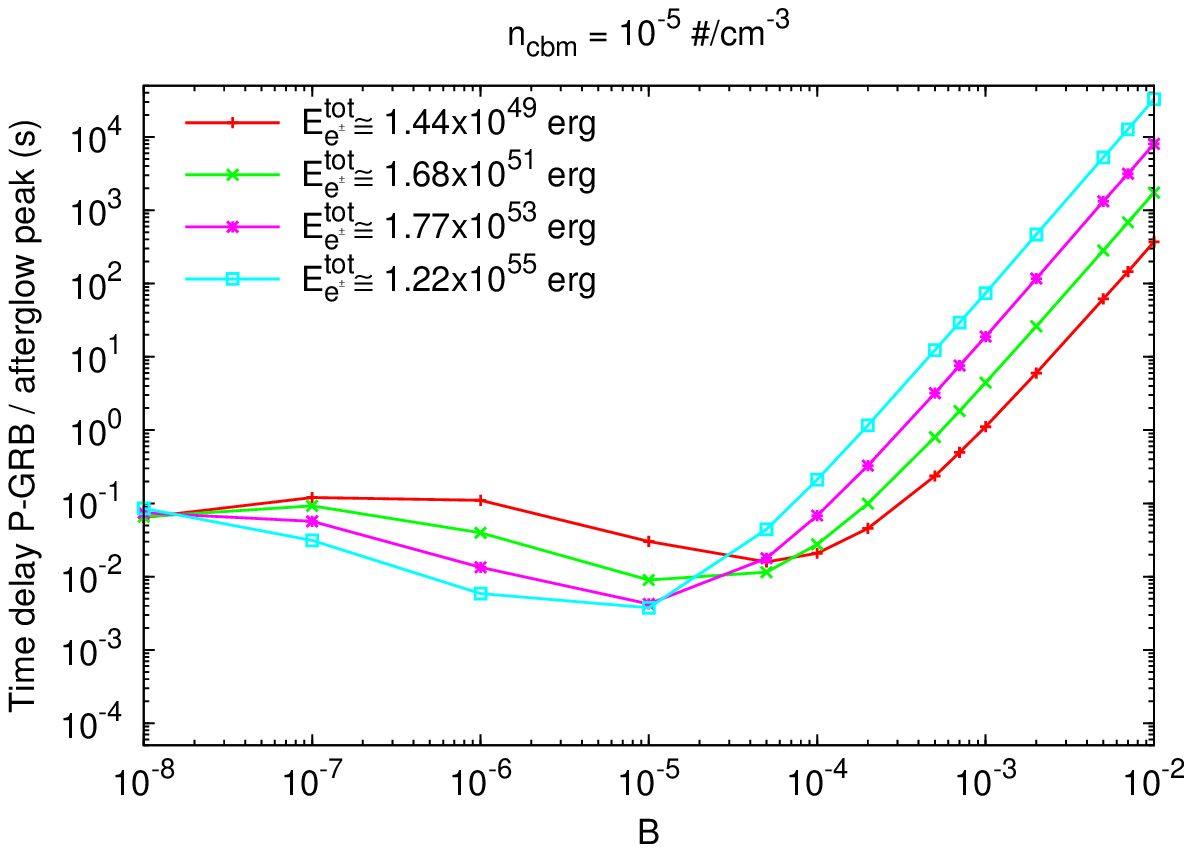}
\caption{For $4$ different values of $E^{tot}_{e^\pm}$, we plot as a function of $B$ the arrival time separation $\Delta t_a$ between the P-GRB and the peak of the extended afterglow (i.e.\ the ``quiescent time between the ``precursor'' and the main GRB event), measured in the source cosmological rest frame. The computation has been performed assuming a constant value of the CBM density in four different cases: $n_{cbm}=1.0$ particles/cm$^3$, $n_{cbm}=1.0\times 10^{-3}$ particles/cm$^3$, $n_{cbm}=1.0\times 10^{-5}$ particles/cm$^3$, $n_{cbm}=1.0\times 10^{-7}$ particles/cm$^3$. The points represent the actual numerically computed values, connected by straight line segments. See details in Ref.~\refcite{2009AIPC.1132..199R}.}
\label{dta}
\end{figure}

In Fig.~\ref{dta} we plot, as a function of $B$, the arrival time separation $\Delta t_a$ between the P-GRB and the peak of the extended afterglow measured in the cosmological rest frame of the source. Such a time separation $\Delta t_a$ is the ``quiescent time'' between the precursor (i.e.\ the P-GRB) and the main GRB event (i.e.\ the peak of the extended afterglow). The plot is drawn for the same four different values of $E_{e^\pm}^{tot}$ of Fig.~\ref{ftemp-fgamma-bcross}. The arrival time of the peak of the extended afterglow emission depends on the detailed profile of the CBM density. In this plot a constant CBM density has been assumed in four different cases: $n_{cbm}=1.0$ particles/cm$^3$, $n_{cbm}=1.0\times 10^{-3}$ particles/cm$^3$, $n_{cbm}=1.0\times 10^{-5}$ particles/cm$^3$, $n_{cbm}=1.0\times 10^{-7}$ particles/cm$^3$. We can see that, for $3.0\times 10^{-4} \lesssim B < 10^{-2}$, which is the condition for P-GRBs to be ``precursors'' (see above), $\Delta t_a$ increases both with $B$ and with $E_{e^\pm}^{tot}$. We can have $\Delta t_a > 10^2$ s and, in some extreme cases even $\Delta t_a \sim 10^3$ s. For $B \lesssim 3.0\times 10^{-4}$, instead, $\Delta t_a$ shows a behavior which qualitatively follows the opposite of $\gamma_\circ$ (see the middle panel of Fig.~\ref{ftemp-fgamma-bcross}).

\begin{figure}[t]
\centering
\includegraphics[width=\hsize,clip]{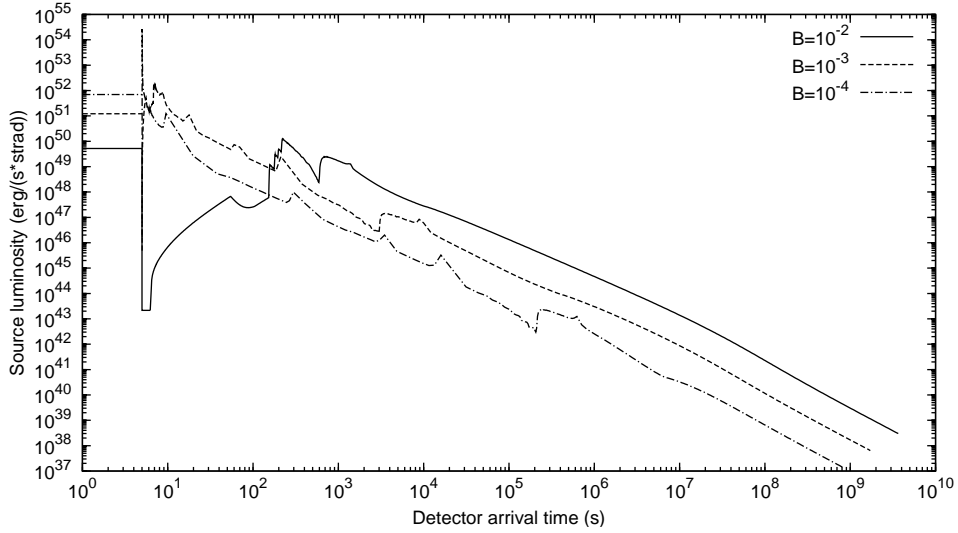}
\caption{We plot three theoretical extended afterglow bolometric light curves together with the corresponding P-GRB peak luminosities (the horizontal segments). The computations have been performed assuming the same $E^{tot}_{e^\pm}$ and CBM structure as the GRB 991216 and three different values of $B$. The P-GRBs have been assumed to have the same duration in the three cases, i.e.\ $5$ s. For $B$ decreasing, the extended afterglow light curve squeezes itself onto the P-GRB curve (Simulations by L. Caito, see details in Ref.~\refcite{2009AIPC.1132..199R}).}
\label{multi_b}
\end{figure}

Finally, in Fig.~\ref{multi_b} we present three theoretical extended afterglow bolometric light curves together with the corresponding P-GRB peak luminosities for three different values of $B$. The duration of the P-GRBs has been assumed to be the same in the three cases (i.e.\ $5$ s). The computations have been performed assuming the same $E^{tot}_{e^\pm}$ and the same detailed CBM density profile as GRB 991216 \cite{2003AIPC..668...16R}. In this picture we clearly see how, for $B$ decreasing, the extended afterglow light curve ``squeezes'' itself onto the P-GRB and the P-GRB peak luminosity increases. We are currently trying to identify some GRBs having this feature.

The radiation viewed in the co-moving frame of the accelerated baryonic matter is assumed for simplicity to have a thermal spectrum and to be produced by the interaction of the CBM with the front of the expanding baryonic shell \cite{2004IJMPD..13..843R}. In Ref.~\refcite{2005ApJ...634L..29B} it was shown that, although the instantaneous spectrum in the co-moving frame of the optically thin fireshell is thermal, the shape of the final instantaneous spectrum in the laboratory frame is nonthermal. In fact, as explained in Ref.~\refcite{2004IJMPD..13..843R}, the temperature of the fireshell is evolving with the co-moving time and, therefore, each single instantaneous spectrum is the result of an integration of hundreds of thermal spectra with different temperatures over the corresponding EQTS. This calculation produces a nonthermal instantaneous spectrum in the observer frame \cite{2005ApJ...634L..29B}.

Another distinguishing feature of the GRB spectra which is also explained within the fireshell model is their hard to soft transition during the evolution of the event \cite{1997ApJ...479L..39C,1999PhR...314..575P,2000ApJS..127...59F,2002A&A...393..409G}. In fact the peak of the energy distributions $E_p$ drift monotonically to softer frequencies with time \cite{2005ApJ...634L..29B}. This feature explains the change in the power-law low energy spectral index \cite{1993ApJ...413..281B} $\alpha$ which at the beginning of the prompt emission of the burst ($t_a^d=2$ s) is $\alpha=0.75$, and progressively decreases for later times \cite{2005ApJ...634L..29B}. In this way the link between $E_p$ and $\alpha$ identified in Ref.~\refcite{1997ApJ...479L..39C} is explicitly shown. This is due to the decrease of the Lorentz gamma factor and of the temperature in the co-moving frame \cite{2005ApJ...634L..29B,2009AIPC.1132..199R}.

The time-integrated observed GRB spectra show a clear power-law behavior. Within a different framework (see e.g.\ Ref.~\refcite{1983ASPRv...2..189P} and references therein) it has been argued that it is possible to obtain such a power-law spectra from a convolution of many non-power-law instantaneous spectra monotonically evolving in time. This result was recalled and applied to GRBs \cite{1999ARep...43..739B} assuming for the instantaneous spectra a thermal shape with a time varying temperature. It was shown that the integration of such energy distributions over the observation time gives a typical power-law shape consistent with the GRB spectra.

\begin{figure}[t]
\centering
\includegraphics[width=\hsize,clip]{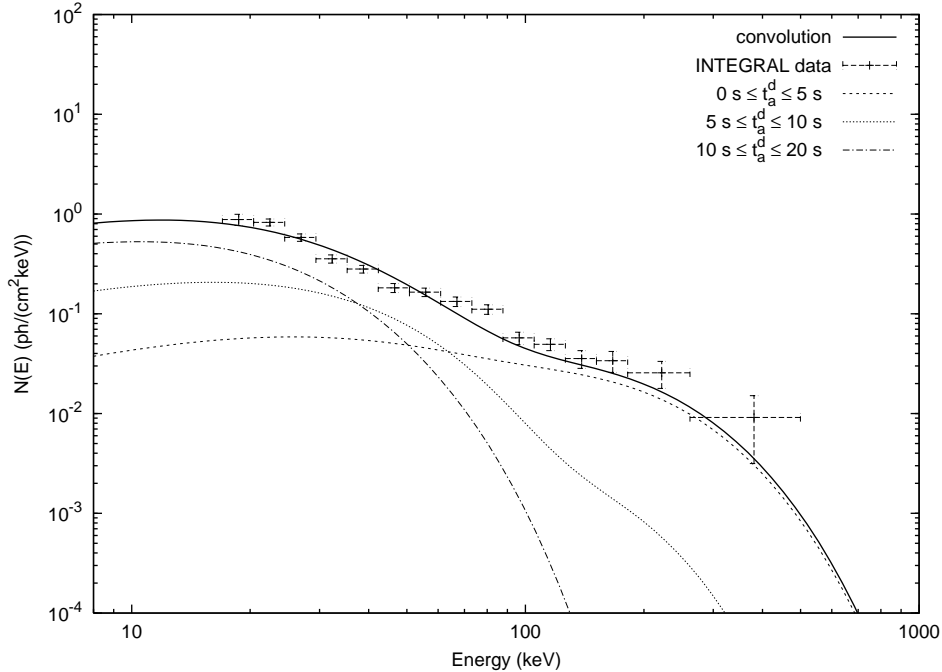}
\caption{Three theoretically predicted time-integrated photon number spectra $N(E)$, computed for GRB 031203 \cite{2005ApJ...634L..29B}, are shown here for $0 \le t_a^d \le 5$ s, $5 \le t_a^d \le 10$ s and $10 \le t_a^d \le 20$ s (dashed and dotted curves), where $t_a^d$ is the photon arrival time at the detector \cite{2001ApJ...555L.107R,2005ApJ...634L..29B}. The hard to soft behavior is confirmed. Moreover, the theoretically predicted time-integrated photon number spectrum $N(E)$ corresponding to the first $20$ s of the ``prompt emission'' (black bold curve) is compared with the data observed by INTEGRAL \cite{2004Natur.430..646S}. This curve is obtained as a convolution of 108 instantaneous spectra, which are enough to get a good agreement with the observed data. See details in Ref.~\refcite{2005ApJ...634L..29B}.}
\label{031203_spettro}
\end{figure}

Our specific quantitative model is more complicated than the one considered in Ref.~\refcite{1999ARep...43..739B}: the instantaneous spectrum in the Fireshell model is not a black body. Each instantaneous spectrum is obtained by an integration over the corresponding EQTS \cite{2004ApJ...605L...1B,2005ApJ...620L..23B}, themselves a convolution, weighted by appropriate Lorentz and Doppler factors, of $\sim 10^6$ thermal spectra with time varying temperature. Therefore, the time-integrated spectra are not plain convolutions of thermal spectra: they are convolutions of convolutions of thermal spectra \cite{2004IJMPD..13..843R,2005ApJ...634L..29B}. In Fig.~\ref{031203_spettro} we present the photon number spectrum $N(E)$ time-integrated over the $20$ s of the entire duration of the prompt event of GRB 031203 observed by INTEGRAL \cite{2004Natur.430..646S}: we obtain a typical nonthermal power-law spectrum which turns out to be in good agreement with the INTEGRAL data \cite{2004Natur.430..646S,2005ApJ...634L..29B} and gives evidence of the possibility that the observed GRB spectra originate from a thermal emission \cite{2005ApJ...634L..29B}.

Before ending this discussion, we mention that, using the diagrams given in Figs.~\ref{ftemp-fgamma-bcross}--\ref{dta}, in principle one can compute the two free parameters of the fireshell model, namely $E^{tot}_{e^\pm}$ and $B$, from the ratio between the total energies of the P-GRB and of the extended afterglow and from the temporal separation between the peaks of the corresponding bolometric light curves. Correspondingly, it is also possible to evaluate the temperature at decoupling boosted by the Lorentz gamma factor as well as the temperature in the comoving frame of the fireshell. None of these quantities depends on the cosmological model. Therefore, one can in principle use this method to compute the GRB intrinsic luminosity and temperature at transparency, and make GRBs the best cosmological distance indicators. The increase in the number of observed sources, as well as the more accurate knowledge of their CBM density profiles, will possibly make this procedure  viable to test cosmological parameters, in addition to the Amati relation \cite{2008MNRAS.391..577A,2008A&A...487L..37G}.

\section{Progress in theoretical physics}

After describing this general picture of the canonical GRB scenario, it is appropriate to turn to some progress in theoretical physics motivated by the quest of clarifying some of the main features of the Fireshell model. We will then return to discuss our understanding of some specific GRBs and introduce a new class of GRBs.

\subsection{Progress in understanding the thermalization process of the electron-positron-baryon plasma}\label{sec:therm}

I already mentioned in Sec.~\ref{sec:intro} that a crucial hypothesis underlying the concept of the dyadosphere, as well as the dynamical analysis of the PEM and PEMB pulse of the fireshell, was the assumption of thermal equilibrium with photons and baryons in the $e^+e^-$ plasma. I also expressed how this assumption has not been adopted in alternative models, e.g.\ the one by Cavallo and Rees \cite{1978MNRAS.183..359C}, where the $e^+e^-$ plasma degrades due to bremsstrahlung down to temperatures below $kT\sim m_e c^2$ and the $e^+e^-$ disappear. The two different concepts of fireball and fireshell then follow. It was then necessary to address a theoretical analysis to justify our assumption based on first principles.

\begin{figure}[t]
\centering
\includegraphics[width=\hsize,clip]{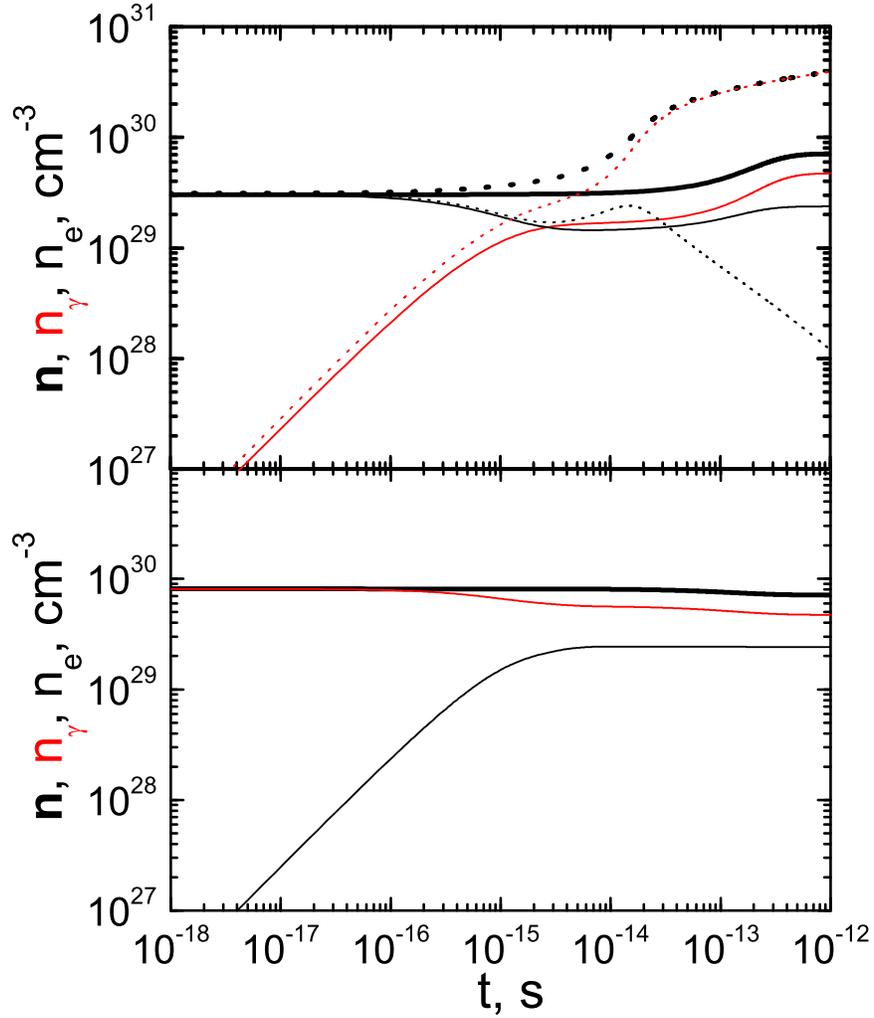}
\caption{Dependence on time of concentrations of pairs (black), photons (red)
and both (thick) when all interactions take place (solid). The upper (lower)
figure corresponds to the case where initially there are mainly pairs
(photons). Dotted curves in the upper figure show concentrations when inverse
triple interactions are neglected. In this case an enhancement of the pairs
occurs with the corresponding increase in photon number and thermal
equilibrium is never reached.}
\label{3p}
\end{figure}

In order to establish which scenario is correct we studied the relaxation of an electron-positron plasma to thermal equilibrium in 
Ref.~\refcite{2007PhRvL..99l5003A, 2009PhRvD..79d3008A}. There the relativistic Boltzmann equations with exact QED collisional integrals taking
into account all relevant two-particle (Compton scattering, Bhabha and Moller scattering, pair creation and annihilation in two photons, electron-proton Coulomb scattering) and three-particle interactions (relativistic bremsstrahlung, double Compton scattering, three photon annihilation and radiative pair creation), see Table~\ref{reactions}, were solved numerically. It was confirmed that a metastable state called ``kinetic equilibrium" \cite{1997ApJ...486..903P} exists in such a plasma, which is
characterized by the same temperature of all particles, but nonzero chemical
potentials. Such a state occurs when the detailed balance of all two-particle
reactions is established. It was pointed out in Refs.~\refcite{2007PhRvL..99l5003A,
2009PhRvD..79d3008A} that direct and inverse three-particle interactions are essential in
bringing the electron-positron plasma to thermal equilibrium. In particular, when inverse three-particle interactions are switched off artificially the thermal equilibrium is never reached, see Fig.~\ref{3p}.

In Ref.~\refcite{2010PhRvE..81d6401A} with Aksenov and Vereshchagin we computed numerically the relaxation time scales for optically thick
electron-positron plasma over a wide range of temperatures and baryon loadings using the kinetic code we developed in
Ref.~\refcite{2007PhRvL..99l5003A, 2009PhRvD..79d3008A}. These time scales were previously estimated in the literature by order of magnitude arguments using the reaction rates of the dominant processes \cite{1981PhFl...24..102G,
1983MNRAS.202..467S}. We have shown that these numerically obtained time scales differ from previous estimations by several orders of magnitude.

Temperatures in the range considered in 
Ref.~\refcite{2007PhRvL..99l5003A,
2009PhRvD..79d3008A, 2010PhRvE..81d6401A}
\begin{equation}\label{temprange}
    0.1<\frac{kT}{m_ec^2}<10
\end{equation}
were selected in order to avoid production of other particles such as muons and neutrinos \cite{2010PhR...487....1R}.

In the uniform isotropic pair plasma relativistic Boltzmann equation for the
distribution function $f_i$ of the particle species $i$ has the following form:
\begin{equation}\label{Boltzmann_class}
\frac1c\frac{d}{d t}f_i(\mathbf{p}_i,t)=
    \sum_q\left(\eta^q_i-\chi^q_i f_i(\mathbf{p}_i,t)\right)\, ,
\end{equation}
where the sum is taken over all two- and three-particle reactions $q$, and
$\eta^q_i$ and $\chi^q_i$ are, respectively, the emission and absorption
coefficients.

\begin{table}
\tbl{Physical processes included in simulations}
{\begin{tabular}{ll}
  \toprule
  Binary interactions & Radiative and \\
  & pair producing variants
  \\ \colrule
  M{\o}ller and Bhaba & Bremsstrahlung \\
  $e_1^\mp e_2^\mp\rightarrow {e_1^\mp}' {e_2^\mp}'$
  & $e_1^\mp e_2^\mp\leftrightarrow {e_1^\mp}' {e_2^\mp}'\gamma$ \\
  $e_1^\mp e_2^\pm\rightarrow {e_1^\mp}' {e_2^\pm}'$
  & $e_1^\mp e_2^\pm\leftrightarrow {e_1^\mp}' {e_2^\pm}'\gamma$ \\
  \hline
  Single Compton & Double Compton \\
  $\gamma e^\mp\rightarrow \gamma' {e^\mp}'$
  & $\gamma e^\mp\leftrightarrow \gamma' {e^\mp}'\gamma'$
  \\ \hline
  Pair production & Radiative pair production \\
  and annihilation & and 3-photon annihilation \\
  $\gamma\gamma'\leftrightarrow e^\mp e^\pm$
  & $\gamma\gamma'\leftrightarrow e^\mp e^\pm\gamma''$ \\
  & $e^\mp e^\pm\leftrightarrow \gamma\gamma'\gamma''$
  \\ \hline
  & $e^\mp\gamma\leftrightarrow {e^\mp}'{e^\mp}''e^\pm$ \\ \hline
  Coulomb scattering &  \\
  $p_1p_2\rightarrow p_1'p_2'$
  & $p e^\mp\leftrightarrow p' {e^\mp}'\gamma$ \\
  $p e^\mp\leftrightarrow p' {e^\mp}'$
  & $p \gamma\leftrightarrow p' e^\mp e^\pm$ \\ \botrule
\end{tabular}}
\label{reactions}
\end{table}

\begin{figure}[t]
\centering
\includegraphics[width=0.49\hsize,clip]{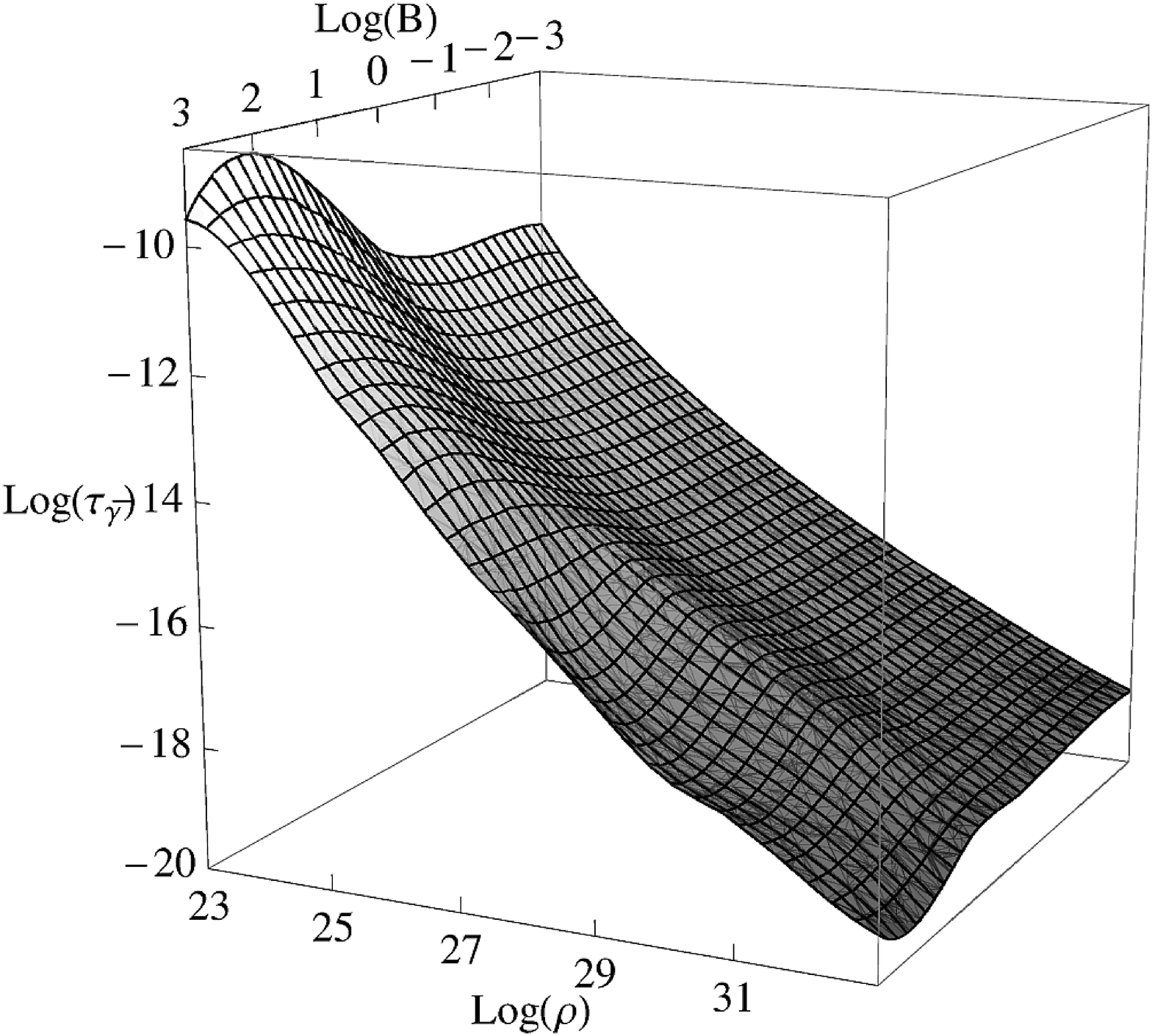}
\includegraphics[width=0.49\hsize,clip]{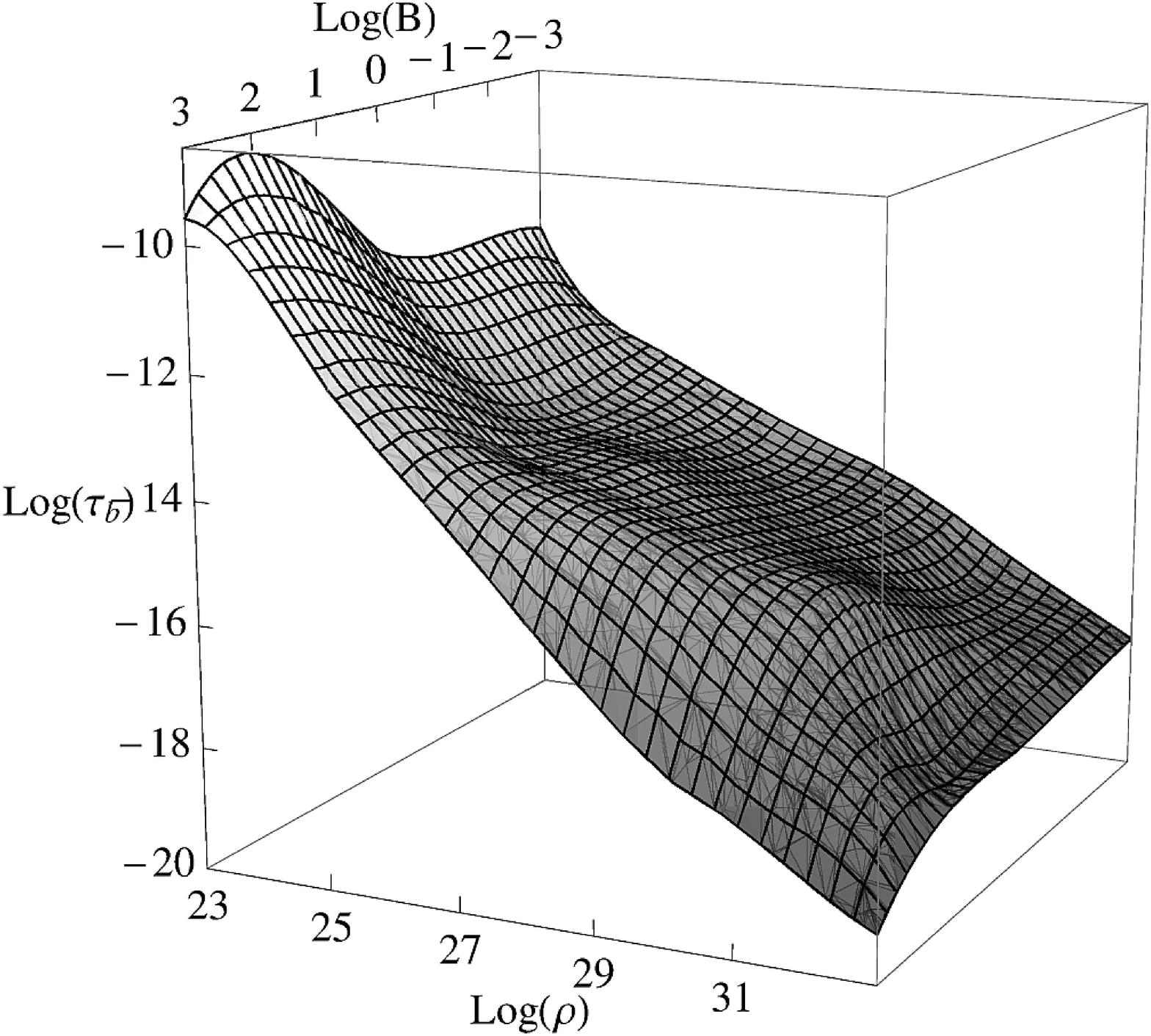}
\caption{\textbf{Left:} The thermalization time scale of the
electron-positron-photon component of plasma as a function of the
total energy density and the baryonic loading parameter. The
energy density is measured in $\mathrm{erg}\cdot\mathrm{cm}^{-3}$,
time is seconds. \textbf{Right:} The final thermalization time scale of a
pair plasma with baryonic loading as a function of the total
energy density and the baryonic loading parameter. The energy
density is measured in $\mathrm{erg}\cdot\mathrm{cm}^{-3}$, time
is seconds.}
\label{timescales}
\end{figure}

We developed a relativistic kinetic code for treating the
electron-positron-proton plasma in the framework of the kinetic Boltzmann
equations in the  homogeneous and isotropic cases \cite{2007PhRvL..99l5003A,2009PhRvD..79d3008A}. 
We focus only on the electromagnetic interactions. The
two basic parameters of the plasma are the total energy density
$\rho$ and the baryonic loading parameter
$B\equiv\frac{\rho_b}{\rho_\gamma+\rho_{e^-}+\rho_{e^+}}$,
  where $\rho_B$, $\rho_\gamma$, $\rho_{e^-}$, $\rho_{e^+}$ are the 
energy densities of baryons, photons, electron, and positrons. We
choose the range of plasma parameters relevant to GRB sources
\begin{equation}
  10^{23}\leq\rho\leq10^{33}\mathrm{erg}\cdot\mathrm{cm}^{-3}\, ,
  \mbox{ }
  10^{-3}\leq B\leq10^{3}\, .
\end{equation}
The corresponding temperatures in the thermal equilibrium state are
$0.1\leq T\leq10$ MeV. Firstly, if we neglect
neutrino channels as well as the creation and annihilation of
baryons and the weak interactions, we have no additional
particles except for the $\gamma$, 
$e^+e^-$, $p$. Secondly, given the smallness of
the plasma parameter $g=(n_e\lambda_D^3)^{-1}$, where $\lambda_D$
is the Debye length, we can use only one-particle distribution functions
$f(\mathbf{p},t)$ in this most interesting temperature range.
The range of the parameter $B$ includes cases
of almost purely electron-positron plasma, and almost purely
electron-proton plasma.

We started from an arbitrary non-equilibrium state for fixed sets of
parameters $\rho$ and $B$. We then solved Eqs.~(\ref{Boltzmann_class}) numerically until the
 plasma steady state is reached. As a result, we evaluated the corresponding time
scales for the plasma relaxation to the equilibrium, reproduced in
Fig.~\ref{timescales}. We have different times scales for the electron-positron
component and a time scale for the entire plasma due to the large
differences in masses of electrons and  protons. The
final time scales can be estimated as
$\tau_\mathrm{th}\simeq\max(\tau_{3p},\min(\tau_{ep},t_{pp}))$,
where $\tau_{3p}$ is the three-particle interaction time scale, $\tau_{ep}$ is the electron-proton
elastic scattering time scale, and $t_{pp}$ is the proton-proton elastic
scattering time scale, respectively. The exact numerical
coefficient in the relation between the calculated time scale and the 
estimated formula varies over a wide range of values \cite{2010PhRvE..81d6401A}.

The validity of our assumption in the canonical GRB scenario has thus been proved. This work can be used as an example of how rigorous fundamental results in relativistic plasma physics have been obtained motivated by the study of GRBs.

\subsection{Progress in the physics of neutron stars: Local vs.\ global charge neutrality and critical fields}\label{sec:loc_glob}

Recent progress has been made in the generalization of the Thomas-Fermi model to special relativistic regimes as well as to general relativistic regimes (see e.g.\ Refs.~\refcite{2009arXiv0903.3727P,2009arXiv0911.4622R,2009arXiv0911.4627R,ruedasubprl2,2010arXiv1012.0154R}). We have generalized classical results obtained by Feynman, Metropolis and Teller \cite{1949PhRv...75.1561F} and, by the introduction of scaling laws, the classical results obtained by Popov and collaborators (e.g.\ Ref.~\refcite{1976JETPL..24..163M,1977JETP...45..436M}) in heavy nuclei to massive cores of $\sim M_\odot$. On the other hand, we have found special justification when applied to astrophysical systems leading to a consistent treatment of white dwarfs (see e.g.\ Ref.~\refcite{2010arXiv1012.0154R}) and to a deeper understanding of neutron star physics (see e.g.\ Ref.~\refcite{2009arXiv0911.4627R,ruedasubprl2}).

V.~S.~Popov et al.\ in Ref.~\refcite{2009arXiv0903.3727P} described heavy nuclei as a degenerate system of $N_n$ neutrons, $N_p$ protons and $N_e$ electrons constrained to a constant nuclear density distribution for the protons and then solved the corresponding relativistic Thomas-Fermi equation. We have first generalized the work of V.~S.~Popov \cite{1976JETPL..24..163M,1977JETP...45..436M,1971JETP...32..526P,1972SvPhU..14..673Z} and W.~Greiner \cite{1969ZPhy..218..327P,1982PhT....35h..24G} by eliminating their constraint between the total number of protons and the total number of baryons, $N_p\approx A/2$, clearly not valid for heavy nuclei. We have self-consistently enforced the condition of beta equilibrium in a new relativistic Thomas-Fermi equation. Using then the existence of scaling laws these results have been extended from heavy nuclei to the case of nuclear matter cores of stellar dimensions. In both these treatments a zero Fermi energy of the electrons, $E_e^F=0$, was assumed.

We have recently generalized this dual approach in M. Rotondo et al.\cite{2009arXiv0911.4622R} by considering first the case of compressed atoms and then, using the existence of scaling laws, the case of compressed nuclear matter cores of stellar dimensions with a positive value of their electron Fermi energies. This approach allows a precise treatment of the electrodynamical interactions within a compressed atom with all their relativistic corrections. Hence, a self-consistent equation of state for compressed nuclear matter is derived which at high densities validates the equation of state due to E.~E.~Salpeter\cite{1961ApJ...134..669S} and overcomes some of its difficulties, like the appearance of negative pressure at low densities. 

We have recently applied such an equation of state to the study of the general relativistic white-dwarf equilibrium configurations in J.~A.~Rueda et al.\cite{2010arXiv1012.0154R}. The contributions of quantum statistics and of the weak and electromagnetic interactions have been further generalized there by considering the contribution of the general relativistic equilibrium of white dwarf matter. This is expressed by the simple formula $\sqrt{|g_{00}|}\mu_{\rm{ws}}=$ constant, linking the chemical potential of the Wigner-Seitz cell $\mu_{\rm{ws}}$ with the general relativistic gravitational potential $g_{00}$ at each point of the configuration. The configuration outside each Wigner-Seitz cell is strictly neutral and therefore no global electric field is necessary to ensure the equilibrium of the white dwarf. These equations modify those used by Chandrasekhar by properly accounting for the Coulomb interaction between the nuclei and the electrons as well as inverse beta decay. They also generalize the work of Salpeter by considering a unified self-consistent approach to the Coulomb interaction in each Wigner-Seitz cell. The consequences for the numerical value of the Stoner-Chandrasekhar-Landau mass limit as well as for the mass-radius relation of white dwarfs were then derived \cite{2010arXiv1012.0154R}. This leads to the possibility of a direct confrontation of these results with observations. This is currently of great interest because of the cosmological implications of the type Ia supernovae \cite{1993ApJ...413L.105P,1998AJ....116.1009R,1999ApJ...517..565P,2004ApJ...607..665R} and because of the low mass white dwarf companion of the pulsar PSRJ1141-6545 \cite{kramerprivate2010} as well as the role of white dwarfs in SNe, soft gamma-ray repeaters (SGRs) and anomalous X-ray pulsars (AXPs) \cite{2010Sci...329..817A} as an explicit alternative to magnetars \cite{2011arXiv1102.0653M}.

\begin{figure}[t]
\centering
\includegraphics[width=\hsize,clip]{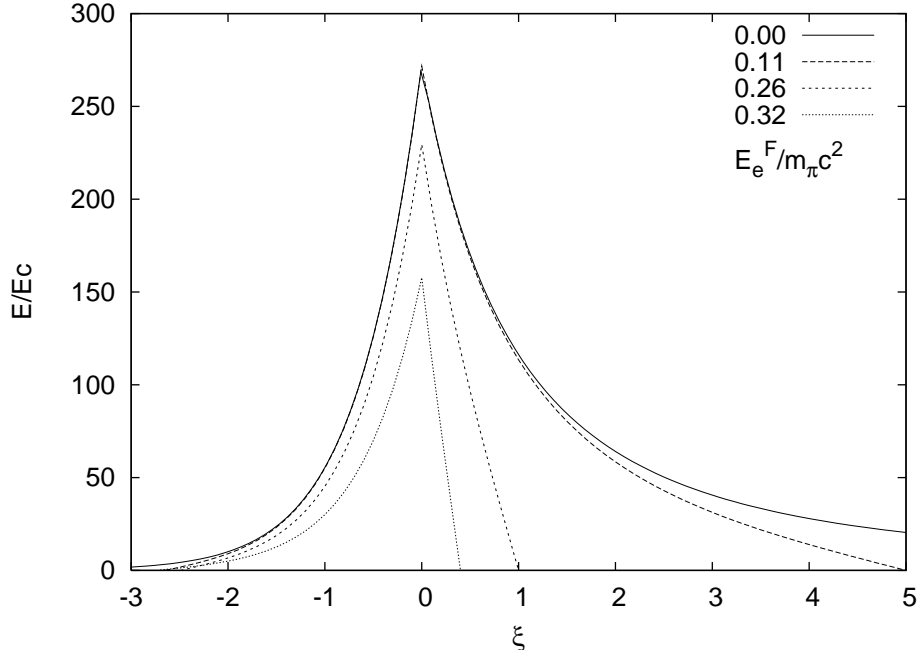}
\caption{The electric field in units of the critical field for vacuum polarization $E_c=m_e^2c^3/(e\hbar)$ is plotted as a function of the coordinate $\xi$, for different values of the electron Fermi energy in units of the pion rest energy. The solid line corresponds to the case of zero electron Fermi energy. An increase in the value of the electron Fermi energy leads to a reduction in the  peak value of the electric field.}
\label{fig:Ecompressed}
\end{figure}

\begin{figure}[t]
\centering
\includegraphics[width=\hsize,clip]{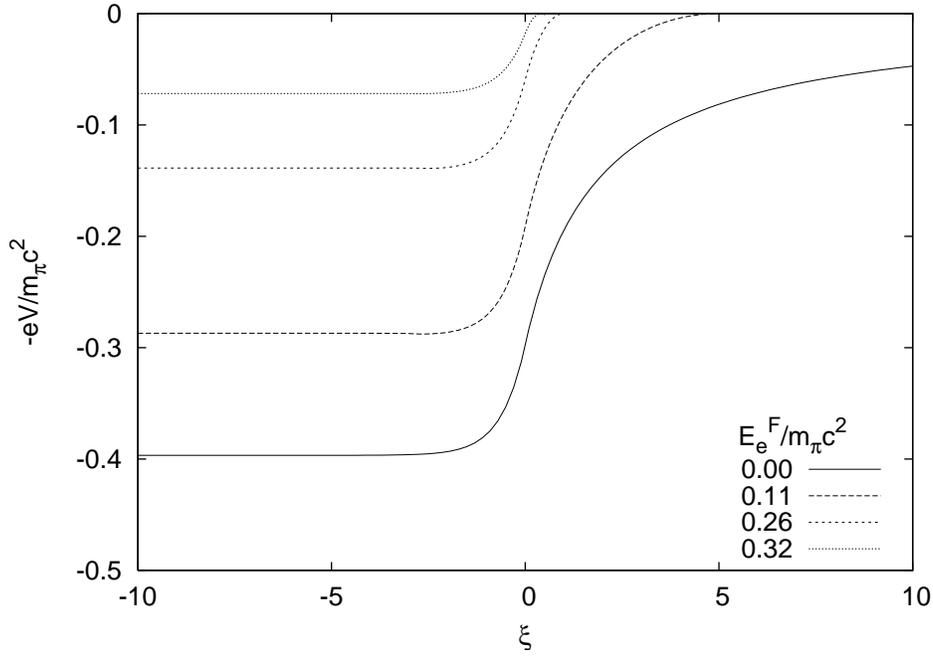}
\caption{The electron Coulomb potential energies in units of the pion rest energy in a nuclear matter core of stellar dimensions with $A\simeq 10^{57}$ or $M_{core} \sim M_{\odot}$ and $R_c \approx 10^6$ cm are plotted as a function of the dimensionless variable $\xi$ for different values of the electron Fermi energy, also in units of the pion rest energy. The solid line corresponds to the case of zero electron Fermi energy. Increasing the value of the electron Fermi energy, the electron Coulomb potential energy depth is reduced.}
\label{fig:eVcompressed}
\end{figure}

These results have been extrapolated to the case of nuclear matter cores of stellar dimensions for $A\approx(m_{\rm Planck}/m_n)^3 \sim 10^{57}$ or $M_{core}\sim M_{\odot}$ (see Ref.~\refcite{2009arXiv0911.4622R} for details). The possibility of obtaining for these systems a self-consistent solution characterized by global but not local charge neutrality was explored there. The results generalize the considerations presented by Popov et al.\ in Ref.~\refcite{2009arXiv0903.3727P} corresponding to a nuclear matter core of stellar dimensions with zero Fermi energy of the electrons. An entire family of configurations exists with values of the electron Fermi energy ranging from zero to a maximum value $(E_e^F)_{max}$ which is reached when the Wigner-Seitz cell coincides with the core radius. The configuration with $E_e^F=(E_e^F)_{max}$ corresponds to the configuration with $N_p=N_e$ and $n_p=n_e$: for this limiting value of the Fermi energy  the system satisfies both global and local charge neutrality and, correspondingly, no electrodynamical structure is present in the core. The other configurations generally have overcritical electric fields close to their surface. The configuration with $E_e^F=0$ has the maximum value of the electric field at the core surface, well above the critical value $E_c$ (see Figs.~\ref{fig:eVcompressed} and \ref{fig:Ecompressed}). All these cores with overcritical electric fields are stable against the vacuum polarization process due to the Pauli blocking by the degenerate electrons \cite{2010PhR...487....1R}.

The above problem is theoretically well defined, and represents a necessary step in order to approach the more complex problem of a neutron star core and its interface with the neutron star crust. 

Neutron stars are composed of two sharply different components: the liquid core at nuclear and/or supra-nuclear densities consisting of neutrons, protons and electrons and a crust of degenerate electrons in a lattice of nuclei and possibly of free neutrons due to neutron drip when this process occurs (see e.g.\ Ref.~\refcite{1971NuPhA.175..225B}). Consequently, the boundary conditions for the electrons at the surface of the neutron star core will generally have a positive value of the electron Fermi energy in order to take into account the compressional effects of the neutron star crust on the core (see e.g.\ Ref.~\refcite{2009arXiv0911.4627R}). The case of zero electron Fermi energy corresponds to the limiting case of the absence of the crust. 

These considerations have been generalized in Ref.~\refcite{2009arXiv0911.4622R} by looking for a violation of the local charge neutrality condition over the entire configuration, still keeping its overall charge neutrality. This effect cannot occur locally, and requires a global description of the equilibrium configuration.

I illustrated\cite{ruedasubprl2} this novel approach with Rotondo, Rueda and Xue by considering the simplest, nontrivial self-gravitating system of degenerate neutrons, protons and electrons in beta equilibrium in the framework of relativistic quantum statistics and the Einstein-Maxwell equations. The impossibility of imposing the condition of local charge neutrality on such systems has been proven in complete generality. The crucial role of the constancy of the generalized electron Fermi energy was then emphasized and consequently the coupled system of the general relativistic Thomas-Fermi equations and the Einstein-Maxwell equations was solved. An explicit solution corresponding to a violation of the local charge neutrality condition over the entire star, still satisfying global charge neutrality when electromagnetic, weak and general relativistic effects are taken into account, is given there.

The results presented in Ref.~\refcite{2009arXiv0911.4622R} on nuclear matter cores of stellar dimensions evidence the possibility of having the existence of critical electromagnetic fields at the core surface. Such an analysis has been extended further by considering the case of a neutron star. At nuclear and supranuclear densities a core described by a self-gravitating system of degenerate neutrons, protons and electrons within the framework of relativistic quantum statistics and Einstein-Maxwell equations was considered\cite{2009arXiv0911.4627R}. At densities lower than the nuclear density such a core is surrounded by a crust. A globally neutral neutron star configuration was examined there in contrast with the traditional ones constructed by imposing local charge neutrality. To illustrate the application of this approach the Baym, Bethe and Pethick \cite{1971NuPhA.175..225B} strong interaction model of the baryonic matter in the core and in the white dwarf-like material of the crust was adopted there. The existence of an overcritical electric field at the boundary of the core as predicted in Ref.~\refcite{2009arXiv0911.4622R} was confirmed there. The electric field extends over a thin shell of thickness $\sim \hbar/(m_e c)$ between the core and the crust and becomes largely overcritical in the limit of decreasing values of the crust mass (see Refs.~\refcite{2009arXiv0911.4627R} for details). 

All the new gravito-electrodynamical effects discussed here deserve further analysis in view of the recent developments in high-energy astrophysics pointing to the relevance of overcritical electric fields in neutron stars and black holes \cite{2010PhR...487....1R}.

The extension of all the above considerations by describing the strong interaction between nucleons through sigma-omega-rho meson exchange in the context of the extended Walecka model, all properly expressed within general relativity, is currently being developed. In this case it is shown that, exactly as in the non-interacting case, the thermodynamic equilibrium condition given by the constancy of the Fermi energy of each particle-species can be properly generalized to include the contribution of all fields \cite{pugliese2011inprep}.

We were clearly motivated in this theoretical development by the study of GRBs. The gravitational collapse of a neutron star endowed with a critical electric field clearly leads to a violent pair creation process. The gravitational energy extraction process due to the presence of an electric field as envisaged in Fig.~\ref{fig3l2} now becomes  naturally implementable. The gravitational collapse process is triggered by the neutrons and the protons reaching of a fully relativistic regime. Consequently, their gravitational collapse leads to an increase of the charge-to-mass ratio of the core and an increase of the electric field given in Fig.~\ref{fig:Ecompressed} well above the critical value and over the Pauli blocking of the degenerate electrons of the core. This process is currently being studied. Due to the increase of density in the gravitational collapse in presence of an overcritical field, some striking analogies have surfaced with the physics of ultra high energy heavy ion collisions.

\subsection{Progress in the physics of dyadosphere: The dyado-torus of a rotating black hole}\label{sec:dt}

In parallel with the above investigations, we have generalized the concept of the dyadosphere to the more general Kerr-Newman geometry where the corresponding dyadosphere region is instead only axially symmetric. This approach leads to the concept of the ``dyadotorus,''
recently introduced in Ref.~\refcite{ruKerr}. The details have been presented in Ref.~\refcite{2009PhRvD..79l4002C}. The goal was to identify the region in the Kerr-Newman geometry where vacuum polarization
processes may occur, leading to the creation of $e^--e^+$ pairs, generalizing the original
concept of the ``dyadosphere'' initially introduced for Reissner-Nordstr\"{o}m
geometries \cite{1998bhhe.conf..167R,1998A&A...338L..87P}. The topology of the axially symmetric dyadotorus was studied for
selected values of the electric field, and its electromagnetic energy was estimated using three different methods all of which give the same result.

\begin{figure}[t]
\centering
\includegraphics[width=0.65\hsize,clip]{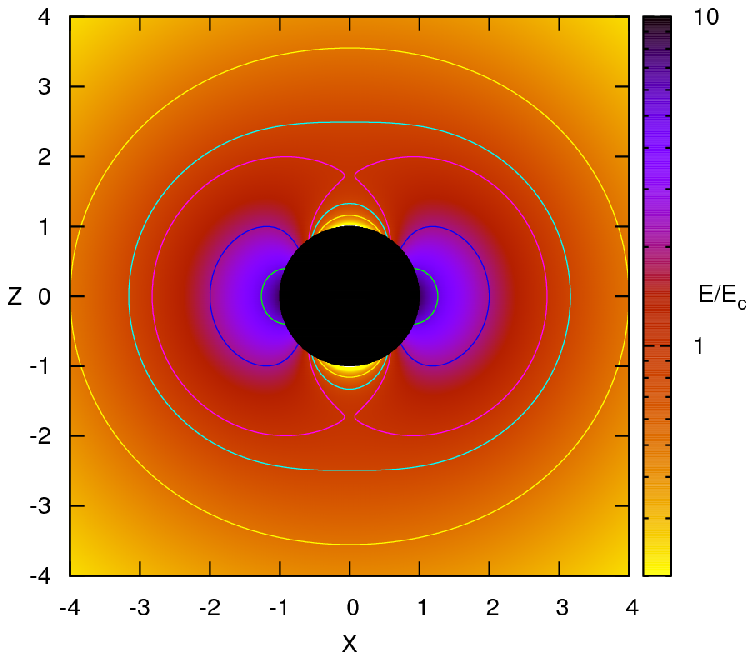}
\includegraphics[width=0.65\hsize,clip]{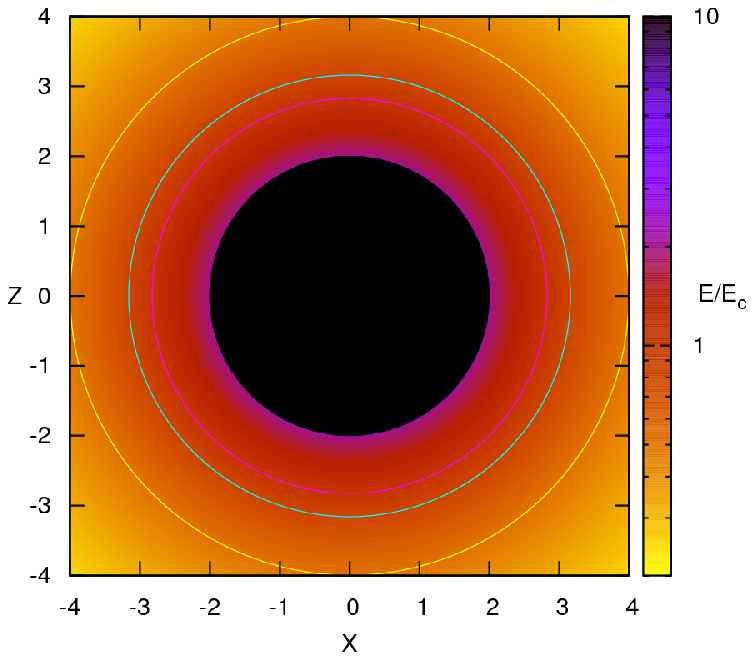}
\caption{The projections of the dyadotorus on the $X$-$Z$ plane corresponding to different values of the ratio $|{\bf E}|/E_c\equiv k$ are shown in the upper panel for $\mu=10$ and $\lambda=1.49\times 10^{-4}$.  The corresponding plot for the dyadosphere with the same mass energy and charge-to-mass ratio is shown in the lower panel for comparison. See details in Ref.~\refcite{2009PhRvD..79l4002C}.}
\label{fig:2bis}
\end{figure}

Vacuum polarization processes can occur in the overcritical field of a Kerr-Newman black hole inside the dyadotorus. Such a region has an invariant character, i.e.\ its existence does not depend on the observer measuring the electromagnetic field: therefore, it is a true physical region.

Some pictorial representations of the boundary surface similar to those commonly used in the literature have been presented in Ref.~\refcite{2009PhRvD..79l4002C} employing Cartesian-like coordinates (i.e.\ interpreting the Boyer-Lindquist radial and angular coordinates as ordinary spherical coordinates on flat space) as well as Kerr-Schild coordinates.
The dyadotorus has been also shown on the corresponding embedding diagram, which gives the correct geometry allowing the visualization of spacetime curvature.

We have then estimated the electromagnetic energy contained in the dyadotorus by using three different approaches, all of which give rise to the same final expression for the energy.
The first one follows the standard approach consisting of using the (unnormalized) timelike Killing vector through the Boyer-Lindquist constant time slice of the Kerr-Newman spacetime (see e.g.\ Ref.~\refcite{2002PhLB..545..233R}), the second one follows a recent observer dependent definition by Katz, Lynden-Bell and Bi{\v c}\'ak \cite{2006CQGra..23.7111K,2007PhRvD..75b4040L} for axially
symmetric asymptotically flat spacetimes, for which we have used the Painlev\'e-Gullstrand geodesic family of infalling observers through the Painlev\'e-Gullstrand constant time slice, and the last one adopts the pseudotensor theory (see e.g.\ Ref.~\refcite{1996GReGr..28.1393A}).
We have found by rough estimates that the extreme Kerr-Newman black hole leads to larger values of the electromagnetic energy as compared to a Reissner-Nordstr\"om black hole with the same total mass and charge.

It is appropriate to recall that the release of energy via the electron-positron pairs in the dyadotorus is the most powerful way to extract energy from black holes and in every way corresponds to a new form of energy: the ``blackholic'' energy \cite{ruKerr}.
This is a new form of energy different from the traditional ones known in astrophysics. The thermonuclear energy has been recognized to be energy source of main sequence stars lasting for $10^9$ years \cite{1968QB464.B46......}, the gravitational energy released by accretion processes in neutron stars and black holes has explained the energy observed in binary X-ray sources on time scales of $10^6-10^8$ years \cite{2003IJMPA..18.3127G}.
The ``blackholic'' energy appears to be the energy source for the most transient and most energetic events in the universe, the GRBs \cite{ruKerr}.

We now turn to the understanding of specific GRBs.

\section{On a new class of ``disguised'' short GRBs within the Fireshell model}\label{disguised}

I proceed now to some recent applications of the Fireshell model to infer properties of GRBs and of their progenitors.

In the current literature, ``long'' GRBs are traditionally related to the idea of a single progenitor, identified as a ``collapsar'' \cite{1993ApJ...405..273W}. Similarly, short GRBs are assumed to originate from binary mergers formed by white dwarfs, neutron stars, and black holes in all possible combinations. It also has been suggested that short and long GRBs originate from different galaxy types. In particular, short GRBs are proposed to be associated with galaxies with low specific star forming rate \cite{2009ApJ...690..231B}. Some evidence against such a scenario have been advanced, due to the small sample size and the different estimates of the star forming rates \cite{2009ApJ...691..182S}.

I give some specific examples of how the understanding of GRB structure and of its relation to the CBM distribution, within the fireshell model, leads to a more complex and interesting perspective than the one in the ``majority view'' of the current literature.

\begin{figure}[t]
\centering
\includegraphics[width=\hsize,clip]{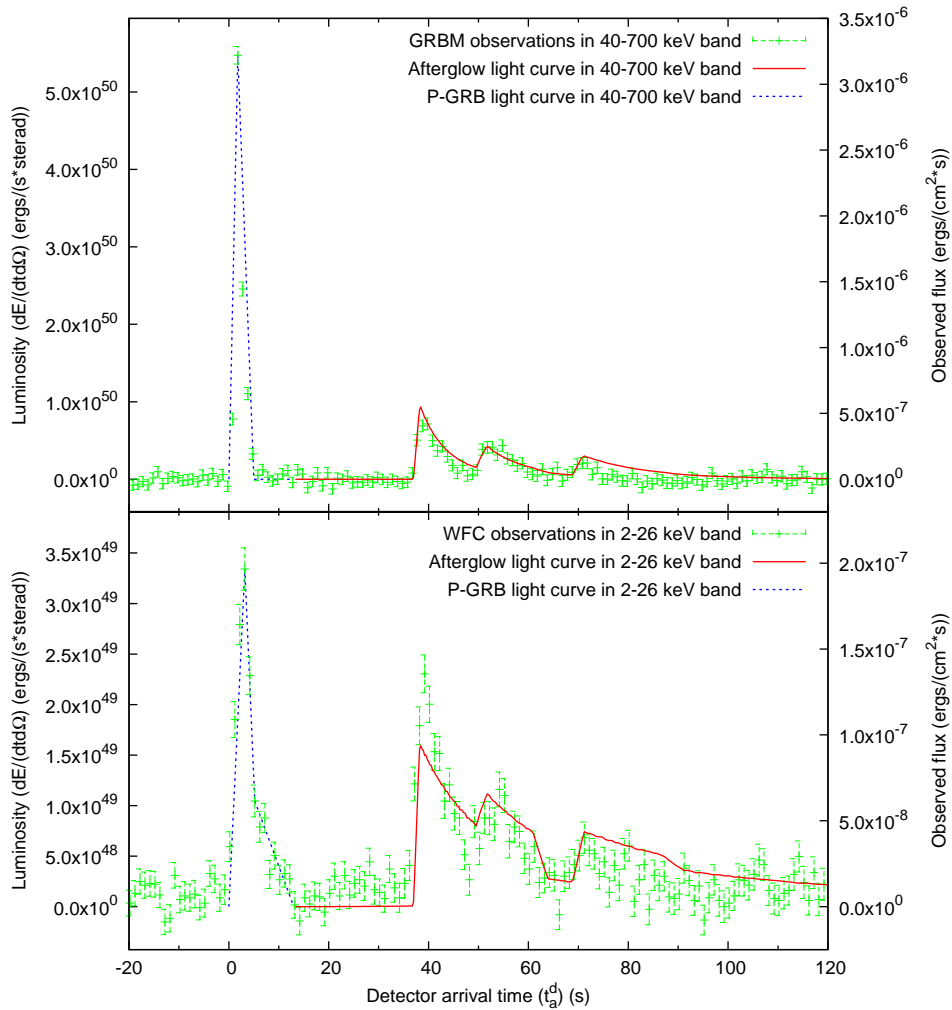}
\caption{The ``canonical GRB'' light curve computed theoretically for the prompt emission of GRB 970228. The BeppoSAX GRBM ($40$--$700$ keV, above) and WFC ($2$--$26$ keV, below) light curves (data points) are compared with the extended afterglow peak theoretical ones (solid red lines). The onset of the extended afterglow coincides with the end of the P-GRB (represented qualitatively by the dotted blue lines). For this source we have $B\simeq 5.0\times 10^{-3}$ and $\langle n_{cbm} \rangle \sim 10^{-3}$ particles/cm$^3$. The total time-integrated energy emitted in the P-GRB is therefore much lower than the one emitted in the extended afterglow, but this last one has its peak luminosity ``deflated'' by the low CBM density. See details in Ref.~\refcite{2007A&A...474L..13B}.}
\label{970228_fit_prompt}
\end{figure}

\begin{figure}[t]
\centering
\includegraphics[width=\hsize,clip]{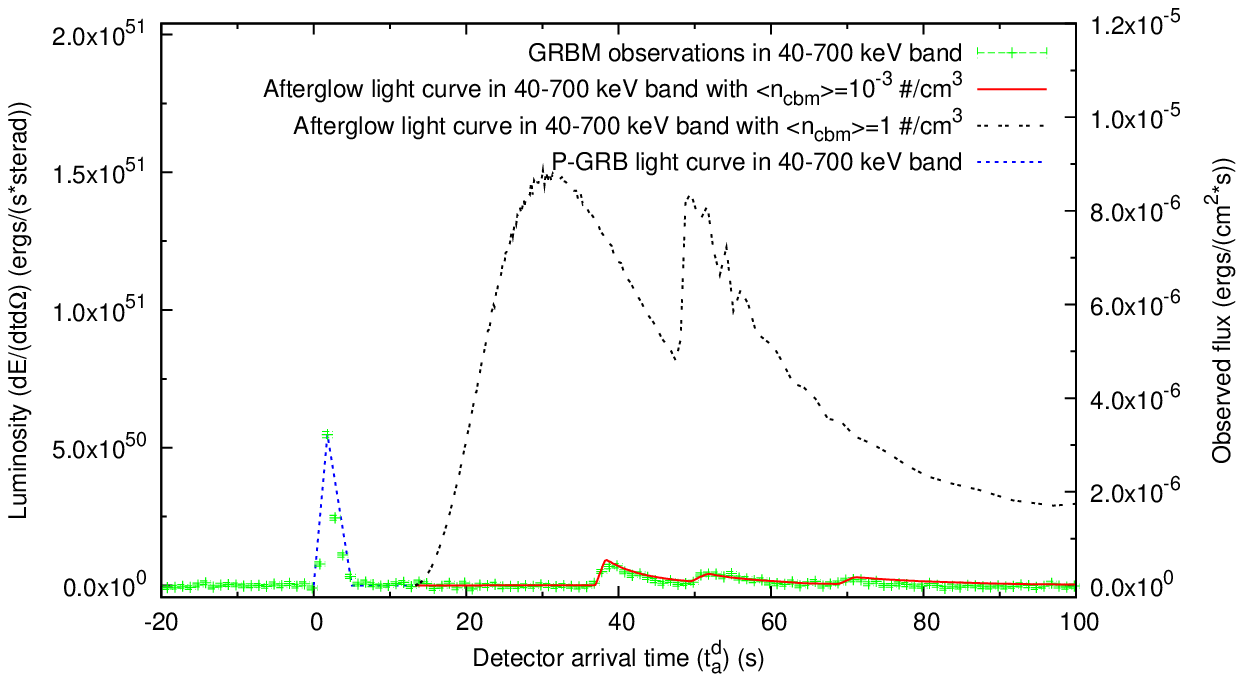}
\caption{The theoretical fit of the BeppoSAX GRBM observations (solid red line, see Fig.~\ref{970228_fit_prompt}) is compared with the extended afterglow light curve in the $40$--$700$ keV energy band obtained rescaling the CBM density to $\langle n_{cbm} \rangle = 1$ particle/cm$^3$ keeping constant its shape and the values of the fundamental parameters of the theory $E_{e^\pm}^{tot}$ and $B$ (black double-dotted line). The P-GRB duration and luminosity (blue dotted line), depending only on $E_{e^\pm}^{tot}$ and $B$, are not affected by this process of rescaling the CBM density. See details in Ref.~\refcite{2007A&A...474L..13B}.}
\label{picco_n=1}
\end{figure}

In the context of the fireshell model, we have considered a new class of GRBs pioneered by Norris and Bonnell \cite{2006ApJ...643..266N}. This class is characterized by an occasional softer extended emission after an initial spike-like emission. The softer extended emission has a peak luminosity lower than the one of the initial spike-like emission. As shown in the prototypical case of GRB 970228 \cite{2007A&A...474L..13B} we developed with Maria Grazia Bernardini, and then in GRB 060614 \cite{2009A&A...498..501C} and in GRB 071227 \cite{2010A&A...521A..80C} we developed with Letizia Caito, we have identified the initial spike-like emission with the P-GRB and the softer extended emission with the peak of the extended afterglow (see Fig.~\ref{970228_fit_prompt}). That the time-integrated extended afterglow luminosity is much larger than the P-GRB one is crucial. This unquestionably identifies these sources as canonical GRBs with $B > 10^{-4}$ (see Figs.~\ref{ftemp-fgamma-bcross}-\ref{f2}). The consistent application of the fireshell model allows us to infer the CBM filamentary structure and average density, which, in that specific case, is $n_{cbm} \sim 10^{-3}$ particles/cm$^3$ \cite{2007A&A...474L..13B}. This low CBM density value explains the peculiarity of the low extended afterglow peak luminosity and its more protracted time evolution (see Fig.~\ref{picco_n=1}). These features are not intrinsic to the progenitor, but depend uniquely on the peculiarly low value of the CBM density.

This led us to expand the traditional classification of GRBs to three classes:

\begin{itemize}
\item \textbf{``Genuine'' short GRBs:} These are the ones in which the total energy emitted in the P-GRB is greater than the one emitted in the extended afterglow. They are characterized by a small value of the fireshell baryon loading $B \lesssim 10^{-5}$.
\item \textbf{``Disguised'' short GRBs:} These are the ones in which the total energy emitted in the P-GRB is smaller than that emitted in the extended afterglow, but the P-GRB peak luminosity is greater than that of the extended afterglow. They are characterized by a high value of the fireshell baryon loading $B \gtrsim 3.0 \times 10^{-4}$, and a low value of the CBM average density $n_{cbm} \sim 10^{-3}$ particles/cm$^3$ which ``deflates'' the extended afterglow peak luminosity.
\item \textbf{The remaining ``long duration'' ones:} These are the ones in which the total energy emitted in the P-GRB is smaller than that emitted in the extended afterglow, and the P-GRB peak luminosity is smaller than that of the extended afterglow. They are characterized by a high value of the fireshell baryon loading $B \gtrsim 3.0 \times 10^{-4}$, and a high value of the CBM average density $n_{cbm} \geq 1$ particle/cm$^3$.
\end{itemize}

A CBM density $n_{cbm} \sim 10^{-3}$ particles/cm$^3$ is typical of a galactic halo environment, and GRB 970228 was indeed found to be in the halo of its host galaxy \cite{1997Natur.387R.476S,1997Natur.386..686V}. We therefore proposed that the progenitors of this new class of ``disguised'' short GRBs are merging binary systems, formed by neutron stars and/or white dwarfs in all possible combinations, which spiraled out from their birth place into the halo \cite{2007A&A...474L..13B,2009A&A...498..501C,KMG11}. This hypothesis can also be supported by other observations (e.g.\ in the optical band), clearly showing an offset of the GRB position from the center of the host galaxy (see e.g.\ the case of GRB 050509b \cite{2006ApJ...638..354B}). Vice versa, there is also evidence for the opposite correlation, directly relating the soft tail peak luminosity to the CBM density: GRBs displaying a more luminous prolonged soft tail appear to have a systematically smaller offset from the center of their host galaxy\cite{2008MNRAS.385L..10T,2010ApJ...708....9F,2010arXiv1005.1068B}.

I turn now to some specific examples of ``disguised'' short GRBs.

\section{Application to GRB 970228}\label{sec970228}

The GRB 970228 has been the first ``afterglow'' ever detected \cite{1997Natur.387..783C}. It was quite surprising that, as we returned to its analysis years later with M.G. Bernardini in Ref.~\refcite{2007A&A...474L..13B}, it was indeed discovered that this GRB became the prototype for the disguised short GRB class identified by Norris \& Bonnell \cite{2006ApJ...643..266N}.

The GRB 970228 was detected by the Gamma-Ray Burst Monitor (GRBM, $40$--$700$ keV) and Wide Field Cameras (WFC, $2$--$26$ keV) on board BeppoSAX on February $28.123620$ UT \cite{1998ApJ...493L..67F}. The burst prompt emission is characterized by an initial $5$ s strong pulse followed, after $30$ s, by a set of three additional pulses of decreasing intensity \cite{1998ApJ...493L..67F}. Eight hours after the initial detection, the NFIs on board BeppoSAX were pointed at the burst location for a first target of opportunity observation and a new X-ray source was detected in the GRB error box\cite{1997Natur.387..783C}. A fading optical transient has been identified in a position consistent with the X-ray transient \cite{1997Natur.386..686V}, coincident with a faint galaxy with redshift $z=0.695$ \cite{2001ApJ...554..678B}. Further observations by the Hubble Space Telescope clearly showed that the optical counterpart was located in the outskirts of a late-type galaxy with an irregular morphology \cite{1997Natur.387R.476S}.

The BeppoSAX observations of the GRB 970228 prompt emission revealed a discontinuity in the spectral index between the end of the first pulse and the beginning of the three additional ones \cite{1997Natur.387..783C,1998ApJ...493L..67F,2000ApJS..127...59F}. The spectrum during the first $3$ s of the second pulse is significantly harder than during the last part of the first pulse \cite{1998ApJ...493L..67F,2000ApJS..127...59F}, while the spectrum of the last three pulses appear to be consistent with the late X-ray afterglow \cite{1998ApJ...493L..67F,2000ApJS..127...59F}. This was soon recognized by Ref.~\refcite{1998ApJ...493L..67F,2000ApJS..127...59F} as pointing to an emission mechanism producing the X-ray afterglow already taking place after the first pulse.

As recalled above, the simultaneous occurrence of an extended afterglow with total time-integrated luminosity larger than that of the P-GRB, but with a smaller peak luminosity, was indeed explainable in terms of a peculiarly small average value of the CBM density and not due to the intrinsic nature of the source. We have shown\cite{2007A&A...474L..13B} that GRB 970228 is a very clear example of this situation. We identify the initial spikelike emission with the P-GRB, and the late soft bump with the peak of the extended afterglow. GRB 970228 shares the same morphology and observational features with the sources analyzed by Norris \& Bonnell \cite{2006ApJ...643..266N}, e.g.\ GRB 050709 \cite{2005Natur.437..855V}, GRB 050724 \cite{2006A&A...454..113C} as well as GRB 060614 \cite{2006Natur.444.1044G}. Therefore, we proposed GRB 970228 as the prototype for this new GRB class\cite{2007A&A...474L..13B}.

\subsection{The analysis of the GRB 970228 prompt emission}\label{theo}

\begin{figure}[t]
\centering
\includegraphics[width=\hsize,clip]{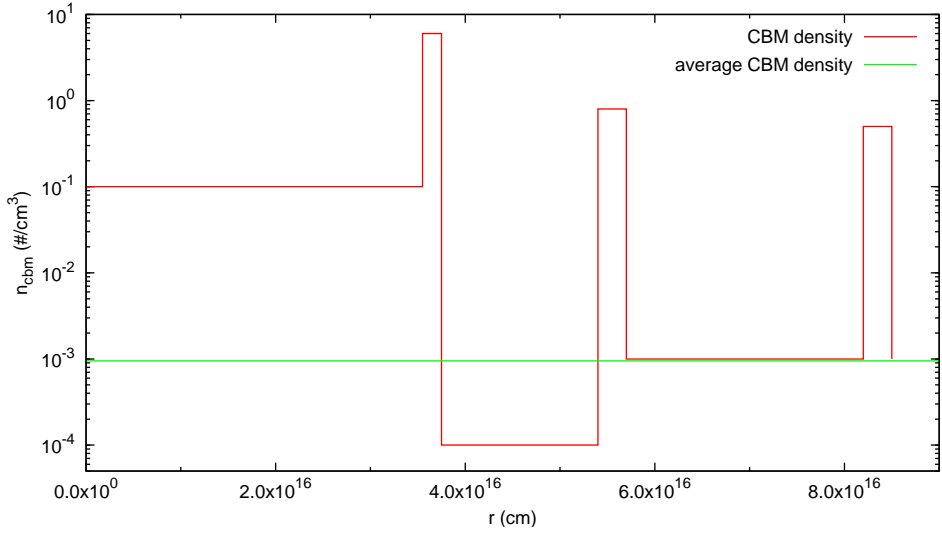}
\caption{The CBM density profile we assumed to reproduce the last three pulses of the GRB 970228 prompt emission (red line), together with its average value $\langle n_{cbm} \rangle = 9.5\times 10^{-4}$ particles/cm$^3$ (green line). Details in Ref.~\refcite{2007A&A...474L..13B}.}
\label{mask}
\end{figure}

In Fig.~\ref{970228_fit_prompt} I recall the theoretical fit of the BeppoSAX GRBM ($40$--$700$ keV) and WFC ($2$--$26$ keV) light curves of the GRB 970228 prompt emission \cite{1998ApJ...493L..67F} performed by Bernardini et al. \cite{2007A&A...474L..13B}. We identified the first main pulse with the P-GRB and the three additional pulses with the extended afterglow peak emission, consistent with the above mentioned observations by Ref.~\refcite{1997Natur.387..783C} and Ref.~\refcite{1998ApJ...493L..67F}. The last three such pulses have been reproduced assuming three overdense spherical CBM regions (see Fig.~\ref{mask}) with a very good agreement (see Fig.~\ref{970228_fit_prompt}).

We therefore obtain for the two parameters characterizing the source in our model $E_{e^\pm}^{tot}=1.45\times 10^{54}$ erg and $B = 5.0\times 10^{-3}$ \cite{2007A&A...474L..13B}. This implies an initial $e^+e^-$ plasma created between the radii $r_1 = 3.52\times10^7$ cm and $r_2 = 4.87\times10^8$ cm with a total number of $e^+e^-$ pairs $N_{e^\pm} = 1.6\times 10^{59}$ and an initial temperature $T = 1.7$ MeV \cite{2007A&A...474L..13B}. The theoretically estimated total isotropic energy emitted in the P-GRB is $E_{P\hbox{\small-}GRB}=1.1\% E_{e^\pm}^{tot}=1.54 \times 10^{52}$ erg, in excellent agreement with the energy observed in the first main pulse ($E_{P\hbox{\small-}GRB}^{obs} \sim 1.5 \times 10^{52}$ erg in $2-700$ keV energy band, see Fig.~\ref{970228_fit_prompt}), as expected due to their identification \cite{2007A&A...474L..13B}.

After the transparency point at $r_0 = 4.37\times 10^{14}$ cm from the progenitor, the initial Lorentz gamma factor of the fireshell is $\gamma_0 = 199$. On average, during the extended afterglow peak emission phase we have for the CBM $\langle {\cal R} \rangle = 1.5\times 10^{-7}$ and $\langle n_{cbm} \rangle = 9.5\times 10^{-4}$ particles/cm$^3$. This very low average value for the CBM density is compatible with the observed occurrence of GRB 970228 in its host galaxy's halo \cite{1997Natur.387R.476S,1997Natur.386..686V,2006MNRAS.367L..42P} and it is crucial in explaining the light curve behavior.

The values of $E_{e^\pm}^{tot}$ and $B$ we determined are univocally fixed by two tight constraints. The first one is the total energy emitted by the source all the way up to the latest afterglow phases (i.e.\ up to $\sim 10^6$ s). The second one is the ratio between the total time-integrated luminosity of the P-GRB and the corresponding one of the whole extended afterglow (i.e.\ up to $\sim 10^6$ s). In particular, in GRB 970228 such a ratio turns out to be $\sim 1.1\%$, typical of a ``long'' GRB (see Figs.~\ref{ftemp-fgamma-bcross}-\ref{f2}).

\subsection{Rescaling the CBM density}\label{rescale}

I recall now an explicit example in order to probe the crucial role of the average CBM density in explaining the relative intensities of the P-GRB and of the extended afterglow peak in GRB 970228 (for details see Ref.~\refcite{2007A&A...474L..13B}). We keep fixed the basic parameters of the source, namely the total energy $E_{e^\pm}^{tot}$ and the baryon loading $B$, therefore keeping fixed the P-GRB and the extended afterglow total time-integrated luminosities. Then in a gedanken experiment we rescale the CBM density profile given in Fig.~\ref{mask} by a constant numerical factor in order to raise its average value to the standard one $\langle n_{cbm} \rangle = 1$ particle/cm$^3$. We then compute the corresponding light curve, shown in Fig.~\ref{picco_n=1}.

We notice a clear enhancement of the extended afterglow peak luminosity with respect to the P-GRB one, when compared to and contrasted with the observational data presented in Fig.~\ref{970228_fit_prompt}. The two light curves actually cross at $t_a^d \simeq 1.8\times 10^4$ s since their total time-integrated luminosities must be the same. The GRB ``rescaled'' to $\langle n_{cbm} \rangle = 1$ particle/cm$^3$ appears to be totally similar to, e.g., GRB 050315 \cite{2006ApJ...645L.109R} and GRB 991216 \cite{2003AIPC..668...16R,2004IJMPD..13..843R,2005AIPC..782...42R}.

It is appropriate to emphasize that, although the two underlying CBM density profiles differ by a constant numerical factor, the two extended afterglow light curves in Fig.~\ref{picco_n=1} do not. This is because the absolute value of the CBM density at each point affects in a nonlinear way all the subsequent evolution of the fireshell due to the feedback on its dynamics \cite{2005ApJ...633L..13B}. Moreover, the shape of the surfaces of equal arrival time of the photons at the detector (EQTS) is strongly elongated along the line of sight \cite{2005ApJ...620L..23B}. Therefore, this is a good example of how photons coming from the same CBM density region are observed over a very long arrival time interval during the extended afterglow.

\subsection{GRB 970228 and the Amati relation}\label{amati_rel}

It is appropriate now to turn to the ``Amati relation'' \cite{2002A&A...390...81A,2006MNRAS.372..233A} between the isotropic equivalent energy emitted in the prompt emission $E_{iso}$ and the peak energy of the corresponding time-integrated spectrum $E_{p,i}$ in the source rest frame. It has been shown by Ref.~\refcite{2002A&A...390...81A,2006MNRAS.372..233A} that this correlation holds for almost all the ``long'' GRBs which have a redshift and a measured $E_{p,i}$ , but not for the ones classified as ``short'' \cite{2006MNRAS.372..233A}.

It clearly follows from our treatment that for the ``canonical GRBs'' with large values of the baryon loading and high $\left\langle n_{cbm}\right\rangle$, which presumably are most of the GRBs for which the correlation holds, the leading contribution to the prompt emission is the extended afterglow peak emission. The case of the ``fake'' short GRBs is completely different: it is crucial to consider both components separately since the P-GRB contribution to the prompt emission in this case is significant.

\begin{figure}[t]
\centering
\includegraphics[width=\hsize,clip]{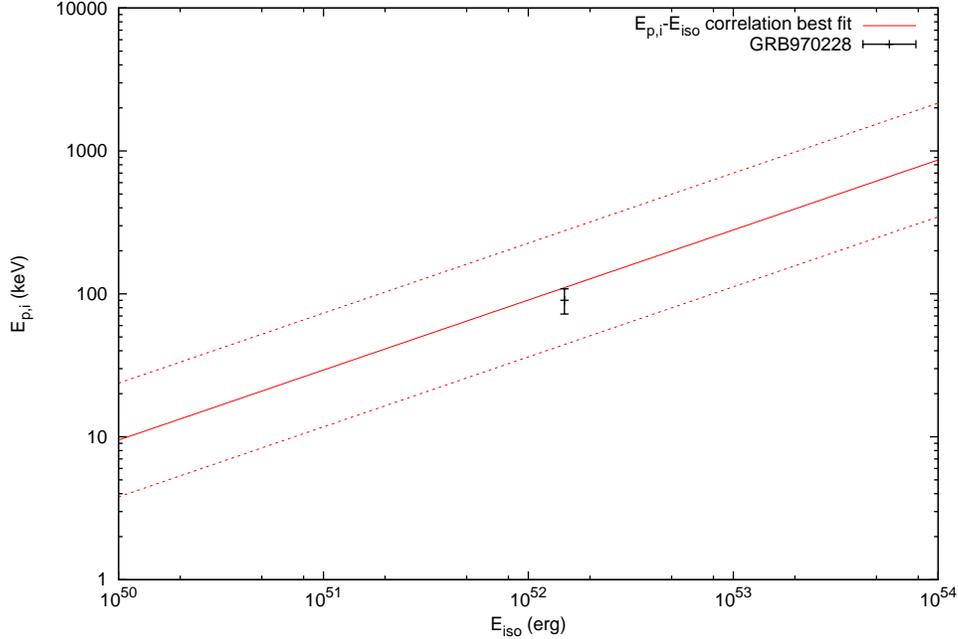}
\caption{The estimated values for $E_{p,i}$ and $E_{iso}$ obtained by our analysis (black dot) compared with the ``Amati relation'' \cite{2002A&A...390...81A}: the solid line is the best fitting a power law \cite{2006MNRAS.372..233A} and the dashed lines delimit the region corresponding to a vertical logarithmic deviation of $0.4$ \cite{2006MNRAS.372..233A}. The uncertainty in the theoretical estimated value for $E_{p,i}$ has been assumed conservatively to be $20\%$. See details in Ref.~\refcite{2008AIPC..966....7B}.}
\label{amati_2}
\end{figure}

To test this scenario, we evaluated from our fit of GRB 970228 $E_{iso}$ and $E_{p,i}$ only for the extended afterglow peak emission component, i.e.\ from $t_a^d= 37$ s to $t_a^d= 81.6$ s. We found an isotropic energy emitted in the $2$--$400$ keV energy band $E_{iso}=1.5 \times 10^{52}$ erg, and $E_{p,i}=90.3$ keV \cite{2008AIPC..966....7B}. As is clearly shown in Fig.~\ref{amati_2}, the sole extended afterglow component of GRB 970228 prompt emission is in perfect agreement with the Amati relation \cite{2008AIPC..966....7B}.

From the energetics of this source, as well as from its location in the galactic halo and the low density of the CBM, we conclude that the progenitors of GRB 970228 are merging binary neutron stars.

\section{Application to GRB 060614}\label{060614}

I turn now to GRB 060614 \cite{2006Natur.444.1044G,2007A&A...470..105M} for at least three different reasons.

The first novelty is that it is the first clear example of a nearby ($z=0.125$) long GRB not associated with a bright Ib/c supernova (SN) \cite{2006Natur.444.1050D,2006Natur.444.1053G}. It has been estimated that, if present, the SN-component should be about $200$ times fainter than the archetypal SN 1998bw associated with GRB 980425; moreover, it would also be fainter (at least $30$ times) than any stripped-envelope SN ever observed \cite{2006AJ....131.2233R}.

Within the standard scenario, long duration GRBs ($T_{90} > 2$ s) are thought to be produced by SN events during the collapse of massive stars in star forming regions (``collapsar'')\cite{1993ApJ...405..273W}. The observations of broad-lined and bright type Ib/c SNe associated with GRBs are often reported to favor this scenario (see Ref.~\refcite{2006ARA&A..44..507W} and references therein). The \emph{ansatz} of the collapsar model has been that every long GRB should have a SN associated with it \cite{2007ApJ...655L..25Z}. Consequently, in all nearby long GRBs ($z \leq 1$), SN emission should be observed.

For these reasons the case of GRB 060614 is of great relevance. Some obvious hypotheses have been proposed and ruled out: the chance superposition with a galaxy at low redshift \cite{2006Natur.444.1053G} and strong dust obscuration and extinction \cite{2006Natur.444.1047F}. Appeal has been made to the possible occurrence of an unusually low luminosity stripped-envelope core-collapse SN \cite{2006Natur.444.1050D}.

The second novelty of GRB 060614 is that it challenges the traditional separation between long soft GRBs and short hard GRBs. Traditionally \cite{1992grbo.book..161K,1992AIPC..265..304D}, the ``short'' GRBs have $T_{90} < 2$ s, present an harder spectrum and negligible spectral lag, and are assumed to originate from the merging of two compact objects, i.e.\ two neutron stars or a neutron star and a black hole (see e.g.\ Refs.~\refcite{1984SvAL...10..177B,1986ApJ...308L..43P,1986ApJ...308L..47G,1989Natur.340..126E,2005RvMP...76.1143P,2006RPPh...69.2259M} and references therein). GRB 060614 lasts about one hundred seconds ($T_{90}=(102 \pm 5)$ s)\cite{2006Natur.444.1044G}, it fulfills the $E_{p}^{rest}$-$E_{iso}$ correlation \cite{2007A&A...463..913A}, and therefore traditionally it should be classified as a ``long'' GRB. However, its morphology is different from typical long GRBs \cite{2007ApJ...655L..25Z,2005Natur.437..822P}. Its optical afterglow luminosity is intermediate between the traditional long and short ones \cite{2008arXiv0804.1959K}. Its host galaxy has a moderate specific star formation rate ($R_{Host}\approx2M_{s}y^{-1}(L^{*})^{-1}$, $M_{vHost}\approx-15.5$)\cite{2006Natur.444.1047F,2006Natur.444.1050D}. The spectral lag in its light curves is very small or absent \cite{2006Natur.444.1044G}. All these features are typical of short GRBs.

A third novelty of GRB 060614 is that its $15$--$150$ keV light curve presents a short, hard and multi-peaked episode (about $5$ s). The episode is followed by a softer, prolonged emission that manifests a strong hard to soft evolution in the first $400$ s of data \cite{2007A&A...470..105M}. The total fluence in the $15$--$150$ keV energy band is $F=(2.17\pm0.04)\times10^{-5}$ erg/cm$^2$, the 20\% emitted during the initial spikelike emission, where the peak luminosity reaches the value of $300$ keV before decreasing to $8$ keV during the BAT-XRT overlap time (about $80$ s).

These apparent contradictions find a natural explanation in the framework of the ``fireshell'' model, as explicitly shown with L. Caito\cite{2009A&A...498..501C}. Within the fireshell model, the occurrence of a GRB-SN is not a necessity.

\subsection{The fit of the observed luminosity}\label{fit}

\begin{figure}[t]
\centering
\includegraphics[width=\hsize,clip]{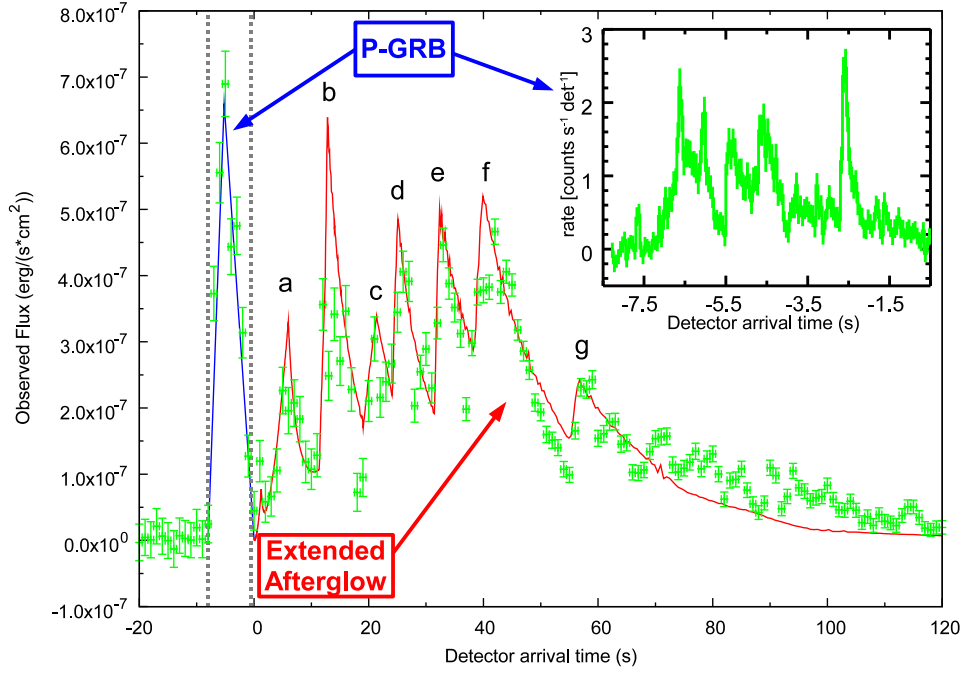}
\caption{The BAT $15$--$150$ keV light curve (green points) at $1$ s time resolution compared with the corresponding theoretical extended afterglow light curve we compute (red line). The onset of the extended afterglow is at the end of the P-GRB (qualitatively sketched in blue lines and delimited by dashed gray vertical lines). Therefore the zero of the temporal axis is shifted by $5.5$ s with respect to the BAT trigger time. The peaks of the extended afterglow light curves are labeled to match them with the corresponding CBM density peak in Fig.~\ref{f4a}. In the upper right corner there is an enlargement of the P-GRB at $50$ms time resolution (reproduced from Ref.~\refcite{2007A&A...470..105M}) showing its structure. See details in Ref.~\refcite{2009A&A...498..501C}.}
\label{f2a}
\end{figure}

\begin{figure}[t]
\centering
\includegraphics[width=\hsize,clip]{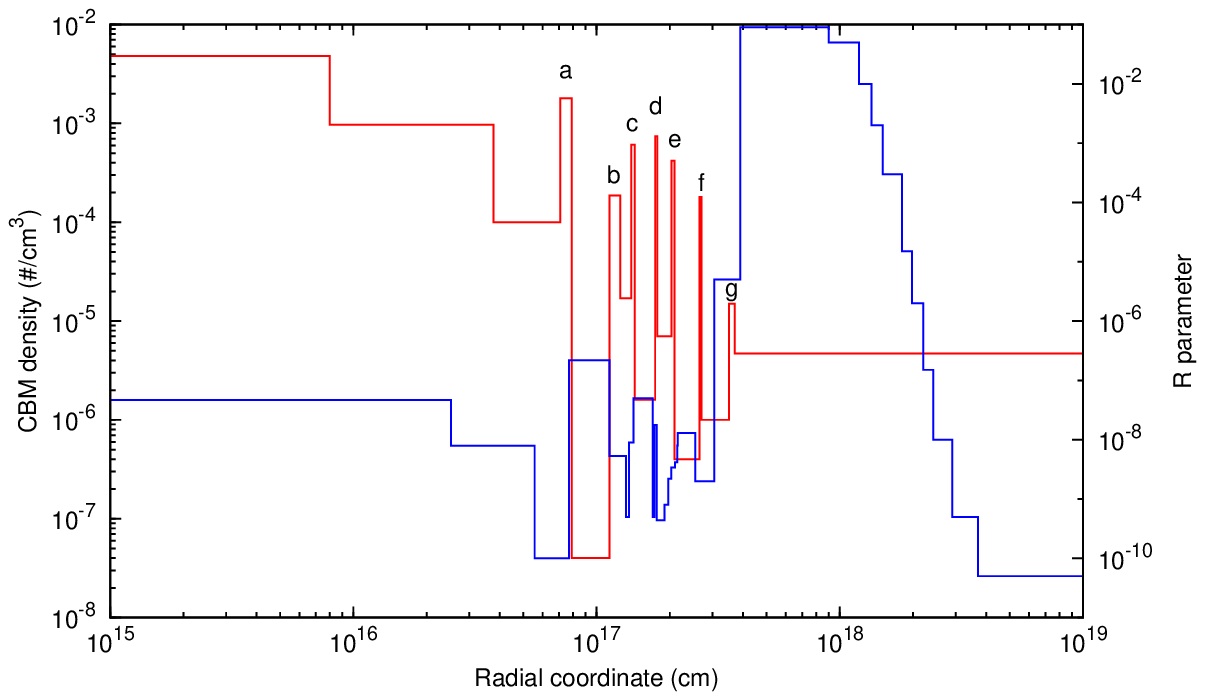}
\caption{The effective CBM density (red line) and the ${\cal R}$ parameter (blue line) versus the radial coordinate of the shell. The CBM density peaks are labeled to match them with the corresponding extended afterglow light curve peaks in Fig.~\ref{f2a}. They correspond to filaments of characteristic size $\Delta r \sim 10^{15}$ cm and density contrast $\Delta n_{cbm}/\langle n_{cbm} \rangle \sim 20$ particles/cm$^3$. See details in Ref.~\refcite{2009A&A...498..501C}.}
\label{f4a}
\end{figure}

We have proceeded to interpret GRB 060614 as a ``disguised'' short GRB\cite{2009A&A...498..501C}. We have performed the analysis of the observed light curves in the $15$--$150$ keV energy band, corresponding to the $\gamma$-ray emission observed by the BAT instrument on the Swift satellite, and in the $0.2$--$10$ keV energy band, corresponding to the X-ray component from the XRT instrument on the Swift satellite. The optical emission represents less than 10\% of the total energy of the GRB and is therefore neglected. From this fit (see Fig.~\ref{f2a}) we have derived the total initial energy $E_{tot}^{e^\pm}$, the value of $B$ as well as the effective CBM distribution (see Fig.~\ref{f4a}). We find $E_{tot}^{e^\pm}=2.94\times10^{51}$ erg, which accounts for the bolometric emission of both the P-GRB and the extended afterglow. Such a value is compatible with the observed $E_{iso} \simeq 2.5\times10^{51}$ erg \cite{2006Natur.444.1044G}. The value of $B$ is $B=2.8\times10^{-3}$ corresponds to a canonical GRB with a very clear extended afterglow which energetically dominates the P-GRB. From the model, having determined $E_{tot}^{e^\pm}$ and $B$, we can compute the theoretically expected P-GRB energetics $E_{P\hbox{\small-}GRB}$ \cite{2001ApJ...555L.113R}. We obtain $E_{P\hbox{\small-}GRB} \simeq 1.15 \times 10^{50}$ erg, which is in good agreement with the observed $E_{iso,1p} \simeq 1.18\times10^{50}$ erg \cite{2006Natur.444.1044G}. The Lorentz gamma factor at the transparency point is $\gamma_\circ=346$, one of the highest of all the GRBs we have examined.

In Fig.~\ref{f2a} we plot the comparison between the BAT observational data of the GRB 0606014 prompt emission in the $15$--$150$ keV energy range and the P-GRB and extended afterglow light curves computed within our model. The temporal variability of the extended afterglow peak emission is due to the inhomogeneities in the effective CBM density (see Figs.~\ref{f2a}, \ref{f4a}). Toward the end of the BAT light curve, the good agreement between the observations and the fit is affected by the Lorentz gamma factor decrease and the corresponding increase of the maximum viewing angle. The source visible area becomes larger than the typical size of the filaments. This invalidates the radial approximation we use for the CBM description. To overcome this problem it is necessary to introduce a more detailed three-dimensional CBM description, in order to avoid an over-estimated area of emission and, correspondingly, to describe the sharpness of some observed light curves. We are still working on this issue \cite{2002ApJ...581L..19R,C07,Venezia_Flares,G07}.

We turn now to the crucial determination of the CBM density, which is derived from the fit. At the transparency point it has the value $n_{cbm} = 4.8 \times 10^{-3}$ particles/cm$^3$ (see Fig.~\ref{f4a}). This density is compatible with the typical values of the galactic halos. During the peak of the extended afterglow emission the effective average CBM density decreases reaching $\left\langle n_{cbm}\right\rangle = 2.25 \times 10^{-5}$ particles/cm$^3$, possibly due to an ongoing fragmentation of the shell \cite{2007A&A...471L..29D} or due to a fractal structure in the CBM. The ${\cal R}$ value on average was $\left\langle {\cal R}\right\rangle = 1.72 \times 10^{-8}$. Note the striking analogy of the numerical value and the overall radial dependence of the CBM density in the present case of GRB 060614 when compared and contrasted with the ones of GRB 970228 \cite{2007A&A...474L..13B}. More details can be found in Ref.~\refcite{2009A&A...498..501C}.

In view of the low CBM density and the overall energetics, which is considerably smaller than for  GRB 970228, we conclude that the progenitor of GRB 060614 is a binary merger formed by a neutron star and a white dwarf. Both GRB 970228 and GRB 060614 are clear counterexamples to the proposal that binary mergers lead necessarily to short GRBs \cite{2006RPPh...69.2259M,2009ARA&A..47..567G}. In both cases they lead to long GRBs disguised as short ones.

\section{GRB 071227: an additional case of a \textit{disguised} short burst}\label{sec071227}

GRB 071227 presents some intriguing anomalies and has been shown to represent a third case of a disguised short GRB \cite{2010A&A...521A..80C}. As for the ``Norris and Bonnel'' GRBs, its BAT light curve shows in the $15$--$150$ keV range a multi-peaked structure lasting $T_{90}=(1.8\pm0.4)$ s, followed by an extended but much softer emission up to $t_0+100$ s \cite{2009A&A...498..711D}. A fading X-ray ($0.3$--$10$ keV) and a faint optical afterglow have also been identified. The optical afterglow emission allowed the measurement of its redshift, $z=0.383$, and therefore of its isotropic equivalent energy, $E_{iso}=5.8\times10^{50}$ erg in $20$--$1300$ keV \cite{2009A&A...498..711D}. The observed X-ray and optical afterglow is superimposed on the plane of the host galaxy, at $(15.0\pm2.2)$ kpc from its center.

On the basis of these characteristics, GRB 071227 has been classified as a short burst. This statement is supported by other main features: {\bf 1)} If we consider the first and apparently predominant short-duration episode, it does not satisfy the Amati relation between the isotropic equivalent radiated energy of the prompt emission $E_{iso}$ and the cosmological rest-frame $\nu F_{\nu}$ spectrum peak energy $E_{p,i}$ \cite{2002A&A...390...81A,2006MNRAS.372..233A,2007A&A...463..913A,2009A&A...508..173A}. {\bf 2)} The spectral lag of the first spike-like emission in the $25$--$50$ keV to $100$--$350$ keV bands is consistent with zero \cite{2007GCN..7156....1S}. {\bf 3)} Multiwavelength observations performed over many days have shown that there is no association with a Ib/c hypernova, the type of SN generally observed with GRBs, even if it is a nearby burst and its isotropic energy is compatible with that of other GRBs associated with them \cite{2009A&A...498..711D}, although the upper limits are not deep enough to rule out a low-energetic core-collapse event. Nevertheless, the explosion of this burst in a star-forming region of a spiral galaxy and its prolonged tail of emission make it most likely to be a long burst.

With L. Caito\cite{2010A&A...521A..80C} we show that all these ambiguities and peculiarities can be explained in the framework of the fireshell model if we assume GRB 071227 to be a \textit{disguised} short burst, in which the first spike-like emission coincides with the P-GRB and the prolonged softer tail with the peak of the extended afterglow emitted in a low CBM density region. We show, moreover, that this tail satisfies the Amati relation, and this is consistent with our interpretation.

\subsection{The interpretation of the GRB 071227 light curves}\label{071227_fit}

\begin{figure}[t]
\centering
\includegraphics[width=\hsize,clip]{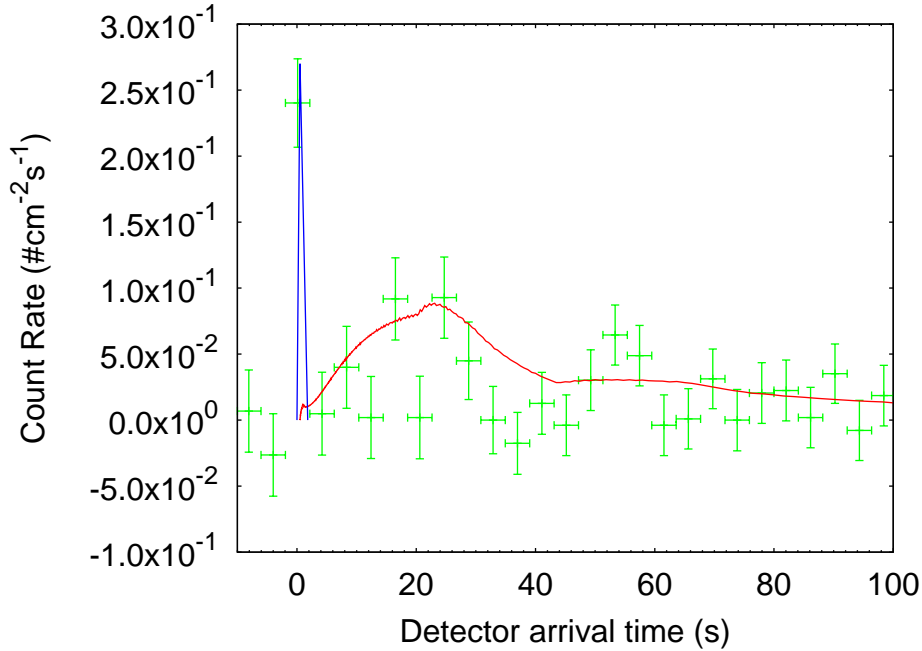}
\caption{The BAT $15$--$50$ keV light curve (green points) compared with the corresponding theoretical extended afterglow light curve (red line). The P-GRB is qualitatively sketched by the blue line. See details in Ref.~\refcite{2010A&A...521A..80C}.}
\label{071227_f2}
\end{figure}

\begin{figure}[t]
\centering
\includegraphics[width=\hsize,clip]{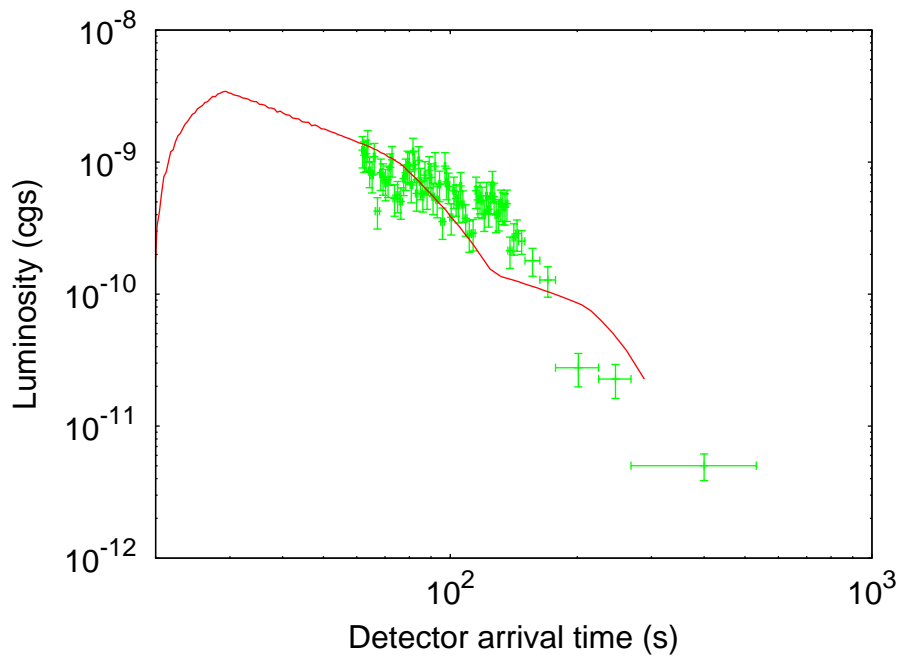}
\caption{The XRT $0.3$--$10$ keV light curve (green points) compared with the corresponding theoretical extended afterglow light curve we obtain (red line). The X-ray data corresponding to the first $60$ seconds are not available since XRT starts to observe only $60$ seconds after the BAT trigger. We stop our analysis at $\gamma \sim 5$ when our relativistic dynamical model can no longer be applied. See details in Ref.~\refcite{2010A&A...521A..80C}.}
\label{071227_f3}
\end{figure}

We have analyzed\cite{2010A&A...521A..80C} the observed light curves of this burst in the $15$--$50$ keV bandpass, corresponding to the lowest band of the gamma-ray emission, detected by the BAT instrument on the Swift satellite, and in the $0.3$--$10$ keV energy band, corresponding to the X-ray component from the XRT instrument. To model the CBM structure, we assume that $n_{cbm}$ is a function only of the radial coordinate, $n_{cbm}=n_{cbm}(r)$ (radial approximation). The CBM is arranged in spherical shells of width $\sim 10^{15}$--$10^{16}$ cm arranged in such a way that the corresponding modulation of the emitted flux closely resembles the observed shape. We assumed that the first short spike-like emission represents the P-GRB and the gamma-ray tail is the peak of the extended afterglow. We therefore began the simulation in such a way that the extended afterglow light curve begins in coincidence with the peak of the P-GRB (about $1$s), as shown in Fig.~\ref{071227_f2}. To reproduce the observational data and the energetics observed for the P-GRB emission ($E_{iso}\sim1.0\times10^{51}$ erg), we require the initial conditions $E_{tot}^{e^\pm}= 5.04\times10^{51}$ erg and $B=2.0\times10^{-4}$. In Figs.~\ref{071227_f2} and \ref{071227_f3} we plot the comparison of the GRB 071227 BAT and XRT data with the theoretical extended afterglow light curves. We obtained a good result in the prompt emission for the $15$--$50$ keV bandpass (see Fig.~\ref{071227_f2}), while for $0.3$--$10$ keV we only succeeded in reproducing the first decaying part of the XRT light curve (see Fig.~\ref{071227_f3}). We assumed this to correspond to the possible onset of the ``plateau'' phase of the extended afterglow \cite{2006ApJ...642..389N}.

\begin{figure}[t]
\centering
\includegraphics[width=\hsize,clip]{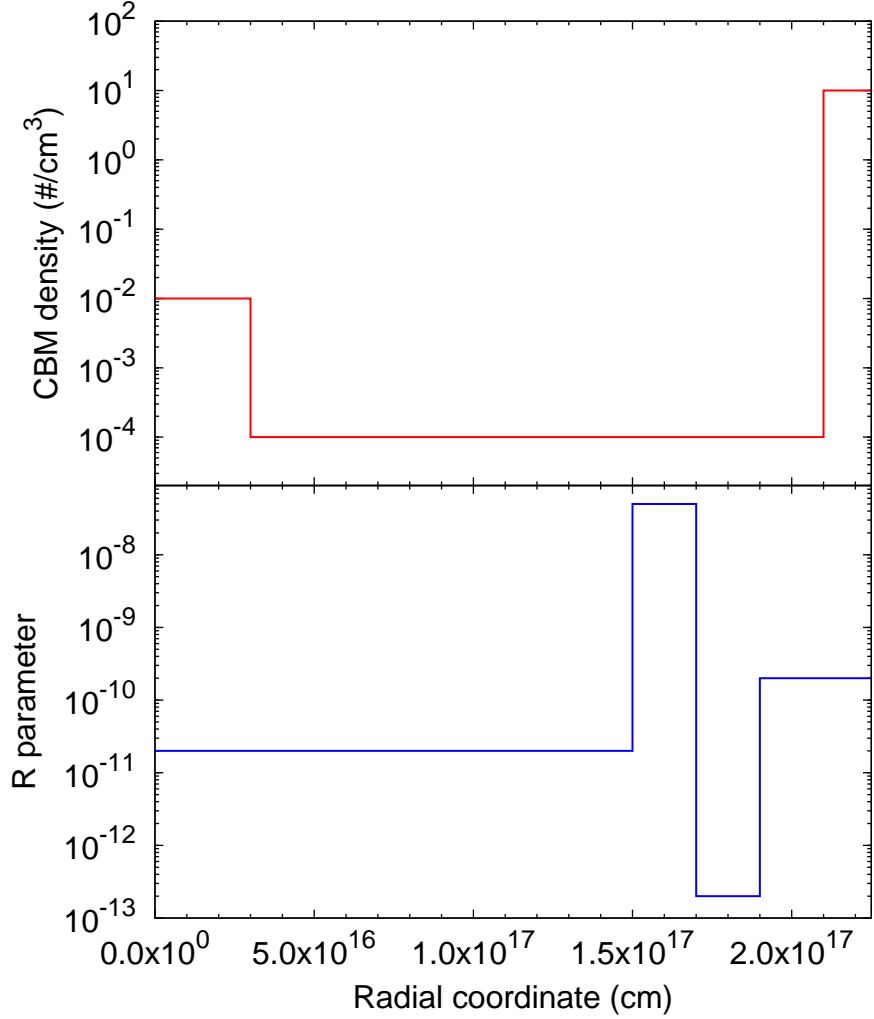}
\caption{The CBM particle number density $n_{cbm}$ (upper panel, red line) and the $\mathcal{R}$ parameter (lower panel, blue line) as functions of the radial coordinate. See details in Ref.~\refcite{2010A&A...521A..80C}.}
\label{071227_f4}
\end{figure}

From our simulation\cite{2010A&A...521A..80C}, the amount of energy stored in the P-GRB is found to be about $20$\% of the total energetics of the explosion. Hence, this burst cannot be a short burst within the fireshell scenario. The baryon loading obtained ($B=2.0\times10^{-4}$) remains in the range of long duration GRBs. This is a very critical value, because it is very close to the crossing point of the plot of the energetics of GRBs as a function of $B$ (see above). This is the lowest baryon loading that we have ever found in our analysis within the fireshell scenario. From our analysis, we found a peculiar result for the average CBM density. We obtained a density of $n_{cbm}=1.0\times10^{-2}$ particles/cm$^3$ at the beginning of the process, later decreasing to $n_{cbm}=1.0\times10^{-4}$ particles/cm$^3$. This low average density, inferred from the analysis, is responsible for the strong deflation of the gamma-ray tail. However, at the radius of about $2.0\times10^{17}$ cm, the density becomes higher and reaches the value of $n_{cbm}=10$ particles/cm$^3$ (the complete profiles of $n_{cbm}$ and $\mathcal{R}$ as functions of the radial coordinate are given in Fig.~\ref{071227_f4}). This is compatible with the observations. The observed X-ray and optical afterglow of GRB 071227 is indeed superimposed on the plane of the host galaxy, at $(15.0\pm2.2)$ kpc from its center \cite{2009A&A...498..711D}. An interesting possibility observed by D. Arnett (private communication) is that this very low density ``cavity'' could be formed in the coalescing phase of a binary formed by a neutron star and a white dwarf. Accurate studies of compact object mergers have shown the distribution of merger locations for different host galaxies \cite{2006ApJ...648.1110B,2010arXiv1005.1068B}. In starburst galaxies, most of the mergers are expected to be found within hosts, while in elliptical galaxies a substantial fraction of mergers take place outside hosts. Spiral galaxies, hosting both young and old stellar populations, represent the intermediate case between the preceding two. This result is therefore compatible with our hypothesis about the binary nature of the progenitor of GRB 071227. Although they did not consider the case of binary systems formed by a neutron star and a white dwarf, the progenitor of GRB 071227 would fit into the tight binary scenario described in Ref.~\refcite{2006ApJ...648.1110B}.

These results clearly imply that GRB 071227 is another example of a disguised short burst\cite{2010A&A...521A..80C}.

\subsection{GRB 071227 within the Amati relation}\label{071227_Amati}

\begin{figure}[t]
\centering
\includegraphics[width=\hsize,clip]{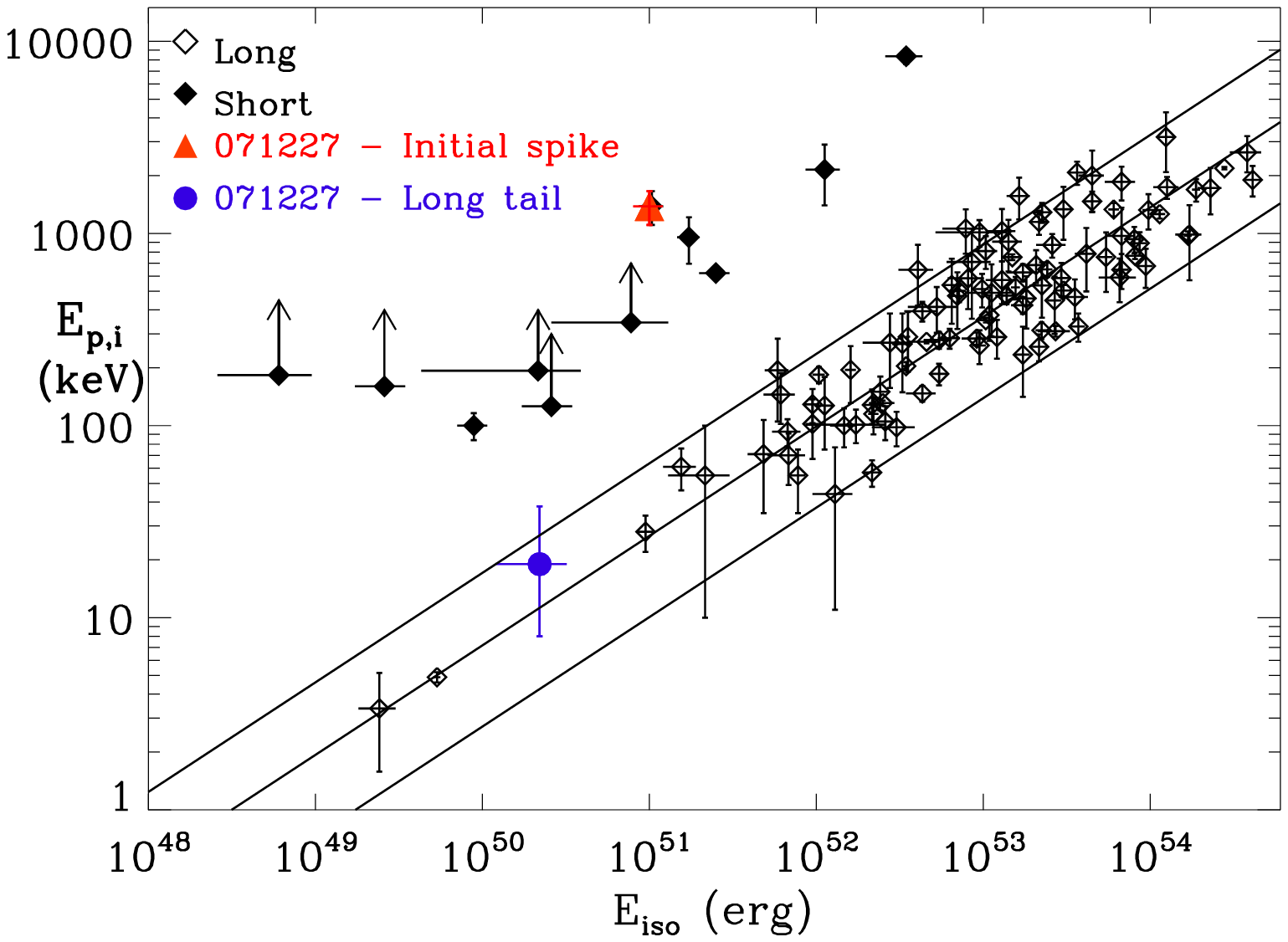}
\caption{Location of the initial short spike and soft long tail of GRB 071227 in the $E_{p,i}$-$E_{iso}$ plane. The data points of long GRBs are from Ref.~\refcite{2008MNRAS.391..577A,2009A&A...508..173A}, the data points and limits of short GRBs are from Ref.~\refcite{2006MNRAS.372..233A,2009A&A...508..173A,2008A&A...491..183P}. The continuous lines show the best-fit power law and the 2$\sigma$ confidence region of the correlation, as determined by Ref.~\refcite{2008MNRAS.391..577A}. See details in Ref.~\refcite{2010A&A...521A..80C}.}
\label{071227_f5}
\end{figure}

We studied the position of GRB 071227 in the $E_{p,i}$-$E_{iso}$ plane\cite{2010A&A...521A..80C}. At the observed redshift, we assumed a ``flat $\Lambda$-CDM model'' with $H_{0}=70$ Km/s/Mpc and $\Omega_\Lambda=0.73$. For the first, hard spike, lasting about $1.8$s, using Konus/WIND data \cite{2007GCN..7155....1G}, and integrating between $1$ and $10\, 000$ keV, we found that $E_{p,i}=(1384 \pm 277)$ keV and $E_{iso}=(1.0\pm0.2)\times10^{51}$ erg. As shown in Fig.~\ref{071227_f5}, this is inconsistent with the $E_{p,i}$-$E_{iso}$ relation, and instead occupies the short-populated region of the plane. For the long tail, lasting about $100$s, we used a band model with $\alpha=-1.5$ and $\beta=-3$, which are typical of soft events. These values are compatible with the low quality statistics of this event. We found that $E_{p,i}=20_{-11}^{+19}$ keV and $E_{iso}=(2.2 \pm 0.1)\times10^{51}$ erg. With these values, the tail of emission is fully consistent with the Amati relation, as for any long GRB (see Fig.~\ref{071227_f5}). This clearly supports our hypothesis about the nature of GRB 071227\cite{2010A&A...521A..80C}.

\section{Analysis of the ``short'' GRB 050509b}\label{appendix_050509b}

I would like to stress, after the considerations we have presented for the last three GRB sources, the consequences for GRB 050509b which was heralded with great prominence as a short GRB on the cover page of \textit{Nature}\cite{2005Natur.437..851G}. This is a particularly exciting system, since the previous work on the disguised short GRBs can be used to properly identify the nature of this intriguing system\cite{2011A&A...529A.130D}.

\begin{figure}[t]
\centering
\includegraphics[width=\hsize,clip]{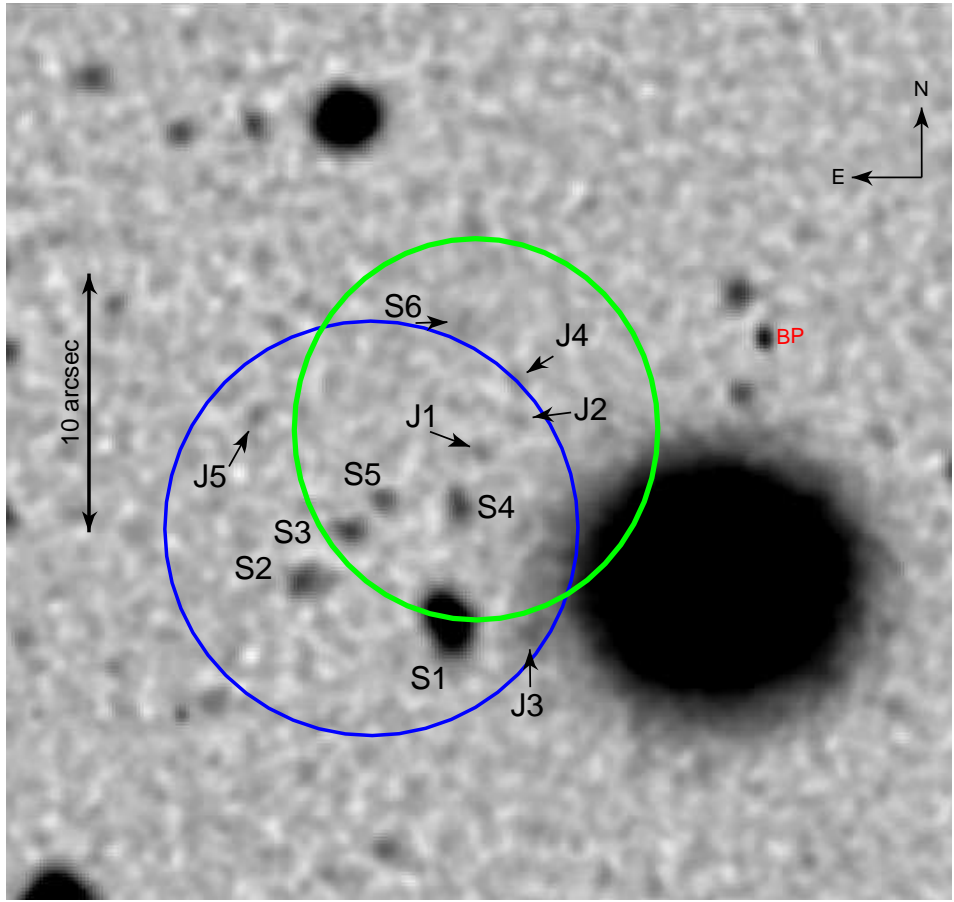}
\caption{Keck LRIS G-band image, zoomed to show the XRT error circle. The larger, blue circle is the revised XRT position from Ref.~\refcite{2005GCN..3395....1R}; the smaller, green circle to the west and north of that is the $2\sigma$ confidence region of the XRT position computed in Ref.~\refcite{2006ApJ...638..354B}. The 11 sources consistent with the Ref.~\refcite{2005GCN..3395....1R} X-ray afterglow localization are labeled in the image. North is up and east is to the left. G1 is  the large galaxy to the west and south of the XRT. Bad pixel locations are denoted with ``BP''. Figure reproduced from Ref.~\refcite{2006ApJ...638..354B} with the kind permission of J. Bloom.}
\label{fbloom}
\end{figure}

The observations of GRB 050509b by BAT and XRT on board the Swift satellite \cite{2004ApJ...611.1005G,2005SSRv..120..165B} have represented a new challenge to the classification of GRBs as long and short, since it is the first ``short'' GRB associated with an afterglow \cite{2005Natur.437..851G}. Its prompt emission observed by BAT lasts 40 milliseconds, but it also has an afterglow in the X-ray band observed by XRT, which begins 100 seconds after the BAT trigger (the time needed to point XRT at the position of the burst) and lasts until $\approx$ 1000 seconds. It is located 40 kpc away from the center of its host galaxy (see Fig.~\ref{fbloom} and Ref.~\refcite{2006ApJ...638..354B}), which is a luminous, non-star-forming elliptical galaxy with redshift $z=0.225$ \cite{2005Natur.437..851G}. Although an extensive observational campaign has been performed using many different instruments, no convincing optical-infrared candidate afterglow nor any trace of any supernova has been found associated with GRB 050509b \cite{2005GCN..3401....1C,2005GCN..3521....1B,2005ApJ...630L.117H,2005A&A...439L..15C,2005GCN..3386....1B,2005GCN..3417....1B,2006ApJ...638..354B}. An upper limit in the $R$-band $18.5$ days after the event onset implies that the peak flux of any underlying supernova should have been $\sim 3$ mag fainter than the one observed for the type Ib/c supernova SN 1998bw associated with GRB 980425, and $2.3$ mag fainter than a typical type Ia supernova (Ref.~\refcite{2005A&A...439L..15C}, see also Ref.~\refcite{2005ApJ...630L.117H}). An upper limit to the brightening caused by a supernova or supernova-like emission has also been established at $8.17$ days after the GRB: $R_c \sim 25.0$ mag \cite{2006ApJ...638..354B}. While some core-collapse supernovae might be as faint as (or fainter than) this limit \cite{2007Natur.449E...1P}, the presence of this supernova in the outskirts of an elliptical galaxy would be truly extraordinary \cite{2005A&A...433..807M,2005PASP..117..773V}.

Unfortunately, we cannot obtain exhaustive observational constraints for this GRB because XRT data are missing in between the first 40 milliseconds and 100 seconds. However, this makes the theoretical work particularly interesting. With G. De Barros \cite{2011A&A...529A.130D} we have inferred from first principles some characteristics of the missing data which are inferred by our model, and consequently we reached an understanding of the source. This is indeed the case, specifically, for the verification of the Amati relation \cite{2002A&A...390...81A,2006MNRAS.372..233A,2009A&A...508..173A} for these sources.

We have proceeded with the identification of the two basic parameters, $B$ and $n_{CBM}$, within two different scenarios \cite{2011A&A...529A.130D}. We first investigate the ``ansatz'' that this GRB is the first example of a ``genuine'' short burst. After disproving this possibility, we show that this GRB is indeed another example of a disguised short burst.

\subsection{Data analysis of GRB 050509b}\label{analysis}

\subsubsection{Scenario 1}\label{anal1}

We first attempted to analyze GRB 050509b under the scenario that assumes it is a ``genuine'' short GRB, namely a GRB in which more than $50\%$ of the total energy is emitted in the P-GRB. This would be the first example of an identified ``genuine'' short GRB.

Within our model, the only consistent solution that does not contradict this assumption leads to the interpretation that all the data belongs to the extended afterglow phase; the BAT data of the prompt emission (see Fig.~2 in Ref.~\refcite{2005Natur.437..851G}) are then the peak of the extended afterglow, and the XRT data represents the decaying phase of the extended afterglow (which in the literature is simply called ``the afterglow,'' see above. 

\begin{figure}[t]
\centering
\includegraphics[width=\hsize,clip]{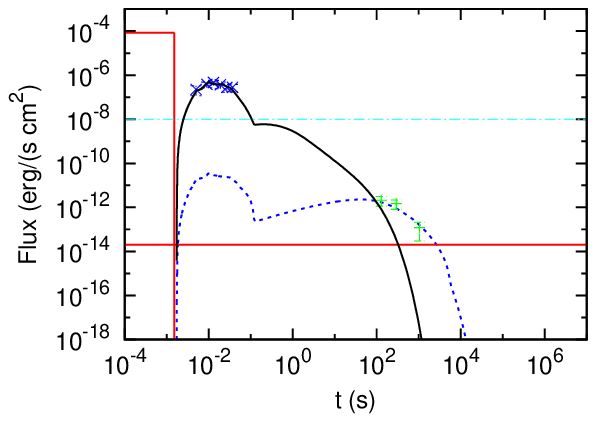}
\caption{Our numerical simulation within scenario 1, assuming that GRB 050509b is a ``genuine'' short GRB, i.e.\ that the P-GRB is energetically predominant over the extended afterglow. The BAT data (crosses) are interpreted as the peak of the extended afterglow. In this case, the predicted P-GRB (solid rectangle) total energy is more than twice that of the extended afterglow. The solid line is the theoretical light curve in the 15--150 keV energy band, and the dashed one is the theoretical light curve in the 0.3--10 keV energy band. The dot-dashed horizontal line represents the BAT threshold and the solid horizontal one represents the XRT threshold. See details in Ref.~\refcite{2011A&A...529A.130D}.}
\label{fig:After}
\end{figure}

In Fig.~\ref{fig:After}, we recall the result of this analysis. We obtained the following set of parameters: $E_{tot}^{e^\pm}=2.8 \times 10^{49}$ erg, $B=1.0 \times 10^{-4}$, and $n_{CBM}=1.0 \times 10^{-3}$ particles/cm$^3$. These parameters would imply, however, that the energy emitted in the P-GRB should be almost $72\%$ of the total value. This P-GRB should have been clearly observable, and has not been detected. Consequently, this scenario is ruled out and we conclude that GRB 050509b cannot be interpreted as a ``genuine'' short GRB.

\subsubsection{Scenario 2}\label{anal2}

\begin{figure}[t]
\centering
\includegraphics[width=\hsize,clip]{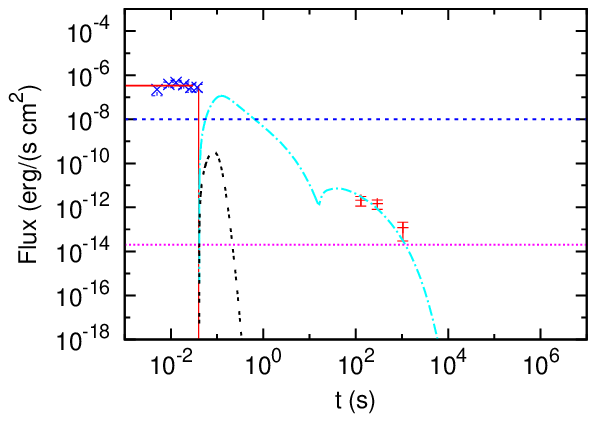}
\caption{Our numerical simulation within scenario 2, assuming that the extended afterglow is energetically predominant over the P-GRB. In this case, the predicted P-GRB (solid rectangle) is less then twice the extended afterglow. We interpret the BAT data (crosses) as the P-GRB and the XRT data as the extended afterglow. The P-GRB has just $28\%$ of the total energy. The double-dashed line is the theoretical light curve in the band $15$--$150$ keV, and the dot-dashed line is the theoretical light curve in the band $0.3$--$10$ keV. The two horizontal lines are from above to below: the BAT threshold and the XRT threshold. See details in Ref.~\refcite{2011A&A...529A.130D}.}
\label{fig:P-GRB}
\end{figure}

We then analyzed GRB 050509b under the alternative scenario that assumes the energy of the extended afterglow is higher than the P-GRB one. Within our model, the only consistent solution leads to the prompt emission observed by BAT (see Fig.~2 in Ref.~\refcite{2005Natur.437..851G}) being interpreted as the P-GRB, and the X-ray decaying afterglow data observed by XRT being interpreted as the extended afterglow.

In Fig.~\ref{fig:P-GRB}, we show the result of this analysis. We obtained the parameters $E_{tot}^{e^\pm}=5.52 \times 10^{48}$ erg, $B=6 \times 10^{-4}$, and an almost constant CBM density $n_{CBM}=1.0 \times 10^{-3}$ particles/cm$^3$. The low value of the number density is justified by the GRB being located $40$ kpc away from the center of the host galaxy (see Fig.~\ref{fbloom})\cite{2006ApJ...638..354B}. The P-GRB has an estimated energy of $E_{P\hbox{\small-}GRB}=28\%E_{tot}^{e^\pm}$, which means that $72\%$ of the energy is released in the extended afterglow. The peak of the extended afterglow, theoretically predicted by our model in Fig.~\ref{fig:P-GRB}, was not observed by BAT, since the energy was below its threshold, and also not observed by XRT, since unfortunately its data collection started only 100 seconds after the BAT trigger. More details regarding the fulfillment of the Amati relation by this source can be found in Ref.~\refcite{2011A&A...529A.130D}.

Following our classification, therefore, due to the values of the baryon loading and of the CBM density, as well as due to the offset with respect to the host galaxy, GRB 050509b is consistent with being another example of a disguised short GRB. This follows the previous identification of GRB 970228 \cite{2007A&A...474L..13B}, GRB 060614 \cite{2009A&A...498..501C} and GRB 071227 \cite{2010A&A...521A..80C}.

\subsubsection{Conclusions on GRB 050509b}

We can then conclude that:
\begin{enumerate}
\item GRB 050509b is not a genuine short burst, but instead a disguised short one, namely a canonical long GRB.
\item In view of the very low density of the CBM and of the clear location outside the host galaxy as well as from the very stringent observational limits on the absence of any supernova, we can safely conclude that the progenitor of this system is a binary merger possibly formed by a neutron star and a white dwarf or possibly two white dwarf components\cite{2011A&A...529A.130D}.
\item Once again, this is an additional example of a GRB not originating from a collapsar.
\end{enumerate}

\section{General conclusions for the disguised short GRBs}\label{sec:disg_concl}

We have just shown how the existence of the disguised short GRBs gives a counterexample both to the collapsar model and to the proposal that binary mergers lead necessarily to short GRBs. It is interesting that the above considerations following from our introduction of the concept of a disguised short burst leads to a third new important conclusion: the possibility that binary mergers can also lead to canonical long duration GRBs. It is clear, in fact, that if a binary system merges within a galaxy, then the average CBM density will be $\sim 1$ particle/cm$^3$ and not the density typical of the halo (see Fig.~\ref{picco_n=1}).

\section{Ongoing progress in understanding GRBs, supernovae and neutron stars}\label{sec:ong_proc}

I would like now to address some ongoing research which promises to make significant progress on:
\begin{enumerate}
\item The analysis of the GRB-supernova connection, which is likely, among others, to lead to the first observations of a newly born neutron star, what I call a ``neo neutron star''.
\item The discovery by the Fermi and Agile satellites of high energy emission from GRBs leading possibly to direct evidence of the ultra high energy baryonic component in GRBs.
\item The identification of the progenitor of the process of gravitational collapse; this has led to introduce the new concept of a ``proto black hole.''
\end{enumerate} 

\subsection{Progress in understanding the GRB-SN association}\label{sec:GRB-SN}

I here recall a most interesting by product of the analysis of GRBs associated with supernovae. We examined GRB 980425 associated with SN 1998bw \cite{2000ApJ...536..778P,2000ApJS..127...59F,2004AdSpR..34.2715R,2005tmgm.meet.2451F,2007ESASP.622..561R
,GraziaMG11}. In this system a peculiar long lasting emission in X-rays was identified and called URCA-1 in view of the fact that its first identification was presented at the Tenth Marcel Grossmann Meeting held in Rio de Janeiro \cite{2005tmgm.meet..369R,2005tmgm.meet.2451F}. We also examined GRB 030329, associated with SN 2003dh, clearly having a similar long lasting feature we named URCA-2 \cite{2004AIPC..727..312B,2005tmgm.meet.2459B,2007ESASP.622..561R}. We have also studied GRB 031203, associated with SN 2003lw, with a long lasting emission defined as URCA-3  \cite{2005ApJ...634L..29B,2007ESASP.622..561R,2008ralc.conf..399R}, see Fig.~\ref{urca123+GRB_full} and Tab.~\ref{tabella}. I would like to review these results and present some general conclusions.

\subsubsection{GRB 980425 / SN 1998bw / URCA-1}\label{980425}

The best fit of the observational data of GRB 980425 we obtained with M.G. Bernardini \cite{2000ApJ...536..778P,2000ApJS..127...59F,2004AdSpR..34.2715R,2005tmgm.meet.2451F,2007ESASP.622..561R
,GraziaMG11} leads to $E_{e^\pm}^{tot}=1.2\times10^{48}$ erg and $B = 7.7\times10^{-3}$. This implies an initial $e^+e^-$ plasma with $N_{e^+e^-} = 3.6\times10^{53}$ and with an initial temperature $T = 1.2$ MeV. After the transparency point, the initial Lorentz gamma factor of the accelerated baryons is $\gamma = 124$. The variability of the luminosity due to the inhomogeneities of the CBM \cite{2002ApJ...581L..19R} is characterized by a density contrast $\delta n / n \sim 10^{-1}$ on a length scale of $\Delta \sim 10^{14}$ cm. We had determined the effective CBM parameters to be: $\langle n_{cbm} \rangle = 2.5\times 10^{-2}$ particle/$cm^3$ and $\langle \mathcal{R} \rangle = 1.2\times 10^{-8}$, where $\mathcal{R}\equiv A_{eff}/A_{vis}$ is the ratio between the effective emitting area $A_{eff}$ of the expanding shell and its entire visible area $A_{vis}$ and takes into account both the effective porosity of the shell and the CBM filamentary structure \cite{2004IJMPD..13..843R,2005IJMPD..14...97R}. Details and fits of the data are presented in Refs.~\refcite{2004AdSpR..34.2715R,2005tmgm.meet.2451F,2007ESASP.622..561R
,GraziaMG11}.

I then review the URCA-1 observations performed by BeppoSAX-NFI in the energy band $2$--$10$ keV \cite{2000ApJ...536..778P}, by XMM-EPIC in the band $0.2$--$10$ keV \cite{2004AdSpR..34.2711P} and by \textit{Chandra} in the band $0.3$--$10$ keV \cite{2004ApJ...608..872K}. The separations between the light curves of GRB 980425 in the $2$--$700$ keV energy band, of SN 1998bw in the optical band \cite{2007astro.ph..2472N,2006Natur.442.1011P}, and of the above mentioned URCA-1 observations are given in Fig.~\ref{urca123+GRB_full}A.

\subsubsection{GRB 030329 / SN 2003dh / URCA-2}\label{030329}

For GRB 030329 \cite{2004AIPC..727..312B,2005tmgm.meet.2459B,2007ESASP.622..561R} we obtained with M.G. Bernardini a total energy $E_{e^\pm}^{tot}=2.12\times10^{52}$ erg and a baryon loading $B = 4.8\times10^{-3}$. This implies an initial $e^+e^-$ plasma with $N_{e^+e^-}=1.1\times10^{57}$ and with an initial temperature $T=2.1$ MeV. After the transparency point, the initial Lorentz gamma factor of the accelerated baryons is $\gamma = 206$. The effective CBM parameters are $\langle n_{cbm} \rangle = 2.0$ particle/$cm^3$ and $\langle \mathcal{R} \rangle = 2.8\times 10^{-9}$, with a density contrast $\delta n / n \sim 10$ on a length scale of $\Delta \sim 10^{14}$ cm. The resulting fit of the observations, both of the prompt phase and of the afterglow, were given in Ref.~\refcite{2005tmgm.meet.2459B,2004AIPC..727..312B}. We compare in Fig.~\ref{urca123+GRB_full}B the light curves of GRB 030329 in the $2$--$400$ keV energy band, of SN 2003dh in the optical band \cite{2007astro.ph..2472N,2006Natur.442.1011P} and of URCA-2 observed by XMM-EPIC in $2$--$10$ keV energy band \cite{2003A&A...409..983T,2004A&A...423..861T}.

\subsubsection{GRB 031203 / SN 2003lw / URCA-3}\label{031203}

\begin{figure}[t]
\centering
\includegraphics[width=\hsize,clip]{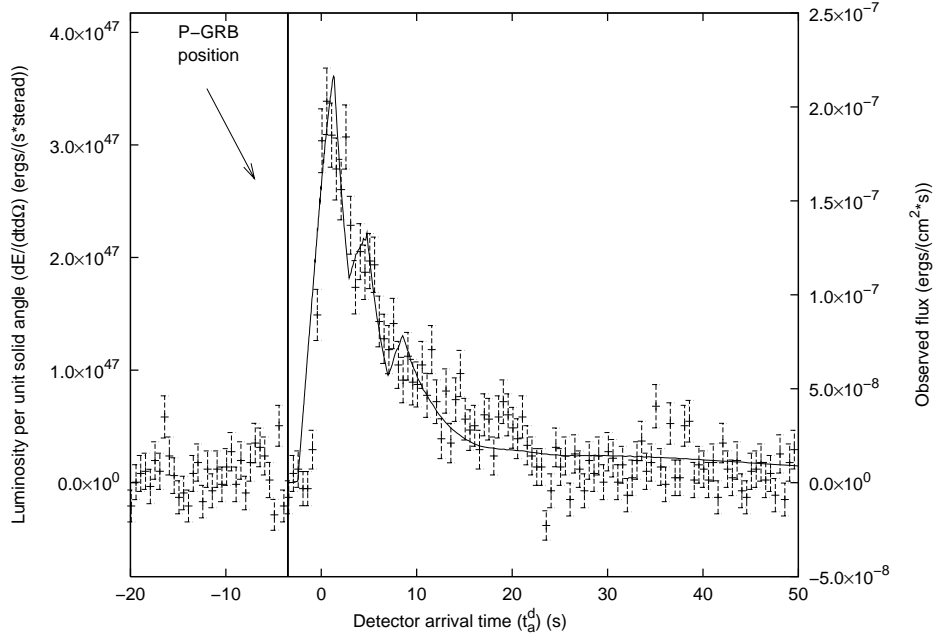}
\caption{Theoretically simulated light curve of the GRB 031203 prompt emission in the $20$--$200$ keV energy band (solid line) is compared with the observed data from Ref.~\refcite{2004Natur.430..646S}. The vertical bold line indicates the time position of P-GRB. See details in Ref.~\refcite{2005ApJ...634L..29B}.}
\label{031203_picco}
\end{figure}

\begin{figure}[t]
\centering
\includegraphics[width=0.8\hsize,clip]{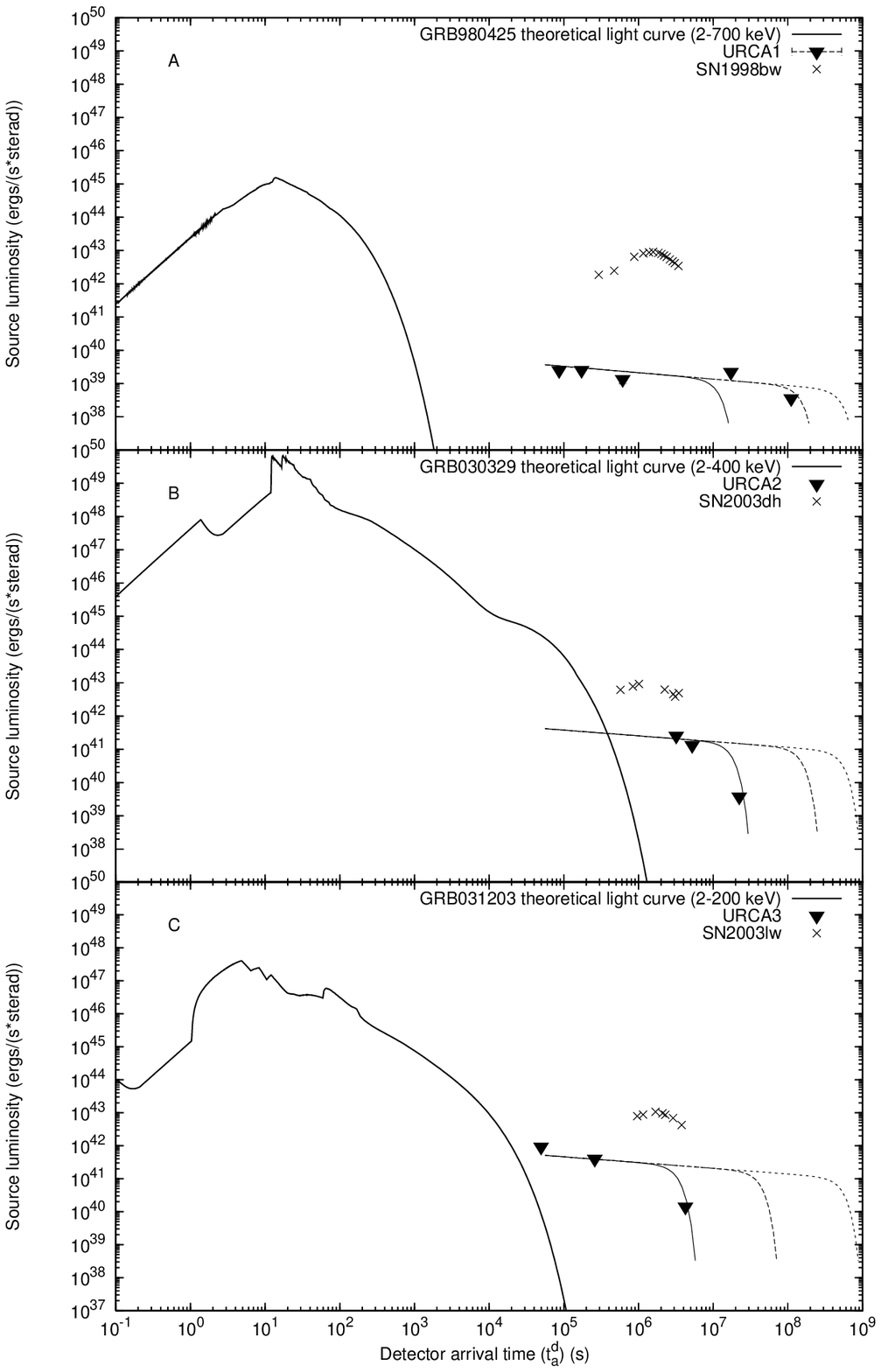}
\caption{Theoretically computed light curves of GRB 980425 in the $2$--$700$ keV band (A)\cite{2004AdSpR..34.2715R,2005tmgm.meet.2451F,GraziaMG11}, of GRB 030329 in the $2$--$400$ keV band (B)\cite{2004AIPC..727..312B,2005tmgm.meet.2459B} and of GRB 031203 in the $2$--$200$ keV band (C)\cite{2005ApJ...634L..29B,2007ESASP.622..561R,2008ralc.conf..399R} are plotted, together with the URCA observational data and qualitative representative curves for their emission, fitted with a power law followed by an exponentially decaying part. The luminosity of the SNe in the $3000$--$24000$ {\AA} range is also plotted \cite{2007astro.ph..2472N,2006Natur.442.1011P}. See details in Ref.~\refcite{2007ESASP.622..561R}.}
\label{urca123+GRB_full}
\end{figure}

The analysis of GRB 031203 with M.G. Bernardini \cite{2005ApJ...634L..29B,2008ralc.conf..399R,2007ESASP.622..561R} leads to a total energy $E_{e^\pm}^{tot}=1.85\times10^{50}$ erg and to a baryon loading $B = 7.4\times10^{-3}$. This implies an initial $e^+e^-$ plasma with $N_{e^+e^-}=3.0\times 10^{55}$ and with an initial temperature $T=1.5$ MeV. After the transparency point, the initial Lorentz gamma factor of the accelerated baryons is $\gamma = 132$. The effective CBM parameters are $\langle n_{cbm} \rangle = 1.6\times 10^{-1}$ particle/$cm^3$ and $\langle \mathcal{R} \rangle = 3.7\times 10^{-9}$, with a density contrast $\delta n / n \sim 10$ on a length scale of $\Delta \sim 10^{15}$ cm. Particularly important in this source was the analysis of the luminosity in selected energy bands as well as of the instantaneous and time integrated spectra (see Figs.~\ref{031203_picco}--\ref{031203_spettro}). We recall here that such instantaneous spectra present a very clear hard-to-soft behavior, and that the corresponding time integrated spectrum is in very good agreement with the observed one. In Fig.~\ref{urca123+GRB_full}C we compare the light curves of GRB 031203 in the $2$--$200$ keV energy band, of SN 2003lw in the optical band \cite{2007astro.ph..2472N,2006Natur.442.1011P} and of URCA-3 observed by XMM-EPIC in the $0.2$--$10$ keV energy band \cite{2004ApJ...605L.101W} and by \textit{Chandra} in the $2$--$10$ keV energy band \cite{2004Natur.430..648S}.

\subsubsection{Discussion}

\begin{table}
\tbl{}
{\begin{tabular}{ccccc}
\toprule
GRB & 980425 & 030329 & 031203 & 060218$^f$\\[6pt]
\colrule
$\begin{array}{c}E_{e^\pm}^{tot}\\ \mathrm{(erg)}\end{array}$ & $1.2\times 10^{48}$ & $2.1\times 10^{52}$ & $1.8\times 10^{50}$ & $1.8\times 10^{50}$\\[6pt]
$B$ & $7.7\times10^{-3}$ & $4.8\times10^{-3}$ & $7.4\times10^{-3}$ & $1.0\times10^{-2}$\\[6pt]
$\gamma_0$ & $124$ & $206$ & $133$ & $99$\\[6pt]
$\begin{array}{c}E_{SN}^{bolom}\\ \mathrm{(erg)^a}\end{array}$ & $2.3\times 10^{49}$ & $1.8\times 10^{49}$ & $3.1\times 10^{49}$ & $9.2\times 10^{48}$\\[6pt]
$\begin{array}{c}E_{SN}^{kin}\\ \mathrm{(erg)^b}\end{array}$ & $1.0\times 10^{52}$ & $8.0\times10^{51}$ & $1.5\times10^{52}$ & $2.0\times10^{51}$\\[6pt]
$\begin{array}{c}E_{URCA}\\ \mathrm{(erg)^c}\end{array}$ & $3\times 10^{48}$ & $3\times10^{49}$ & $2\times10^{49}$ & $?$\\[6pt]
$\displaystyle\frac{E_{e^\pm}^{tot}}{E_{URCA}}$ & $0.4$ & $6\times 10^{2}$ & $8.2$ & $?$\\[6pt]
$\displaystyle\frac{E_{SN}^{kin}}{E_{URCA}}$ & $1.7\times10^{4}$ & $1.2\times10^{3}$ & $3.0\times10^{3}$ & $?$\\[6pt]
$\begin{array}{c}R_{NS}\\ \mathrm{(km)^d}\end{array}$ & $ 8$ & $14$ & $20$ & $?$\\[6pt]
$z^e$ & $0.0085$ & $0.1685$ & $0.105$ & $0.033$\\
\botrule
\end{tabular}}
\begin{tabnote}
a) see Ref.~\refcite{2007ApJ...654..385K}; b) Mazzali, P., private communication at MG11 meeting in Berlin, July 2006; c) evaluated fitting the URCAs with a power law followed by an exponentially decaying part; d) evaluated assuming a mass of the neutron star $M=1.5 M_\odot$ and $T \sim 5$--$7$ keV in the source rest frame; e) see Refs.~\refcite{1998Natur.395..670G,2003GCN..2020....1G,2004ApJ...611..200P,2006ApJ...643L..99M}; f) see Ref.~\refcite{2007A&A...471L..29D}.
\end{tabnote}
\label{tabella}
\end{table}

In Table~\ref{tabella} (see details in Ref.~\refcite{2007ESASP.622..561R}) I have summarized the representative parameters for the above four GRB-SN systems, including the very large kinetic energy observed in all SNe \cite{mazzaliVen}. Some general conclusions on these weak GRBs at low redshift associated with SN Ib/c can be established on the grounds of our analysis:\\
{\bf 1)} These results extend the applicability of our fireshell model to this low-energy GRB class at small cosmological redshift, which now spans a range of energy of six orders of magnitude from $10^{48}$ to $10^{54}$ erg \cite{2003AIPC..668...16R,2004AdSpR..34.2715R,2005tmgm.meet.2459B,2004AIPC..727..312B,2005ApJ...634L..29B,2006ApJ...645L.109R}. Distinctive of this class is the very high value of the baryon loading which in one case (GRB 060218)\cite{2007A&A...471L..29D} is very close to the maximum limit compatible with the dynamical stability of the adiabatic optically thick acceleration phase of the GRBs \cite{2000A&A...359..855R}. Correspondingly, the maximum Lorentz gamma factors are systematically smaller than the ones of the more energetic GRBs at large cosmological distances. This in turn implies the smoothness of the observed light curves in the so-called ``prompt phase.''\\
{\bf 2)} These four GRB sources present a large variability in their total energy: a factor $10^4$ between GRB 980425 and GRB 030329. Remarkably, the SN emission both in their very high kinetic energy and in their bolometric energy appears to be almost constant, respectively $10^{52}$ erg and $10^{49}$ erg.\\
{\bf 3)} The URCAs present a remarkably steady behavior around a ``standard luminosity'' and a typical temporal evolution. We have considered the possibility that the URCAs are related to the SN event: either to dissipative processes in the SN ejecta, or to the formation of a neutron star in the SN explosion \cite{2005tmgm.meet..369R}. As an example, we have given an estimate of the thermal emission from the neutron star surface and of the corresponding radius for URCA-1, URCA-2 and URCA-3 (see Table~\ref{tabella}). The different spectral properties of the GRBs and the URCAs have been pointed out \cite{2000ApJ...536..778P}. It will certainly be interesting to compare and contrast the spectra of all URCAs in order to show the expected analogies among them. Observations of the URCA sources on time scales of $0.1$--$10$ seconds would be highly desirable in presence of a strong enough X-ray signal.

The investigation of the thermal evolution of neutron stars is a powerful tool to probe the inner composition of these objects. The cooling of neutron stars has been investigated by many authors, where many different microscopic models were assumed \cite{Schaab1996,Page2004,Page2006,Page2009,Blaschke2000, Grigorian2005,Blaschke2006,Negreiros2010}. Most of the research on the thermal evolution of compact stars focus on objects with ages greater than $10$--$100$ years, which is comprehensible if one considers that the thermal data currently available to us is for pulsars with estimated ages of or greater than $330$ years \cite{Page2004,Page2009}. The identification of the URCAs with young neutron stars, which we have recently called ``neo neutron stars'' (Negreiros et al., in preparation), will allow the study of the thermal evolution of young neutron stars in the little explored time window that spans from ages greater than 1 minute (just after the proto-neutron star regime \cite{Prakash2001}) to ages $\leq 10$--$100$ years, when the neutron star becomes isothermal (see Ref.~\refcite{Gnedin2001} for details). Relevant also are the observations of the isolated Type Ic Supernova SN 1994I \cite{2002ApJ...573L..27I} and SN 2002ap \cite{2004A&A...413..107S} which show late emissions similar to the ones observed in URCA-1, URCA-2, and URCA-3. In our recent work with R. Negreiros we propose a revision of the boundary conditions usually employed in the thermal cooling theory of neutron stars, in order to match the proper conditions of the atmosphere at young ages. We also discuss the importance of the thermal processes taking place in the crust, which also have important effects on the initial stages of thermal evolution.

\begin{figure}[t]
\centering
\includegraphics[width=\hsize,clip]{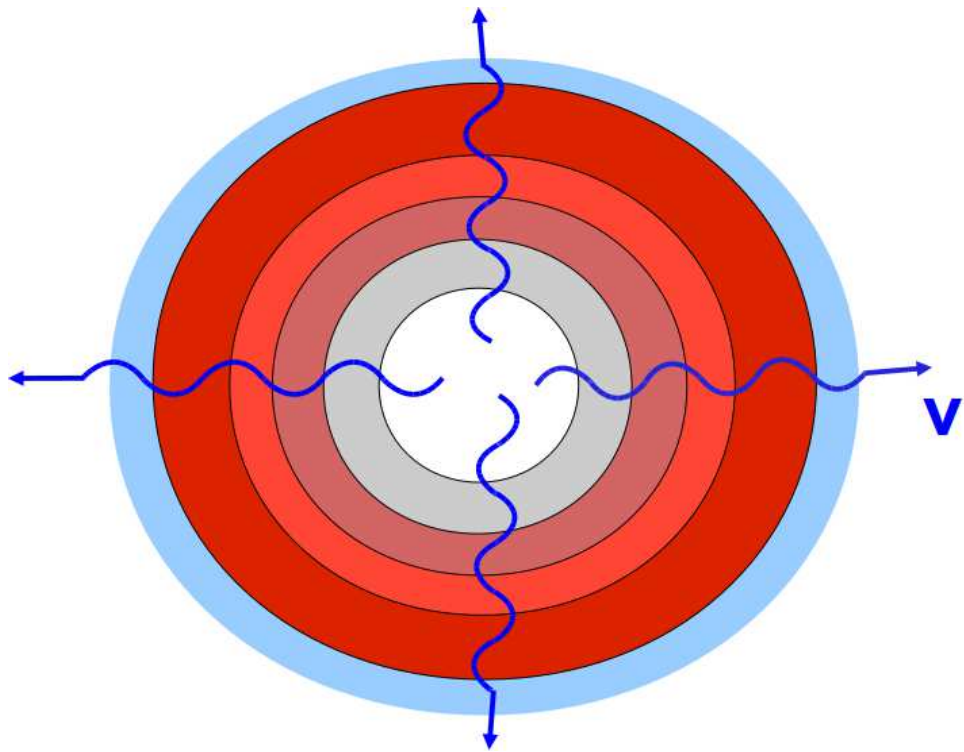}
\caption{Schematic representation of the cooling of a young neutron star. Due to stronger neutrino emissivities, the core of the star cools down more quickly than the crust, causing the star to cool inside out. Darker and lighter areas represent higher and lower temperatures respectively. Details in Negreiros et al., in preparation.}
\label{fig:cool_scheme} 
\end{figure}

There are three important ingredients that govern the thermal evolution of a compact star, these are: 1) the microscopic input, which accounts for the neutrino emissivities, specific heat  and thermal conductivity; 2) the macroscopic structure of the star, namely its mass, radius, pressure profile, crust size, etc.; and 3) the boundary condition at the surface of the star, which provides a relationship between the mantle temperature and that of the atmosphere, the latter being what we ultimately observe. These ingredients have been extensively studied, and a comprehensive review can be found in Ref.~\refcite{Page2006}. As discussed in Ref.~\refcite{Gnedin2001}, during the initial stages of thermal evolution (ages $\leq 10$--$100$ years), the core and the crust of the neutron star are thermally decoupled. This is due to the fact that the high density core is emitting neutrinos at a much higher rate than the crust, which causes it to cool down more quickly. This effectively means that initially the neutron star is cooling ``inside out,'' with the core colder than the outer layers. This scenario is schematically depicted in Fig.~\ref{fig:cool_scheme}. Details may be found in Negreiros et al., in preparation.

All this suggests the exciting possibility that we are for the first time observing nascent, hot neutron stars. This possibility warrants further study in order to obtain a more concrete picture of the thermal evolution of neo-neutron stars. We have also  encouraged dedicated observations of isolated SN in view of the similarities between URCA-1--URCA-3 and the Type Ic Supernova SN 1994I \cite{2002ApJ...573L..27I} and SN 2002ap \cite{2004A&A...413..107S}.

\subsection{The high-energy emission in the fireshell scenario}\label{sec:080916c}

The launch of the AGILE and Fermi satellites opened a new high energy window for GRB science. Thanks to the Large Area Telescope (LAT) \cite{2009ApJ...697.1071A} on-board the Fermi spacecraft, it is possible to detect very high energy photons ($> 100$ MeV) from GRBs. The first results obtained from Fermi is that not all GRBs have this high energy component, but just a fraction of them: currently LAT has detected energetic photons from  $\sim 20$ GRBs among the whole set of $\sim 480$ GRBs detected by the Fermi Gamma-Ray Burst Monitor (GBM) \cite{2009ApJ...702..791M}. However, this last evidence has already been noticed by the EGRET detector on board the Compton Gamma-Ray Observer satellite, which detected high energy photons only from a few GRBs.

\begin{figure}[t]
\centering
\includegraphics[width=\hsize,clip]{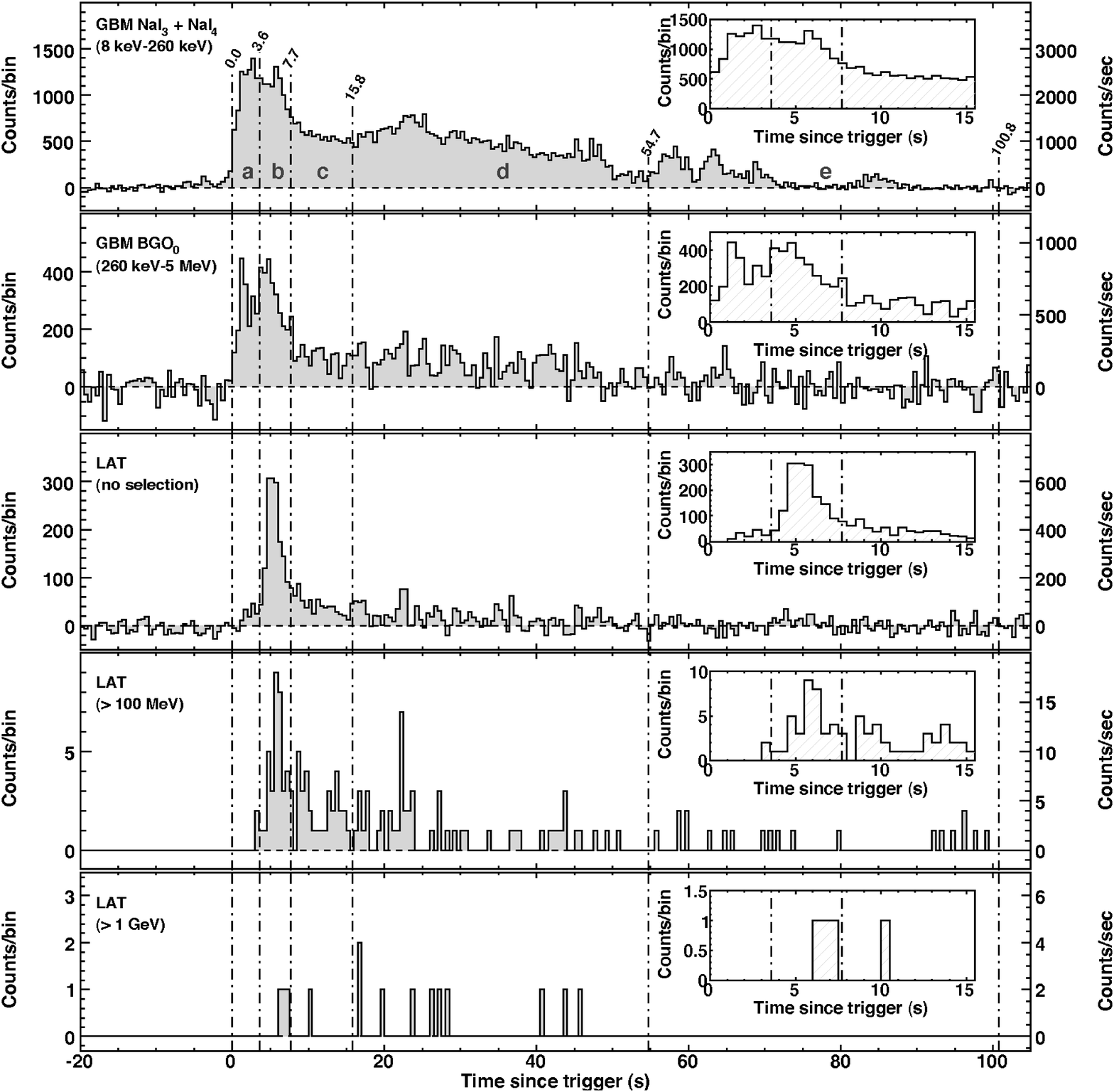}
\caption{Fermi GBM and LAT light curves of GRB 080916C. Details in Ref.~\refcite{2009Sci...323.1688A}. Note the P-GRB emission starting before the LAT data.}
\label{fig_080916C}
\end{figure}

The other very interesting discovery of the Fermi LAT was that this high energy component is delayed compared to the emission observed by the GBM detector. Moreover when the GBM signal fades in the canonical way, the high energy one is prolonged and can last for hundreds of seconds. The case of GRB 080916c \cite{2009Sci...323.1688A} is remarkable (see Fig.~\ref{fig_080916C}): it is currently the most energetic GRB to date, with an estimated isotropic energy of $E_{iso}$ = 8.8 $\times$ 10$^{54}$ erg. 
A significant contribution to this value was provided by the high-energy gamma-rays ($> 100$ MeV) emitted by this GRB. Moreover this high energy emission started $\sim$ 3.6 s after the GBM trigger and persisted up to 1400 s after the GBM trigger. The spectral energy distribution of these high energy photons is best represented by a power-law function whose photon index was found to be the same as the harder component of the GBM Band spectrum. This observed continuity does not appear to be an universal property of all sources presenting a high energy emission.

The main contribution of the fireshell model applied to this family of sources is:\\
{\bf 1)} That the initial phase of GRBs lacking the high energy component represents the P-GRB.\\
{\bf 2)} The corresponding determination of the Lorentz gamma factor of the source, and the strong correlation between the existence of the high energy component and very large values of this factor ($\gamma > 10^3$).\\
{\bf 3)} The possible connection of the high energy component to the ultrarelativistic baryons interacting, via proton-proton collision, with a dense CBM following the classical work of Fermi (see e.g. Ref.~\refcite{FermiAstro}).

I briefly recall some preliminary results on the parameters of GRB 080916c within the fireshell model. The best fit correspond to a value $E_{dya} = 8.8 \times 10^{54}$ erg. From the observed flux, we estimated the energy emitted in the P-GRB, $E_{P\hbox{\small-}GRB} = 2.8 \times 10^{53}$ erg, which, in view of 
Fig.~ \ref{ftemp-fgamma-bcross}, gives a value of $B = 3.3 \times 10^{-4}$. The corresponding value of the average CBM density is $\langle n\rangle = 2.2 \times 10^{-3}$ particles/cm$^{-3}$. From these values, we infer a very high value for the Lorentz gamma factor at the transparency: $\gamma_\circ = 3.17 \times 10^3$.

\begin{figure}[t]
\centering
\includegraphics[width=0.49\hsize,clip]{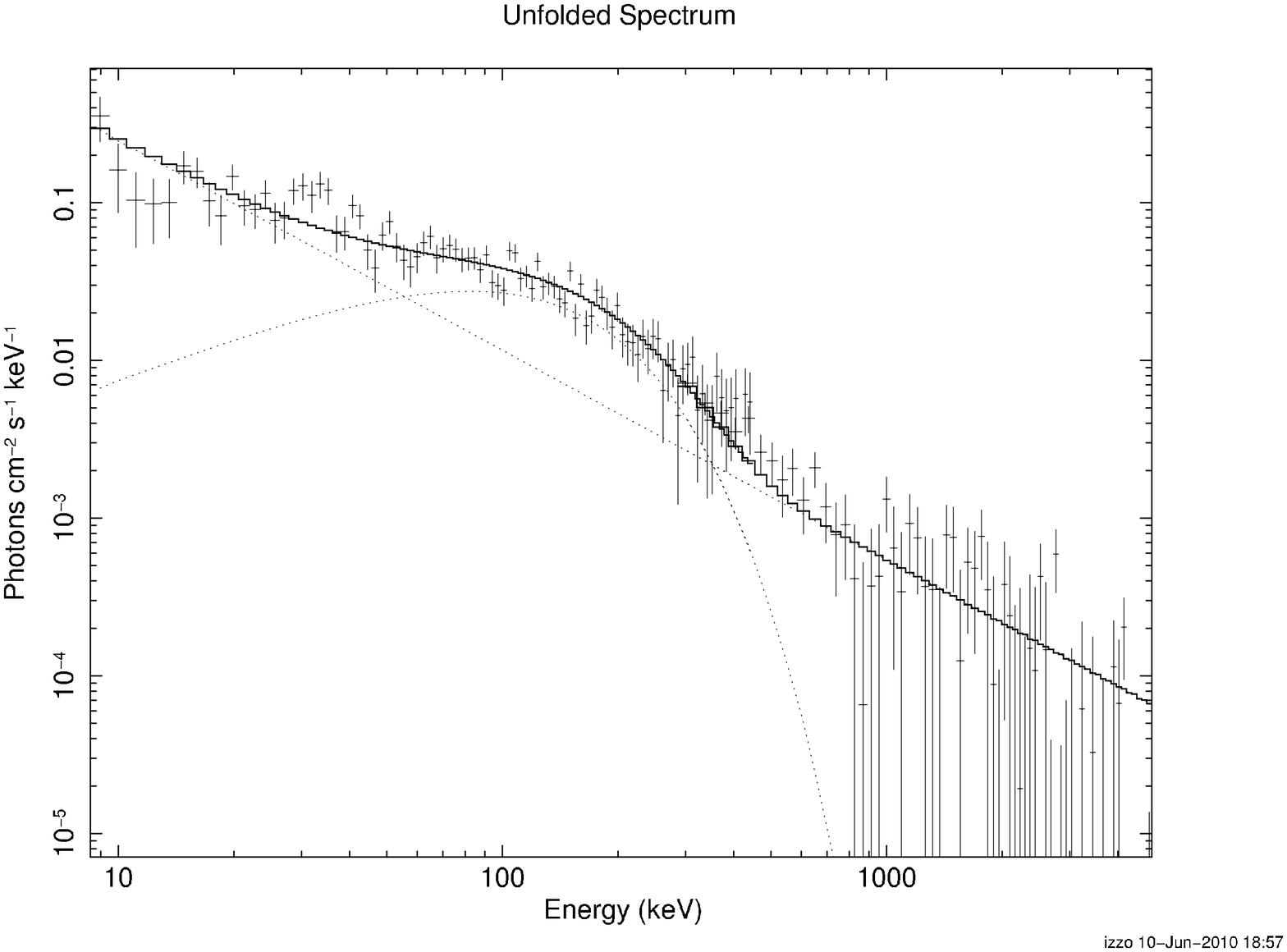}
\includegraphics[width=0.49\hsize,clip]{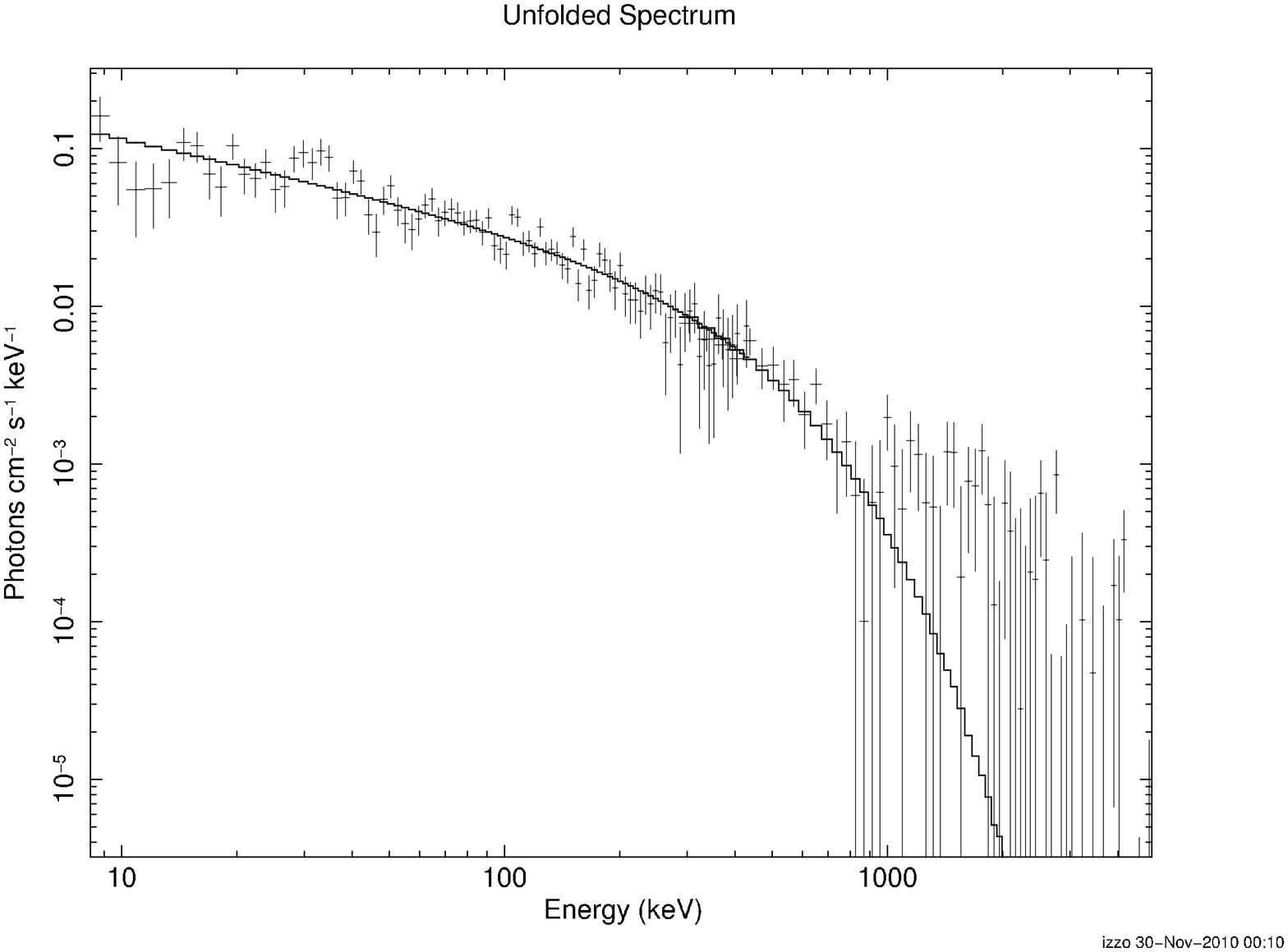}
\caption{Spectral fits of the Fermi GBM and BGO count spectra of the P-GRBs of GRB 080916C in the range (8 keV--5 MeV). We consider in these cases a blackbody spectral model with an extra power-law component, a classical band function and a modified black body spectral model respectively.} 
\label{fig:no4}
\end{figure}

The identification of the first pulse in GRB 080916c with the P-GRB is further supported by its duration: it is comparable with what is expected for a P-GRB, e.g.\ on the order of few seconds \cite{1999A&A...350..334R}. Moreover, a time-resolved spectral analysis of the P-GRB appears to be consistent either with a classical band model, represented by two broken power-laws smoothly connected at a given energy $E_0$ or with a black-body model with an extra power-law component (see Fig.~\ref{fig:no4}).

All the above represents a first step toward the understanding of these GRBs within the fireshell model and opens further studies on the origin of the high energy component by refinements of the energy release process in the comoving frame of the baryon component interacting with the CBM.

\subsection{Possible multiple components in a GRB: the case of GRB 090618}\label{sec:mult_coll}

I have emphasized how the study of GRBs encompasses an unprecedented number of fundamental physics topics, each one probing untested regimes. The physics of gravitational collapse leading to the formation of a black hole needs a new self-consistent treatment in the ultra-relativistic regimes of strong, weak, electromagnetic and gravitational interactions as pointed out in Sec.~\ref{sec:loc_glob}. The outcome of the collapse leads to the formation of an optically thick $e^+ e^-$-baryon plasma, self-accelerating to Lorentz gamma factors in the range $200 < \gamma < 3000$, never reached before in macroscopic objects. These novel ultrarelativistic plasma regimes have been summarized in Sec.~\ref{sec:therm}. The transparency condition of the $e^+ e^-$-baryon plasma, already observed in the cosmological background radiation, gains in GRBs the additional component of $\sim 10^{56-57}$ baryons in the TeV region (see Sec.~\ref{sec_canonical}). The collision of these baryons with the CBM clouds, characterized by dimensions of $10^{15-16}$ cm, leads to high energy collision regimes never reached before in dimensions and energetics (see Sec.~\ref{sec:080916c}). Before concluding this review, I would like to outline how the progress in the last decades of optical, X and gamma-ray observatories has made possible the observation of GRB 090618 offering an unprecedented possibility for testing the theoretical models of the above objects with crucial observations and linking for the first time the observation of GRBs to their immediate progenitor.

\begin{figure}[t]
\includegraphics[width=\hsize,clip]{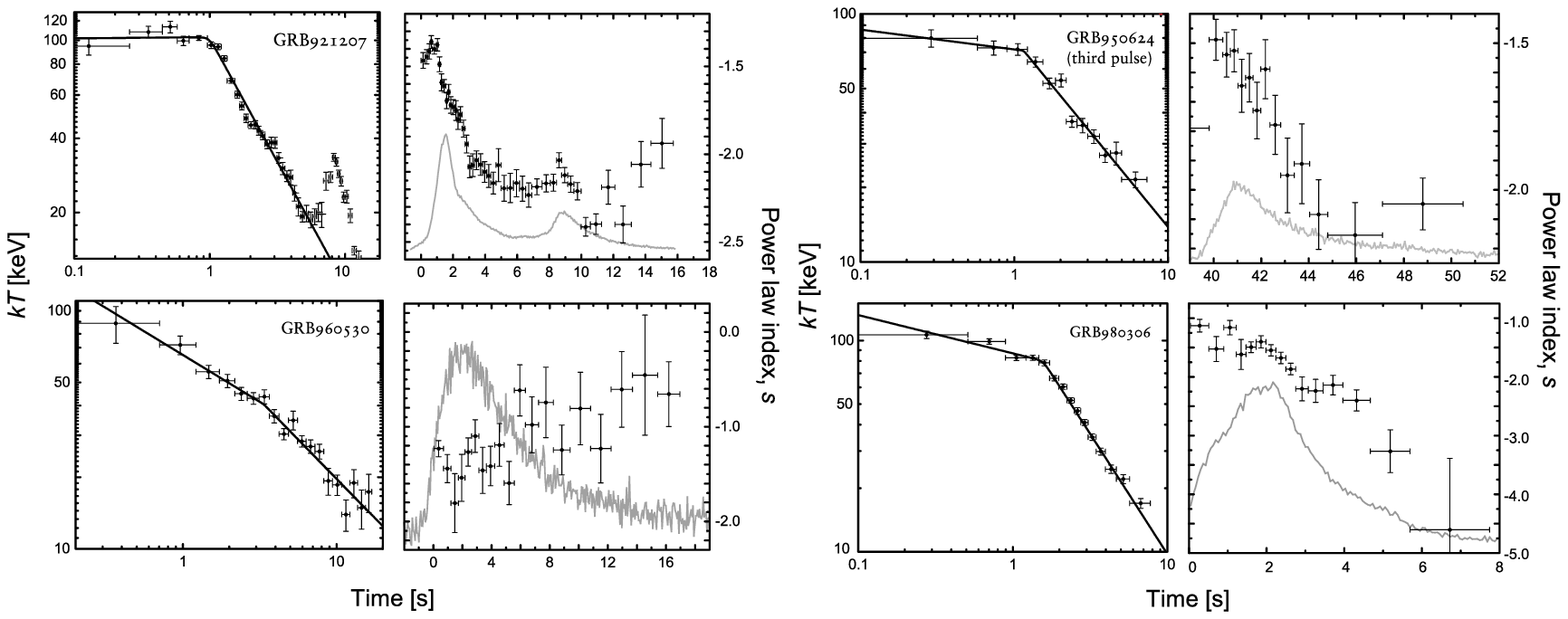}
\caption{Evolution of $kT$ (left) and power-law index $s$ (right). The grey curves are the detector count light-curves, with arbitrary normalization. The time is measured from the trigger, except in the plot of $kT$ for GRB950624, where time is measured from the beginning of that pulse. See details in Ref.~\refcite{2005ApJ...625L..95R}.}
 \label{fig:ryde}
\end{figure}

GRB 090618 is indeed one of the closest ($z = 0.54$) and most energetic ($E_{iso} = 2.70 \times 10^{53}$ erg) GRBs, observed under ideal conditions by the satellites Fermi, Swift, Konus-WIND, AGILE, RT-2 and Suzaku, as well as from on-ground optical observatories. It therefore represents  an ideal case to test our fireshell model. We have analyzed the emission from this GRB with special attention to the thermal and power-law components highlighted by F. Ryde and collaborators. It is interesting in fact that independently of the development of new missions, the BATSE data have continued to attract full scientific interest, even so many years after the end of the mission in 2000. Based on the BATSE data, important inferences for the spectra of the early emission of the GRB have been made in Refs.~\refcite{2004ApJ...614..827R} and \refcite{2006ApJ...652.1400R}. They have convincingly demonstrated that the spectral feature composed of a black body and a power-law behavior together plays an important role in selected episodes in the early part of the GRB emission. They have also shown in some cases a power-law variation of the thermal component as a function of time, following a $t^{-\alpha}$ behavior (see Fig.~\ref{fig:ryde}).

\begin{figure}[t]
\includegraphics[width=\hsize,clip]{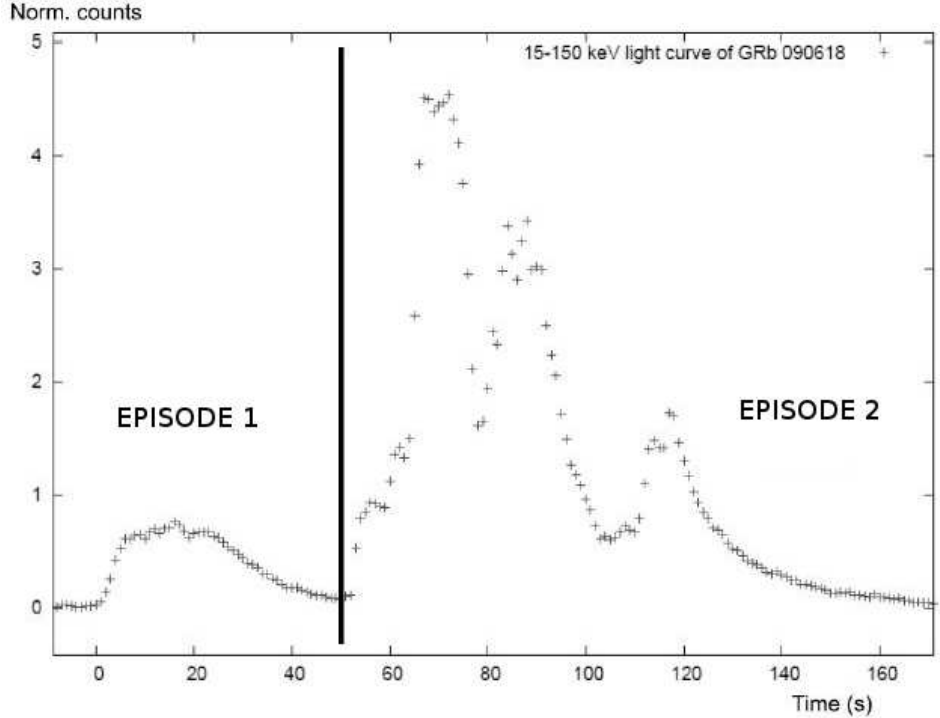}
\caption{The two episode nature of GRB 090618. Details in Ref.~\refcite{COSPAR}.}
\label{fig:cospar}
\end{figure}

\begin{figure}[t]
\centering
\includegraphics[width=\hsize,clip]{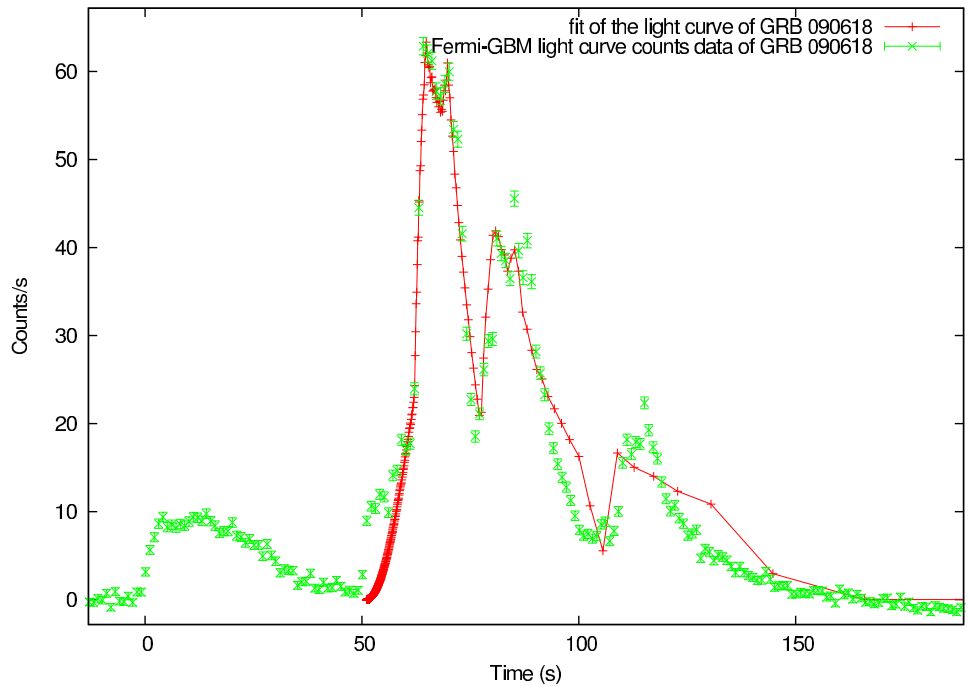}
\caption{Simulated light curve of the extended afterglow of GRB 090618. See details in Ref.~\refcite{TEXAS}}
\label{fig:firesh}
\end{figure}

We have proceeded to the determination of the fundamental parameters of GRB 090618 in the fireshell model, including the identification of the P-GRB, the dyadosphere energy, the baryon loading, the density and porosity of the CBM. We have clearly shown the existence of two different components in GRB 090618: Episode 1 and Episode 2 (see Fig.~\ref{fig:cospar}). We have confirmed that Episode 1, lasting $50$ s, cannot be considered either a GRB or a part of a GRB \cite{TEXAS}. Much like the cases considered by F. Ryde, it shows a very clear thermal component with temperature evolving following a precise power-law behavior between $kT = 54$ keV and $kT = 12$ keV (Izzo et al., in preparation). We have proposed that the radius is increasing between $\sim 12000$ km and $70000$ km, with an estimate mass of $\sim 15 M_{\odot}$. The second component is a canonical long GRB with a Lorentz gamma factor at transparency of $\gamma = 490$, a temperature at transparency of $25.48$ keV and with characteristic size of the clouds, generating the observed luminosity variations, of $R_{cl}$ $\sim$ 10$^{15}$ cm (see Fig.~\ref{fig:firesh}). Episode 1, in our interpretation, is related to the progenitor of the collapsing bare core leading to the black hole formation: what we have defined a ``proto-black hole'' of $\sim 15M_\odot$. For the first time we may be witnessing the process of formation of the black hole from the phases just preceding the gravitational collapse all the way up to the GRB emission.

Of particular interest in this respect are the first two-dimensional hydrodynamical simulations of the progenitor evolution of a $23 M_{\odot}$ star close to core-collapse, leading to a naked core, as shown in the recent work of Arnett and Meakin \cite{2011ApJ...733...78A}. In that work, pronounced asymmetries and strong dynamical interactions between burning shells are seen: the dynamical behavior proceeds to large amplitudes, enlarging deviations from the spherical symmetry in the burning shells. It is of clear interest to find a possible connection between the proto-black hole concept that we have introduced and the Arnett and Meakin results: to compare the radius, the temperature and the dynamics of the core that we have found with the naked core obtained by Arnett and Meakin from the thermonuclear evolution of the progenitor star. There is also the interesting possibility that the CBM clouds observed in GRBs, as well as the baryon loading captured by the $e^+ e^-$-plasma dynamics, are related to the vigorous dynamics in the violent activity of matter ejected in the evolution of the original massive star prior to the formation of the naked core.

\section{Conclusions}

I have summarized the vast scenario made possible by the unprecedented technological possibilities of observing SNe, GRBs, GRBs associated with SNe, all the way to neo-neutron stars and proto-black holes. The corresponding theoretical work is rapidly expanding over new regimes of ultrarelativistic field theories and encompassing all fundamental interactions. A novel unified approach to nuclear physics, neutron stars, white dwarfs and gravitationally collapsed objects is rapidly emerging. All this offers unprecedented possibilities to enlarge our knowledge of fundamental physics and relativistic astrophysics, as well as to reach an understanding of yet unexplained phenomena of fundamental physics.

\section*{Acknowledgements}

It gives me a great pleasure to thank some of my former and current students in the preparation of this manuscript, including M.G. Bernardini, C.L. Bianco (especially for the editing of this manuscript), L. Caito, C. Cherubini, D. Christodoulou, M.G. Dainotti, T. Damour, G. De Barros, S. Filippi, F. Fraschetti, R. Guida, L. Izzo, R.T. Jantzen (also especially for the editing of this manuscript), M. Lattanzi, B. Patricelli, A.V. Penacchioni, L.J. Rangel Lemos, M. Rotondo, J.A. Rueda Hernandez, I Siutsou, G.V. Vereshchagin, L. Vitagliano.

Special thanks for the many discussion and collaboration with A.G. Aksenov, L. Amati, W.D. Arnett, S.K. Chakrabarti, P. Chardonnet, R. Giacconi, H. Gursky, A. Nandi, R. Negreiros, G. Preparata, J. Salmonson, J. Wilson, S.-S. Xue.

\end{document}